\documentstyle[preprint,prc,aps,eqsecnum,epsfig]{revtex}
\tighten
\begin{document}
\draft
\title{The high-precision, charge-dependent Bonn nucleon-nucleon
potential (CD-Bonn)}
\author{R. Machleidt~\footnote{E-mail address: machleid@uidaho.edu}}
\address{Department of Physics, University of Idaho, Moscow,
Idaho 83844, U. S. A.}

\date{\today}

\maketitle

\tableofcontents

\newpage

\begin{abstract}
We present a charge-dependent nucleon-nucleon ($NN$) potential
that fits the world proton-proton data below 350 MeV available in the
year of 2000 with a $\chi^2$ per datum of 1.01 for 2932 data 
and the corresponding neutron-proton data 
with $\chi^2$/datum $= 1.02$ for 3058 data. 
This reproduction of the $NN$ data is more accurate than by any
phase-shift analysis and any other $NN$ potential. 
The charge-dependence of the present potential (that has been dubbed
`CD-Bonn') is based upon the predictions by the Bonn Full Model
for charge-symmetry and charge-independence breaking in all partial
waves with $J\leq 4$. 
The potential is represented in terms of the covariant
Feynman amplitudes for one-boson exchange which are nonlocal.
Therefore, the off-shell behavior of the CD-Bonn potential
differs in a characteristic and well-founded way from 
commonly used local potentials and leads to larger binding
energies in nuclear few- and many-body systems, where underbinding
is a persistent problem.
\end{abstract}


\newpage


\section{Introduction}

In the 1970's and 80's,
a comprehensive fieldtheoretic meson-exchange model for the 
nucleon-nucleon ($NN$) interaction was developed at the 
University of Bonn. 
The final version, published in 1987,
has become known as the Bonn Full Model~\cite{MHE87}.
For a pedagogical review see Ref.~\cite{Mac89}.

In the language of fieldtheoretic perturbation theory, the lowest order
contributions to the $NN$ interaction generated by mesons
are the one-boson exchange diagrams.
Furthermore, there are many irreducible multi-meson exchanges. 
The diagrams of $2\pi$ exchange are most prominent since they 
provide the intermediate-range attraction of the nuclear force.
However, once explicit diagrams of $2\pi$ exchange (with intermediate
$\Delta$ isobars) are used in
a model, then it is vital to also include  the corresponding
diagrams of $\pi\rho$ exchange. There are characteristic (partial)
cancellations between the two groups of diagrams, which
are crucial for a quantitative reproduction of the $NN$ data.
Moreover, the Bonn model contains additional classes of
irreducible $3\pi$ and $4\pi$ exchanges which are
important conceptually rather than quantitatively, since they
appear to indicate convergence of the diagrammatic
expansion chosen by the Bonn group~\cite{MHE87}.

The development of the Bonn Full Model was necessary 
to test reliably the meson-exchange concept for nuclear forces
and to assess systematically the range of its validity.
Thus, the model represents a benchmark for any alternative attempt
(based, e.~g., on quark models, chiral perturbation theory, or other ideas)
to explain the nuclear force.

Due to its comprehensive character, the Bonn model provides a sound basis for 
addressing many important issues.
One of them is the charge dependence of nuclear forces.
The charge-symmetry-breaking (CSB) of the $NN$ interaction 
due to nucleon mass splitting
has been investigated in Ref.~\cite{LM98a}.
It turns out that considerable CSB is generated by
the $2\pi$-exchange contribution to the $NN$ interaction and the
$\pi\rho$ diagrams such that
the CSB difference in the singlet scattering lengths 
can be fully explained from nucleon mass splitting.
Also, noticeable CSB effects occur in $P$ and $D$ waves.
Empirical evidence for CSB is seen in the
the Nolen-Schiffer (NS) anomaly~\cite{NS69}
regarding the energies of neighboring mirror nuclei. 
A recent study~\cite{MPM99} has shown that
the CSB in partial waves with $L>0$
as derived from the Bonn model
is crucial for a quantitative explanation of the NS anomaly.

The charge-independence breaking (CIB) of the $NN$ interaction 
has also been investigated~\cite{LM98b}.
Pion mass splitting is the major cause, and it is well-known
that the one-pion-exchange (OPE) explains about 50\% of the CIB difference
in the singlet scattering lengths. 
However, 
the $2\pi$-exchange model and
the diagrams of three and four irreducible pion exchanges
contribute additional CIB which can amount
up to 50\% of the OPE CIB-contribution, in $S$, $P$, and $D$ waves.
This effect is not negligible.

Other important issues related to the nuclear force are
relativistic effects, medium effects, and many-body forces to be expected in 
the nuclear many-body problem. 
The medium effects on the nuclear force when inserted into nuclear matter have 
been calculated thoroughly. 
A large repulsive contribution to these medium effects comes from
intermediate $\Delta$ isobar states
which also give
rise to energy-dependence. 
On the other hand, isobars 
create many-body forces that are attractive. Thus, large 
cancellations between these two classes of many-body forces/effects occur
and it has been shown that
the net contribution is very small~\cite{Mac99a}.
Relativistic effects, however, may play an important role
in the nuclear many-body problem~\cite{Mac89}.

Multi-meson exchange diagrams are very involved.
Moreover, contributions of this kind are, in general,
energy dependent. This would make the $NN$ potenital---defined
as the sum of irreducible diagrams---energy dependent.
A $NN$ potential that depends on energy creates conceptual and practical
problems when applied in nuclear many-body systems.
For a large class of nuclear structure problems, these complications
are without merit.

For these reasons, already early in the history of the meson theory
of nuclear forces, the so-called one-boson-exchange (OBE) model
was designed which---by definition---includes only single-meson exchanges
(which can be represented in an energy-independent way). 
Usually, the model includes all mesons with masses below the nucleon mass,
i.~e., $\pi$, $\eta$, 
$\rho(770)$, and $\omega(782)$~\cite{PDG98}.
In addition, the OBE model typically introduces a scalar, isoscalar
boson---commonly denoted by $\sigma$ (or $\epsilon$).
Based upon what we discussed above concerning multi-meson
exchange contributions, it is clear now that this $\sigma$
must approximate more than just the $2\pi$ exchange.
In particular, it has to simulate $2\pi+\pi\rho$ exchanges
which are clearly not of purely scalar, isoscalar nature.
Consequently, the $\sigma$ approximation is poor
(as demonstrated in Fig.~11 of Ref.~\cite{MHE87}).
One way to make up for this deficiency is to readjust
the parameters of the $\sigma$ boson in each partial wave.
Moreover, the $2\pi+\pi\rho$ exchanges create---in terms of
ranges---a very broad contribution that cannot be reproduced well
by a single boson-mass; two masses will do better.
The fact that we are dealing here with a very broad mass distribution
is supported
by an entry in the Particle Data Tables~\cite{PDG98} which lists a
$\sigma$ (or $f_0$) with a mass between 400 and 1200 MeV.

Based upon the philosophy just outlined,
we have constructed a $NN$ potential
that is energy-independent and defined in the framework
of the usual (nonrelativistic) Lippmann-Schwinger equation.
Thus, it can be applied in the same way as any other conventional $NN$
potential. The crucial point, however, is that it
reproduces important predictions by the Bonn Full Model, 
while avoiding the problems that the Bonn Full Model creates in
applications.
The charge-dependence (CD) predicted by the Bonn Full Model
is reproduced accurately by the new potential, which is why
we have dubbed it the CD-Bonn potential.
The off-shell behavior of CD-Bonn is based 
upon the relativistic Feynman amplitudes for meson-exchange.
Therefore, the CD-Bonn potential differs off-shell
from conventional $NN$ potentials---a fact that has
attractive consequences in nuclear structure applications.

An earlier version of the CD-Bonn potential---which, however, did
not contain all the charge-dependence---was published in 
Ref.~\cite{MSS96} where the off-shell aspects are discussed
in great detail.

In Sec.~II, we present the potential model.
Charge dependence is discussed in Sec.~III.
The results for $NN$ scattering and the deuteron are
presented in Sec.~IV and V, respectively.
Conclusions are given in Sec.~VI.
The paper has four appendices which contain
mathematical and other details.

\section{The model}

As discussed in the Introduction, the CD-Bonn potential is based upon
meson exchange.
We include all mesons with masses below the nucleon mass, i.~e.,
$\pi$, $\eta$, $\rho(770)$, and $\omega(782)$. Besides this, we introduce
two scalar-isoscalar $\sigma$ bosons.

For the $\eta$ (with a mass of 547.3 MeV), we assume a 
vanishing coupling to the nucleon,
which implies that---{\it de facto}---we drop the $\eta$.
This assumption is supported by semi-empirical evidence
from various sources. Analyzing $NN$ scattering data in terms
of forward dispersion relations, Grein and Kroll~\cite{GK80} determined
the $\eta NN$ coupling constant to be consistent with zero.
Tiator and coworkers~\cite{TBK94} extracted the $\eta$ coupling
from $\eta$ photoproduction data and found $g_\eta^2/4\pi=0.4$.
Such a small coupling constant generates a negligible 
contribution in the $NN$ system.
In the development of the Bonn Full Model 
for the $NN$ interaction~\cite{MHE87},
it was noticed that a good fit of the $NN$ data favors
a vanishing $\eta$ contribution.

In Table~\ref{tab_basicpar}, we list the hadrons involved in our model 
together with their masses and coupling parameters.
For the $\pi NN$ coupling constant, we choose the `small' value
$g_\pi^2/4\pi=13.6$---consistent
with recent determinations by the Nijmegen~\cite{STS93,Tim95}
 and VPI group~\cite{ASW94,AWP94,Pav99}. It is appropriate 
to mention that the precise value of the $\pi NN$ coupling
constant is an unsettled issue at this time, 
and we refer the interested
reader to Refs.~\cite{Mac99,MB99} for a critical discussion
and review of the topic.
For the vector mesons $\rho$ and $\omega$, for which precise
empirical determinations of the coupling constants are difficult
(if not impossible), we use the values from the 
Bonn Full Model~\cite{MHE87}.

We start from the following
Lagrangians that describe the coupling of the mesons of interest
to nucleons:
\begin{eqnarray}
{\cal L}_{\pi^0 NN}&=& - g_{\pi^0} \bar{\psi} i\gamma^{5} 
                           \tau_3 \psi
                           \varphi^{(\pi^0)}
\label{eq_pi0NN}
\\
{\cal L}_{\pi^\pm NN}&=& - \sqrt{2} g_{\pi^\pm} \bar{\psi} i\gamma^{5} 
                           \tau_\pm \psi
                           \varphi^{(\pi^\pm)}
\label{eq_pipmNN}
\\
{\cal L}_{\sigma NN}&=& -g_{\sigma}\bar{\psi}\psi\varphi^{(\sigma)}
\label{eq_sigmaNN}
\\
{\cal L}_{\omega NN}&=& -g_{\omega}\bar{\psi}\gamma^{\mu}\psi
                       \varphi^{(\omega)}_{\mu}
\label{eq_omegaNN}
\\
{\cal L}_{\rho NN} &=& -g_{\rho}
                           \bar{\psi}  \gamma^\mu  
                           \mbox{\boldmath $\tau$} \psi
                           \cdot 
                           \mbox{\boldmath $\varphi$}^{(\rho)}_\mu
                  - \frac{f_\rho}{4M_p}
                           \bar{\psi}  \sigma^{\mu\nu}  
                           \mbox{\boldmath $\tau$} \psi
                           \cdot 
                           (\partial_\mu
                           \mbox{\boldmath $\varphi$}^{(\rho)}_\nu
                           -\partial_\nu
                           \mbox{\boldmath $\varphi$}^{(\rho)}_\mu)
\label{eq_rhoNN}
\end{eqnarray}
where
$\psi$ denotes nucleon fields, $\varphi$ meson fields, and
$\tau_{3,\pm}$ are standard definitions of Pauli matrices  and
combinations thereof for isospin $\frac12$ \cite{BD64}.
$M_p$ is the proton mass which is used as scaling mass in the $\rho NN$
Lagrangian to make $f_\rho$ dimensionless. To avoid the creation
of unmotivated charge dependence, the scaling mass $M_p$ is
used in the $\rho NN$ vertex no matter what nucleons are involved.

In the c.~m.\ system of the two interacting nucleons, the OBE 
Feynman amplitude generated by meson $\alpha$ is,
\begin{equation}
-i\bar{V}_{\alpha}(q',q) =
\frac{\bar{u}_1({\bf q'})\Gamma_1^{(\alpha)} u_1({\bf q}) P_\alpha
\bar{u}_2(-{\bf q'})\Gamma_2^{(\alpha)} u_2(-{\bf q})}
{(q'-q)^2-m_\alpha^2} \; ,
\label{eq_OBEamp}
\end{equation}
where $\Gamma_i^{(\alpha)}$ ($i=1,2$) are vertices
derived from the above Lagrangians, $u_i$ Dirac spinors representing
the interacting nucleons, and $q$ and $q'$ their relative four-momenta
in the initial and final states, respectively; $P_\alpha$
divided by the denominator is  the
appropriate meson propagator.

The one-boson-exchange potential is defined by ($i$ times) the sum over
the OBE Feynman amplitudes of the mesons included in the model 
(Fig.~\ref{fig_OBE}); i.~e.,
\begin{equation}
V({\bf q'},{\bf q}) =
\sqrt{\frac{M}{E'}}
\sqrt{\frac{M}{E}}
\sum_{\alpha=\pi^0,\pi^\pm,\rho,\omega,\sigma_1,\sigma_2} 
\bar{V}_{\alpha}({\bf q'},{\bf q}) 
{\cal F}_{\alpha}^2({\bf q'},{\bf q}; \Lambda_\alpha) \; .
\label{eq_OBEP}
\end{equation}
As customary, we include form factors,
${\cal F}_{\alpha}({\bf q'},{\bf q}; \Lambda_\alpha)$, applied to the
meson-nucleon vertices, and a square-root factor
$M/\sqrt{E'E}$
(with $E=\sqrt{M^2+{\bf q}^2}$,
$E'=\sqrt{M^2+{\bf q'}^2}$, and $M$ the nucleon mass).
The form factors [see Appendix B, Eq.~(\ref{31.13}), for details] 
regularize the amplitudes for large momenta
(short distances) and account for the extended structure of
nucleons in a phenomenological way.
The square root factors make it possible to cast the 
unitarizing,
relativistic, three-dimensional Blankenbecler-Sugar (BbS) 
equation~\cite{BS66}
for the scattering amplitude 
[a reduced version of the four-dimensional Bethe-Salpeter (BS)
equation~\cite{SB51}] 
into the following form (see Appendix A for a proper derivation):
\begin{equation}
 {T}({\bf  q'},{\bf  q})= {V}({\bf  q'},{\bf  q})+
\int d^3k\:
{V}({\bf  q'},{\bf  k})\:
\frac{M}
{{\bf  q}^{2}-{\bf  k}^{2}+i\epsilon}\:
{T}({\bf  k},{\bf  q})
\label{eq_LS}
\end{equation}
Notice that this is the familiar
(non-relativistic) Lippmann-Schwinger
equation. Thus, Eq.~(\ref{eq_OBEP}) defines a relativistic
potential which can be consistently applied in conventional, non-relativistic
nuclear structure, in the usual way.

The Feynman amplitudes, Eq.~(\ref{eq_OBEamp}), are in general nonlocal
expressions; i.~e., Fourier transform into 
configuration space
will yield functions of $r$ and $r'$, the relative
distances between the two in- and out-going nucleons,
respectively.
The square root factors in Eq.~(\ref{eq_OBEP}) create additional non-locality.

While for heavy vector-meson exchange (corresponding to short distances)
non-locality appears quite plausible, we have to stress here
that even the one-pion-exchange (OPE) Feynman 
amplitude is non-local.
This fact is often overlooked.
It is important because the pion creates the dominant part of the tensor force
which plays a crucial role in nuclear structure.

Applying 
the $\pi NN$ Lagrangian Eq.~(\ref{eq_pi0NN})
to the amplitude Eq.~(\ref{eq_OBEamp})
yields the one-pion-exchange (OPE) potential 
(suppressing charge-dependence and isospin factors for the moment)
\begin{eqnarray}
\bar{V}_\pi ({\bf q'}, {\bf q}) & =&  -\frac{g^2_\pi}{4M^2}
\frac{(E'+M)(E+M)}
{({\bf q'}-{\bf q})^2+m_\pi^2}
\left(
 \frac{\mbox{\boldmath $\sigma_{1} \cdot $} {\bf q'}}{E'+M}
-
 \frac{\mbox{\boldmath $\sigma_{1} \cdot $} {\bf q}}{E+M}
\right)
\nonumber \\
 & & \times
\left(
 \frac{\mbox{\boldmath $\sigma_{2} \cdot $} {\bf q'}}{E'+M}
-
 \frac{\mbox{\boldmath $\sigma_{2} \cdot $} {\bf q}}{E+M}
\right) 
\, .
\label{eq_OPEPrel}
\end{eqnarray}
If we would now apply the approximation,
$E'\approx E \approx M$ (static approximation), 
then this simplifies to
\begin{equation}
V_\pi^{(loc)}({\bf k})  =  -\frac{g_{\pi}^{2}}{4M^{2}}
 \frac{{\mbox {\boldmath $(\sigma_{1} \cdot $}} {\bf k)}
{\mbox{\boldmath $(\sigma_{2} \cdot $}} {\bf k)}}
 {{\bf k}^{2}+m_{\pi}^{2}}
\;
\label{eq_OPEPnr}
\end{equation}
with ${\bf k} = {\bf q'} - {\bf q}$.
Fourier transform of this latter expression yields,
\begin{eqnarray}
V_\pi^{(loc)}({\bf r}) & = &
\frac{g^2_\pi}{12\pi} 
\left(\frac{m_\pi}{2M}\right)^2
\left[ 
\left(
\frac{e^{-m_\pi r}}{r}
-\frac{4\pi}{m_\pi^2}\delta^{(3)}({\bf r})
\right)
\mbox{\boldmath $\sigma_{1} \cdot \sigma_{2}$}
\right.
\nonumber \\
& & +
\left.
\left(1+\frac{3}{m_\pi r}+\frac{3}{(m_\pi r)^2}\right)
\frac{e^{-m_\pi r}}{r}
\mbox{\boldmath $S_{12}$} 
\right] 
\: .
\label{eq_OPEPnrr}
\end{eqnarray}
This is the local OPE potential that is used by most practitioners.
However, the important point to notice here is that this local
OPE is not the full, original OPE Feynman amplitude; 
it is an approximation.

The obvious question to raise at this point is: How much does
the local approximation change the original result or, in other
words, how drastic is the local approximation?

For this purpose, we show in
Fig.~\ref{fig_pot3SD1} the half off-shell
$^3S_1$--$^3D_1$
potential that can be produced only by tensor forces.
The on-shell momentum $q'$ is held fixed at 265 MeV
(equivalent to 150 MeV lab.\ energy),
while the off-shell momentum $q$ runs from zero
to 2000 MeV.
The on-shell point ($q=265$ MeV) is marked by a solid dot.
The solid curve is the relativistic OBE amplitude of $\pi
+ \rho$ exchange.
Now, when the relativistic
OPE amplitude, Eq.~(\ref{eq_OPEPrel}), is replaced by the static/local
approximation, Eq.~(\ref{eq_OPEPnr}),
the dashed curve is obtained.
When this approximation is also used for the one-$\rho$
exchange, the dotted curve results.
It is clearly seen that the static/local approximation
does change the potential drastically off-shell:
it makes the tensor force substantially stronger off-shell.

In summary, one characteristic point of the CD-Bonn potential is
that it uses the Feynman amplitudes of meson exchange
in its original form; local approximations are not applied.
This has impact on the off-shell behavior of the potential,
particularly, the off-shell tensor potential. It is well known that
the off-shell behavior of an $NN$ potential is an important
factor in microscopic nuclear structure calculations. 
Therefore, the predictions by the CD-Bonn potential 
for nuclear structure problems differ in a charcteristic way from 
the ones obtained with local $NN$ potentials.
For more discussion of this issue, see Sec.~VI and Refs.~\cite{MSS96,Mac98}.

\section{Charge dependence}

By definition, {\it charge independence} is invariance under any 
rotation in isospin space. 
A violation of this symmetry is referred to as charge dependence
or charge independence breaking (CIB).
{\it Charge symmetry} is invariance under a rotation by 180$^0$ about the
$y$-axis in isospin space if the positive $z$-direction is associated
with the positive charge.
The violation of this symmetry is known as charge symmetry breaking (CSB).
Obviously, CSB is a special case of charge dependence.

CIB of the strong $NN$ interaction means that,
in the isospin $T=1$ state, the
proton-proton ($T_z=+1$), 
neutron-proton ($T_z=0$),
or neutron-neutron ($T_z=-1$)
interactions are (slightly) different,
after electromagnetic effects have been removed.
CSB of the $NN$ interaction refers to a difference
between proton-proton ($pp$) and neutron-neutron ($nn$)
interactions, only. 
For recent reviews on these matters, see Refs.~\cite{MO95,MNS90}.

CIB is seen most clearly in the $^1S_0$ 
$NN$ scattering lengths. 
The latest empirical values for the singlet scattering length $a$ 
and effective range $r$ are:
\begin{eqnarray}
a^N_{pp}=-17.3\pm 0.4 \mbox{ fm} \cite{MNS90}, \hspace*{.8cm} 
    &\hspace*{1.0cm}
    & r^N_{pp}=2.85\pm 0.04 \mbox{ fm} \cite{MNS90}; \\
a^N_{nn}=-18.9\pm 0.4 \mbox{ fm} \cite{How98,Gon99}, \hspace*{.16cm}
   && r^N_{nn} = 2.75\pm 0.11 \mbox{ fm} \cite{MNS90};\\
a_{np}=-23.740\pm 0.020 \mbox{ fm} \cite{KMS84}, 
    && r_{np}=2.77\pm 0.05 \mbox{ fm} \cite{KMS84}.
\end{eqnarray}
The values given for $pp$ and $nn$ 
scattering refer to the nuclear part of the interaction
as indicated by the superscript $N$; i.~e., 
electromagnetic effects have been removed from the experimental
values. 

The above values imply that
charge-symmetry is broken by the following amounts,
\begin{eqnarray}
\Delta a_{CSB}& \equiv& a_{pp}^N-a_{nn}^N = 1.6\pm 0.6 \mbox{ fm}, 
\label{eq_CSBlep1}
\\
\Delta r_{CSB}& \equiv& r_{pp}^N-r_{nn}^N = 0.10\pm 0.12 \mbox{ fm}; 
\label{eq_CSBlep2}
\end{eqnarray}
and, focusing on $pp$ and $np$,
the following CIB is observed:
\begin{eqnarray}
\Delta a_{CIB} \equiv
 a_{pp}^N
 - 
 a_{np}
 &=& 6.44\pm 0.40 \mbox{ fm},
\label{eq_CIBlep1}
\\
\Delta r_{CIB} \equiv
 r_{pp}^N
 - 
 r_{np}
 &=& 0.08\pm 0.06 \mbox{ fm}.
\label{eq_CIBlep2}
\end{eqnarray}
In summary, the $NN$ singlet scattering lengths show a small amount
of CSB and a clear signature of CIB.

The current understanding is 
that---on a fundamental level---the 
charge dependence of nuclear forces is due to
a difference between the up and down quark masses and electromagnetic 
interactions among the quarks. 
As a consequence of this---on the hadronic level---major
causes of CIB are mass differences between hadrons of the same
isospin multiplet, meson mixing, and irreducible meson-photon exchanges.

\subsection{Charge symmetry breaking}

The difference between the masses of  neutron and proton
represents the most basic cause for CSB of the nuclear force. 
Therefore, it is important to have a very thorough 
accounting of this effect. 

The most trivial consequence of nucleon mass splitting is a
difference in the kinetic energies: for the heavier neutrons,
the kinetic energy is smaller than for protons. 
This raises the magnitude of the $nn$ scattering length by
0.26 fm as compare to $pp$. 
The nucleon mass difference also affects the OBE diagrams, Fig.~\ref{fig_OBE},
but only by a negligible amount. In summary, the two most obvious and trivial
CSB effects explain only about 15\% of the empirical $\Delta a_{CSB}$
(cf.\ Table~\ref{tab_CSBlep}). 
Usual models for the nuclear force include only the two CSB effects 
just dicussed and, therefore, do not reproduce the empirical CSB.

However, in Ref.~\cite{LM98a} it was found that
the irreducible diagrams of
two-boson exchange (TBE) create a much larger CSB effect
than the OBE diagrams and, in fact, fully explain
the empirical CSB splitting of the singlet scattering length.
The major part of the CSB effect comes from diagrams of
$2\pi$ exchange where those with $N\Delta$ intermediate states
make the largest contribution. 
The CSB effect from
irreducible diagrams that exchange a $\pi$ and $\rho$ meson
were also included in the study.
The $\pi\rho$ diagrams give rise to non-negligible
CSB contributions that are typically smaller and of opposite 
sign as compared to the $2\pi$ effects. 
The net effect explains $\Delta a_{CSB}$ quantitatively.

The above-mentioned investigation~\cite{LM98a} was based upon the Bonn
Full Model~\cite{MHE87}.
This model uses the $\pi NN$ coupling constant $g^2_\pi/4\pi=14.4$ 
which is not current. For that reason we
have revised the Bonn Full Model using
$g^2_\pi/4\pi=13.6$ and then repeated the CSB calculations
of Ref.~\cite{LM98a}. The total $\Delta a_{CSB}$ predicted
by the revised model is 1.508 fm (about 5\% less than what was
obtained in Ref.~\cite{LM98a} with the original model),
implying a TBE effect of 1.275 fm.

The only reliable empirical information about CSB of the $NN$ interaction
is the scattering length difference in the $^1S_0$ state,
Eq.~(\ref{eq_CSBlep1}). As discussed,
the TBE model of Refs~\cite{MHE87,LM98a} can explain this
entirely from nucleon mass splitting.
For this reason, we have confidence in the CSB predictions
by this model. Therefore, we will use its predictions 
also for energies and states
where no empirical information is available; namely,
higher energies in the $^1S_0$ state and partial 
waves other than $^1S_0$.

Thus, using the revised Bonn Full Model, we have calculated
the difference $nn$ phase shift 
minus $pp$ phase shift without electromagnetic interactions,
$\delta_{nn}-\delta_{pp}$,
that is caused by CSB of the strong nuclear force
due to nucleon mass splitting. The total effect obtained
is listed in the last colum (`Total') of Table~\ref{tab_CSB}
for energies up to 300 MeV and partial wave states in which
these effects are non-negligible.
In that table, 
we also list the very small effects from the OBE diagrams
(Fig.~\ref{fig_OBE}) and the
kinematical effects (column `Kinematics')~\cite{foot1}.
CSB phase shift differences are plotted in Fig.~\ref{fig_CSB}.
It is clearly seen that in most states the TBE effect is the
largest and, therefore, certainly not negligible as compared to
the other CSB effects shown.

Because of the outstanding importance of the CSB effect from TBE, we include
it in our model. 
By doing so, we go beyond
what is usually done in charge-dependent $NN$ potentials.
In most recent models, only the kinematical effects and
the effect of nucleon mass splitting on the OBE diagrams
are included.
However, as discussed, this does not explain the
CSB scattering length difference. Thus, some models leave CSB
simply unexplained~\cite{Sto94}, while other models add a 
purely phenomenological term
to the potential that fits $\Delta a_{CSB}$~\cite{WSS95}.

Before finishing this subsection,
a word is in place concerning other mechanisms that cause
CSB of the nuclear force.
Traditionally, it was believed that 
$\rho^0-\omega$ 
mixing explains essentially all CSB in the nuclear force~\cite{MNS90}.
However, recently some doubt has been cast on this paradigm.
Some researchers~\cite{GHT92,PW93,KTW93,Con94} found that 
$\rho^0-\omega$ exchange may have a substantial
$q^2$ dependence such as to cause this contribution to nearly vanish
in $NN$.
Our finding that the empirically known CSB in the 
nuclear force can be explained solely from nucleon mass splitting 
(leaving essentially no room for additional CSB contributions 
from $\rho^0-\omega$ mixing or other sources) fits well into this 
scenario.
On the other hand, Miller~\cite{MO95} and Coon and coworkers~\cite{CMR97}
have advanced counter-arguments that would restore the traditional role
of $\rho$-$\omega$ exchange.
The issue is unresolved.
Good summaries of the controversial points of view can be found in
Refs.~\cite{MO95,Con97,CS00}.
We do not include $\rho-\omega$ mixing in our model.

Finally, for reasons of completeness, we mention that irreducible
diagrams of $\pi$ and $\gamma$ exchange between
two nucleons create a charge-dependent nuclear force.
Recently, these contributions have been calculated to
leading order in chiral perturbation theory~\cite{Kol98}.
It turns out that to this order the $\pi\gamma$ force is
charge-symmetric (but does break charge independence).

\subsection{Charge independence breaking}

The major cause of CIB in the $NN$ interaction is pion mass splitting.
Based upon the Bonn Full Model for the $NN$ interaction,
the CIB due to pion mass splitting has been calculated 
carefully and systematically in Ref.~\cite{LM98b}.

The largest CIB effect comes from the OPE diagram which
accounts for about 50\% of the empirical $\Delta a_{CIB}$, 
Eq.~(\ref{eq_CIBlep1}), (cf.\ Table~\ref{tab_CIBlep}). 

In $pp$ scattering, the one-pion-exchange potential, $V^{OPE}$,
is given by,
\begin{equation}
V^{OPE}(pp) = V_{\pi^0}
\: ,
\end{equation}
while in $T=1$ $np$ scattering, we have,
\begin{equation}
V^{OPE}(np, T=1) = -V_{\pi^0} + 2 V_{\pi^\pm}
\: .
\end{equation}
If the pion masses were all the same, these would be identical potentials.
However, due to the mass splitting, the $T=1$ $np$ potential is weaker
as compared to the $pp$ one. This causes a difference between $T=1$
$pp$ and $np$ that is known as CIB.
For completeness, we also give the $T=0$ $np$ OPE potential which is
\begin{equation}
V^{OPE}(np, T=0) = -V_{\pi^0} - 2 V_{\pi^\pm}
\: .
\end{equation}

Due to the small mass of the pion, OPE is also a sizable
contribution in all partial waves with $L>0$;
and due to the pion's relatively large mass splitting (3.4\%),
OPE creates relatively large charge-dependent effects 
in all partial waves 
(cf.\ Tables~\ref{tab_CIB1S0} and \ref{tab_CIB}
and Fig.~\ref{fig_CIB1}). Therefore,
all modern phase shift analyses~\cite{ASW94,Sto93} and all
modern $NN$ potentials~\cite{Sto94,WSS95,MSS96}
include the CIB effect created by OPE.

However, pion mass splitting creates further CIB effects through
the diagrams of $2\pi$ exchange and other two-boson exchange diagrams
that involve pions. The evaluation of this
CIB contribution is very involved, but it has been accomplished
in Ref.~\cite{LM98b}. The CIB effect from all the relevant two-boson
exchanges (TBE) 
contributes about 1.3 fm to $\Delta a_{CIB}$.
Concerning phase shift differences,
it is noticable up to $D$ waves and can amount up to 50\% 
of the OPE effect in some states
(cf.\ Tables~\ref{tab_CIB1S0} and \ref{tab_CIB} \cite{foot2}).

Another source of CIB is irreducible $\pi\gamma$ exchange. 
Recently, these contributions have been
evaluated in the framework of chiral perturbation theory
by van Kolck {\it et al.}~\cite{Kol98}. Based upon this work,
we have calculated the impact of the $\pi\gamma$ diagrams on
the $^1S_0$ scattering length and on $np$ phase shifts.
(see column `$\pi\gamma$' in Tables~\ref{tab_CIBlep}, \ref{tab_CIB1S0},
and \ref{tab_CIB}).
In $L>0$ states, the size of this contribution is typically
the same as the CIB effect from TBE.

In the $^1S_0$ state, the $\pi\gamma$ contribution increases the
discrepancy between theory and experiment (cf.\ Table~\ref{tab_CIBlep}).
As a matter of fact, about 25\% of 
$\Delta a_{CIB}$
is not explained. For that reason, a quantitative fit
of the empirical
$\Delta a_{CIB}$
requires a small phenomenological contribution.
The same is true for the difference between the empirical
$np$ and $pp$ phase shifts in the $^1S_0$ state 
(cf.\ Table~\ref{tab_CIB1S0}).

For convenience, the major CIB effects on the strong $NN$ force
are plotted in Fig.~\ref{fig_CIB1}. 
In Fig.~\ref{fig_CIB2} the total CIB phase shift effect caused
by the strong force is compared to the Coulomb effect 
on $pp$ phase shifts ($\delta^C$ denotes the phase
shift in the presence of the Coulomb force, see Appendix 
A.3 for precise definitions
of $\delta$ and $\delta^C$).

From the figures and tables it is evident that TBE and $\pi\gamma$ create
sizable CIB effects in states with $L>0$. 
Therefore, we will include these two effects in our model.
We note that conventional charge-dependent
$NN$ models ignore these two contributions.

\section{Nucleon-nucleon scattering}

We construct three $NN$ interactions:
a proton-proton ($pp$), a neutron-neutron ($nn$),
and a neutron-proton ($np$) potential.
The three potentials are not independent. They are all based
upon the model described in Sec.~II and the differences
between them are determined by CSB and CIB as discussed
in Sec.~III. Thus, when one of the three potentials is fixed,
then the $T=1$ parts of the other two potentials
are also fixed due to CSB and CIB.

We start with the $pp$ potential since the $pp$ data are the most
accurate ones. Data fitting is done in three steps. In the first step,
the $pp$ potential is adjusted to reproduce
closely the $pp$ phase shifts of the Nijmegen
multi-energy $pp$ phase shift analysis~\cite{Sto93}.
This is to ensure that phase shifts are in the right ballpark.
In the second step, the $\chi^2$ 
that results
from applying the Nijmegen $pp$ error matrix~\cite{SS93} 
is minimized. 
The error matrix allows to calculate the $\chi^2$ in regard to
the $pp$ data in an approximate way requirings little computer time.
Finally, in the third and crucial step, the $pp$ potential parameters
are fine-tuned by minimizing the exact $\chi^2$ that results from a direct
comparison with all experimental $pp$ data. 
During these calculations, it was revealed that the Nijmegen $pp$ error matrix
yields very accurate $\chi^2$ up to 75 MeV. 
Therefore, in this final step, we used the error matrix up to 75 MeV and
direct $\chi^2$ calculations above this energy.

The $nn$ potential is constructed by starting from the $pp$
potential, leaving out the Coulomb force, changing the nucleon masses, 
and fine-tuning the parameters
such that the CSB differences listed in Tables~\ref{tab_CSBlep}
and \ref{tab_CSB} are reproduced.

Concerning the $np$ potential, we need to distinguish between the
$T=0$ and $T=1$ states. In $T=1$, we start from the $pp$ potential,
leave out the Coulomb force, change the nucleon masses,
and replace the $pp$ OPE potential by the one
that applies to $np$. This produces a large part of CIB. 
The additional CIB due to TBE and $\pi\gamma$ discussed in Sec.~III
is incorporated by fine-tuning the parameters such that the total
CIB phase shift differences as given in Table~\ref{tab_CIB}
are reproduced.
This fixes the $np$ potential in the $T=1$ states with
$L>0$. The $^1S_0$ $np$ potential is adjusted such as to
minimize the $\chi^2$ in regard to the $np$ data.
The $np$ $T=0$ potential is fixed by going through the
entire three-step procedure: fit of Nijmegen $T=0$ phase shifts,
minimizing the approximate $\chi^2$ obtained from the
Nijmegen error matrix, and
finally minimizing the exact $\chi^2$ that results from a direct 
comparison with all experimental $np$ data. 

The resulting phase shifts for $pp$, $nn$, and $np$ scattering
in partial waves with $J\leq 4$ are given
in Tables~\ref{tab_phpp} -- \ref{tab_phnp0};
$pp$ phase shifts are plotted in Fig.~\ref{fig_phpp1} and
$np$ phase shifts are shown in Fig.~\ref{fig_ph1}.
For $pp$ scattering, we show the phase shifts of the nuclear
plus relativistic Coulomb interaction with respect to
Coulomb wave functions; that is---in the notation of 
Ref.~\cite{Ber88}---we use $V_{C} = \alpha'/r$ 
for the Coulomb potential and calculate the 
phase shifts $\delta^{C}_{C+N}$ ($\equiv \delta^C$ in our notation).
We note that, for the calculation of observables (e.~g., to obtain
the $\chi^2$ in regard to experimental data),
we use electromagnetic phase shifts, {\it as necessary,} which we obtain by adding to
the Coulomb phase shifts the effects from two-photon exchange, vacuum
polarization, and magnetic moment interactions
as calculated by the Nijmegen group~\cite{Ber88,Sto95}.
This is important for $^1S_0$ below 30 MeV and negligible otherwise.
For $nn$ and $np$ scattering, we show the phase shifts of the
nuclear interaction with respect to Riccati-Bessel functions.
All details of our phase shift calculations are given in
Appendix A.3

The low-energy scattering parameters are shown in 
Table~\ref{tab_lep}.
For $nn$ and $np$, the effective range expansion without
any electromagnetic interaction is used. In the case of $pp$
scattering, the quantities
$a_{pp}^C$ and $r_{pp}^C$ 
are obtained by using the effective range expansion
appropriate in the presence of
the Coulomb force 
(see Appendix A.4 for details).
Note that the empirical values for 
$a_{pp}^C$ and $r_{pp}^C$ 
that we quote in Table~\ref{tab_lep} 
were obtained by subtracting from the corresponding 
electromagnetic values
the effects due to two-photon exchange and vacuum polarization.
Thus, the comparison between theory and experiment conducted
in Table~\ref{tab_lep} is adequate.

For the comparison with the $NN$ data,
we consider three databases: 
1992 database, after-1992 data, and 1999 database.
The 1992 database
is identical to the one used by the Nijmegen group for
their phase shift analysis~\cite{Ber90,Sto93}.
It consists of all $NN$ data below 350 MeV published between January 1955
and December 1992 that were not rejected in the Nijmegen
data analysis (for details of the rejection criteria and a complete
listing of the data references, see Refs.~\cite{Ber88,Ber90,Sto93}).
The 1992 database contains 1787 $pp$ data and 2514 $np$ data.

After 1992, there has been a fundamental breakthrough in 
the development of experimental methods for conducting
hadron-hadron scattering experiments.
In particular, the method of internal polarized gas targets
applied in stored, cooled beams is now working perfectly in several hadron
facilities, e.~g., IUCF and COSY.
Using this new technology, IUCF has produced a large number
of $pp$ spin correlation parameters of very high precision.
In Table~\ref{tab_ppdat}, we list the new IUCF data
together with other $pp$ data published between January 1993 and December
1999. Table~\ref{tab_ppdat} lists all published after-1992
$pp$ data below 350 MeV except for one set, namely, 14  
$pp$ differential cross sections at 45 deg (lab.) between 
299.8 adn 406.8 keV
by Dombrowski {\it et al.}~\cite{Do97}; according to the Nijmegen
rejection criteria~\cite{Ber88}, this set is to be discarded.
The total number of (accepted) after-1992 $pp$ data is 1145, which should be
compared to the number of $pp$ data in the 1992 base, namely, 1787. 
Thus, the $pp$ database has increased by about
2/3 since 1992. The importance of the new $pp$ data is further enhanced 
by the fact that they are of much higher quality than the old ones.

Neutron-proton data published between January 1993 and December 1999
and included in our $\chi^2$ calculations
are listed in Table~\ref{tab_npdat}.  There are 544 such data, which is 
a small number as compared to the 2514 $np$ data of the 1992 base.
Note that Table~\ref{tab_npdat} is not a list of all $np$ data published 
after 1992. Not listed are four measurements
of $np$ differential cross sections~\cite{Go94,Be97,Ra98,Fr99}.
We have examined these data and found
in each case that they produced an improbably high $\chi^2$ when
compared to current phase shift analyses~\cite{Sto93,SM99}.
Applying the Nijmegen rejection rule~\cite{Ber88,Sto93},
the data of all four experiments are to be discarded.
We follow this rule here, because we use the Nijmegen
database for the pre-1993 period. When we add data to this base,
then consistency requires that we apply the same selection criteria
used for assembling the older part of the base.
However, we like to stress that we do understand that any discarding
of published data (i.~e., data that have passed the refereeing process)
is a highly questionable procedure. The problem of the $np$
differential cross section data is an unresolved issue that
deserves the full attention of all $NN$ practitioners.
Some aspects of the problem were recently discussed
in Ref.~\cite{Up99}.

Finally, our 1999 database is the sum of the 1992 base 
and the after-1992 data and,
thus, consists of the world $NN$ data below 350 MeV that were published before 
the year of 2000 (and not rejected).

The $\chi^2$/datum produced by the CD-Bonn potential
in regard to the databases defined above are listed 
in Table~\ref{tab_chi2}. For the purpose of comparison, we also give
the corresponding $\chi^2$ values for the Nijmegen phase shift
analysis~\cite{Sto93} and the recent 
Argonne $V_{18}$ potential~\cite{WSS95}.
What stands out in Table~\ref{tab_chi2} are the rather large values
for the $\chi^2$/datum generated by the Nijmegen analysis and the
Argonne potential for the the after-1992 $pp$ data, which are
essentially the new IUCF data. This fact is a clear indication that
these new data provide a very critical test/constraint for any $NN$
model. It further indicates that fitting the pre-1993 $pp$
data does not nessarily imply a good fit of those IUCF data. 
On the other hand, fitting the new IUCF data does imply a good fit of the
pre-1993 data. The conclusion from these two facts is that the new
IUCF data provide information that was not contained in the old database.
Or, in other words, the pre-1993 data were insufficient and still left
too much lattitude for pinning down $NN$ models.
One thing in particular that we noticed is that the $^3P_1$ phase
shifts above 100 MeV have to be lower than the values given in
the Nijmegen analysis.

The bottom line is that, for the 1999 data base (which contains 5990 $pp$
and $np$ data), the CD-Bonn potential yields a $\chi^2$/datum of 1.02,
while the Nijmegen analysis produces 1.04 and the Argonne potential
1.21. We have also compared other recent $NN$ potentials and $NN$ analyses
to the 1999 database and found in all case a $\chi^2$/datum $\geq 1.05$.

{\it Thus we can conclude that the CD-Bonn potential fits the world $NN$ data 
below 350 MeV available in the year of 2000 better than any phase 
shift analysis and any other $NN$ potential.}

\section{The deuteron}

The CD-Bonn potential has been fitted to the empirical value for the
deuteron binding energy $B_d = 2.224575$ MeV~\cite{LA82}
using relativistic kinematics. Once this adjustment has been
made, the other deuteron properties listed in in Table~\ref{tab_deu}
are predictions. For the asymptotic $D/S$ state ratio, we find
$\eta = 0.0256$---in accurate agreement with the empirical
determination by Rodning and Knutson~\cite{RK90}.
The deuteron matter radius is predicted to be $r_d = 1.966$ fm
which agrees well with the value extracted from recent 
hydrogen-deuterium isotope
shift measurements, $r_d = 1.971(6)$ fm~\cite{MSZ95}.
Note that the deuteron effective range $\rho_d \equiv \rho(-B_d,-B_d)$
and the asymptotic $S$ state $A_S$ are not directly observable quantities.
Thus, `empirical' values for $\rho_d$ and $A_S$
quoted in the literature are model dependent.
Therefore, the perfect agreement between our predictions
and the empirical values for $\rho_d$ and $A_S$
is of no fundamental significance.
It only means that all models (including our own) are consistent
with each other.

More interesting is our prediction for the deuteron quadrupole moment
$Q_d = 0.270$ fm$^2$ which is below
the empirical value of 0.2859(3) fm$^2$~\cite{BC79,ER83}.
Our calculation does not include relativistic and meson current corrections
which according to Henning~\cite{Hen93} contribute typically about 0.010 fm$^2$
for the Bonn OBE potentials. This would raise our theoretical
value to $Q_d \approx 0.280$ fm$^2$, still 0.006 fm$^2$ below
experiment. All recent $NN$ potentials that use the `small'
$\pi NN$ coupling constant $g^2_\pi/4\pi = 13.6$ underpredict $Q_d$
by about the same amount. In Refs.~\cite{MS91,Mac99} it was shown
that $Q_d$ depends sensitively on $g_\pi$ and that a value
$g^2_\pi/4\pi \geq 14.0$ would solve the problem.
However, a larger $g_\pi$ is inconsistent with the low-energy
$pp$ $A_y$ data (see Ref.~\cite{Mac99} for a detailed discussion of this issue).
Thus, the accurate explanation of the deuteron quadrupole moment
is an unresolved problem at this time.

In Table~\ref{tab_deu}, we also give the deuteron $D$-state
probability $P_D$. This quantity is not an observable, but it is
of great theoretical interest.
CD-Bonn predicts $P_D = 4.85$\% while local potentials typically
predict $P_D \approx 5.7$ \%, which is clearly reflected in the
deuteron $D$-waves,
Figs.~\ref{fig_dwaves} and \ref{fig_dwaves2}.
The smaller $P_D$ value of CD-Bonn can be traced to the non-localities
contained in the tensor force as discussed in Sec.~II 
and demonstrated in Fig.~\ref{fig_pot3SD1}. 
The CD-Bonn and the Nijmegen-I~\cite{Sto94} potentials have nonlocal
central forces which explains the soft behavior of their deuteron
$S$-waves at short distances that is particularly apparent
in the plot of Fig.~\ref{fig_dwaves2}.
Numerical values of our deuteron waves and a convenient
parametrization are given in Appendix D which also contains
an account of how to conduct deuteron calculations
in momentum space.

\section{Conclusions}

We have constructed charge-dependent $NN$ potentials,
that fit the world proton-proton data below 350 MeV
(2932 data) with a $\chi^2$/datum of 1.01 and the
corresponding neutron-proton data (3058 data) with 
$\chi^2$/datum $= 1.02$.
This reproduction of the $NN$ data is more accurate than by any
other known $NN$ potential or phase-shift analysis.
A particular challenge are the $pp$ spin correlation parameters
that were recently measured at the IUCF Cooler Ring with
very high precision (1126 data below 350 MeV).
Our $pp$ potential reproduces these data with
$\chi^2$/datum $= 1.03$, while the high-quality Nijmegen 
analysis~\cite{Sto93} and the Argonne $V_{18}$ potential~\cite{WSS95} 
produce $\chi^2$/datum of 1.24  and 1.74, respectively, for these data.

The charge-dependence of the present potential (that has been dubbed
`CD-Bonn') is based upon the predictions by the Bonn Full Model
for charge-symmetry and charge-independence breaking in all partial
waves with $J\leq 4$. 
Thus, our model includes considerably more charge-dependence than
other recently developed charge-dependent potentials~\cite{Sto94,WSS95}.
For example, the Nijmegen potentials~\cite{Sto94} include essentially only
charge-dependence due to OPE which produces CIB, but
no CSB. Thus, the Nijmegen group does not offer any genuine 
neutron-neutron potentials.
To have distinct $pp$ and $nn$ potentials 
is important for addressing several interesting issues
in nuclear physics, like the $^3$H-$^3$He binding energy difference
for which the CD-Bonn potential predicts 60 keV 
in agreement with empirical estimates.
Another issue is the Nolen-Schiffer anomaly~\cite{NS69}.
Some potentials that include CSB focus on the $^1S_0$ state only, since 
this is where the most reliable empirical information is.
However, this is not good enough.
In Ref.~\cite{MPM99} it has been shown that CSB in
states with $J>0$ is crucial for the explanation
of the Nolen-Schiffer anomaly.

The CD-Bonn potential is represented in terms of the covariant
Feynman amplitudes for one-boson exchange which are nonlocal.
Therefore, the off-shell behavior of the CD-Bonn potential
differs in a characteristic and well-founded way from the one
of commonly used local potentials.

The simplest system in which off-shell differences between $NN$
potentials can be investigated is the deuteron
(see Ref.~\cite{Pol98} for a thorough study of this issue).
Our plots of the deuteron wave functions,
Figs.~\ref{fig_dwaves} and \ref{fig_dwaves2},
make this point very clear.
Empirical tests of deuteron wave functions can be conducted via
the structure functions
$A(Q^2)$, $B(Q^2)$,
and the tensor polarization in elastic electron-deuteron
scattering 
$T_{20}(Q^2)$ 
or, alternatively, via the three deuteron
form factors
$G_C(Q^2)$, $G_Q(Q^2)$, and $G_M(Q^2)$,
for which the deuteron wave functions are crucial input.
Using the deuteron wave functions derived from the Bonn model,
Arenh\"ovel and coworkers~\cite{ARW00} find a good agreement
between theory and experiment for
$A(Q^2)$, $B(Q^2)$, and $T_{20}(Q^2)$ 
up to $Q^2 = 30$ fm$^{-2}$.
Very recently, the tensor polarization
$T_{20}(Q^2)$ 
has been measured up to
$Q^2 = 45$ fm$^{-2}$
at the Jefferson Laboratory~\cite{Abb00}.
The best reproduction of these new high-precision data
is provided by two calculations that are based upon the
Bonn deuteron wave functions~\cite{PWD99,CK99}.

Another way in which
the off-shell behavior of our potential shows up 
is by yielding larger binding
energies in microscopic calculations of nuclear few- and many-body 
systems~\cite{Eng97}, where underbinding is a persistent problem.
To demonstrate this, we have computed the binding energy of the triton
in a 34-channel, charge-dependent Faddeev calculation. 
The prediction by the CD-Bonn potential is 8.00 MeV.
Local potentials typically predict 7.62 MeV~\cite{Fri93,NKG00}
and the experimental value is 8.48 MeV.
Thus, the nonlocality of the CD-Bonn potential
explains almost 50\% of the gap that persists between the predictions
by local potentials and experiment.
Similar results are obtained for the $\alpha$ particle~\cite{NKG00,NKB00}.
Concerning the small difference that is left between the 
CD-Bonn predictions and experiment,
two comments are in place. First, 
besides the relativistic, nonlocal effects that can be absorbed
into the two-body potential concept, 
there are further relativistic corrections
that come from a relativistic treatment of the three-body system. 
This increases
the triton binding energy by 0.2--0.3 MeV~\cite{SXM92,RT92,SM98,MSS96}.
Second, notice
that the present nonlocal potential includes only the
nonlocalities that come from meson-exchange. 
However, the composite structure (quark substructure) of hadrons
should provide additional nonlocalities~\cite{VS95} which may be even larger.
It is a challenging topic for future research to derive 
these additional nonlocalities, and test their impact on 
nuclear structure predictions.

The trend of the nonlocal Bonn potential to increase binding
energies has also a very favorable impact on predictions for
nuclear matter~\cite{Mac99a,Mac98} and the
structure of finite nuclei~\cite{Jia92,And96,Hol98}.

Due to the very accurate fit of even the latest high-precision $NN$ data;
due to the comprehensive and sophisticated charge-dependence incorporated
in the model; and due to the well-founded off-shell behavior,
the CD-Bonn potential~\cite{foot4} represents a promising starting point
for exact few-body calculations and
microscopic nuclear many-body theory.

\acknowledgments
The author likes to thank Dick Arndt for a personal 
copy of the $NN$ software package SAID.
This work was supported in part by the U.S. National Science Foundation
under Grant No.~PHY-9603097.

\appendix

\section{Two-nucleon scattering in momentum space}
\subsection{Scattering equation}
Two-nucleon scattering is described covariantly by
 the Bethe-Salpeter (BS) equation~\cite{SB51}
which reads in operator notation
\begin{equation}
{\cal T=V+VGT}
\label{21.1}
\end{equation}
with ${\cal T}$ the invariant amplitude for the two-nucleon scattering process,
${\cal V}$ the sum of all connected two-particle irreducible diagrams, and 
${\cal G}$ the relativistic two-nucleon propagator. Since this four-dimensional
integral equation is very difficult to solve, so-called 
three-dimensional reductions have been proposed, which are more amenable to
numerical solution.
Furthermore, it has been shown by Gross~\cite{Gro82} that the full BS equation 
in ladder approximation (that is, the kernel ${\cal V}$ is restricted to 
the exchange of single particles as, e.~g., in the OBE model)
does not have the correct one-body limit (i.~e., 
when one of the particles becomes very massive) while a large family 
of three-dimensional quasi-potential equations does.
 These approximations to the BS equation are also covariant
and satisfy relativistic elastic unitarity.
Three-dimensional reductions are typically derived by replacing 
Eq.~(\ref{21.1}) with two coupled equations~\cite{WJ73}:
\begin{equation}
{\cal T=W+W}g{\cal T}
\label{21.2}
\end{equation}
and
\begin{equation}
{\cal W=V+V(G}-g){\cal W}
\label{21.3}
\end{equation}
where $g$ is a covariant three-dimensional propagator
with the same elastic unitarity cut as ${\cal G}$ in the physical region.
 In general, 
the second term on the r.h.s.\ of Eq.\ (\ref{21.3}) is dropped to obtain a true
simplification of the  problem.

More explicitly, the BS equation for an arbitrary frame reads~\cite{BD64}
\begin{equation}
{\cal T}(q';q|P)={\cal V}(q';q|P)+\int d^{4}k\: {\cal V}(q';k|P)\: {\cal G}(k|P)
\: {\cal T}(k;q|P)
\label{21.4}
\end{equation}
with
\begin{eqnarray}
{\cal G}(k|P)&=&\frac{i}{2\pi} 
\frac{1}{(\frac{1}{2}\not\!P+\not\!k-M+i\epsilon)^{(1)}}
\frac{1}{(\frac{1}{2}\not\!P-\not\!k-M+i\epsilon)^{(2)}}\\
             &=&\frac{i}{2\pi} 
\left[\frac{\frac{1}{2}\not\!P+\not\!k+M}
{(\frac{1}{2}P+k)^{2}-M^{2}+i\epsilon}\right]^{(1)}
\left[\frac{\frac{1}{2}\not\!P-\not\!k+M}
{(\frac{1}{2}P-k)^{2}-M^{2}+i\epsilon}\right]^{(2)}
\label{21.6}
\end{eqnarray}
where
$q$, $k$, and $q'$ are the initial, intermediate, and final relative
four-momenta, respectively, 
and $P=(P_0,{\bf P})$ is the total four-momentum.
For example, in the initial state we have: $q=\frac12(p_1-p_2), P=p_1+p_2,
\mbox{ and } p_{1/2}=\frac12P\pm q$ with $p_1$ and $p_2$ the individual 
four-momenta of particle 1 and 2.
In the center-of-mass (c.m.) frame, we will have
 $P=(\sqrt{s},{\bf 0})$ with $\sqrt{s}$ the total energy.
For all four-momenta, our notation is
$k=(k_{0},{\bf   k})$; $\not\!\!k \equiv\gamma^{\mu}k_{\mu}$. $M$ denotes the
nucleon mass.
 The superscripts in Eq.~(\ref{21.6})
refer to particle (1) and (2). 
At this stage, ${\cal T, V,\mbox{ and } G}$ are operators in spinor 
space, i.~e. they are 
$16\times 16$ matrices which, when sandwiched between Dirac 
spinors, yield the corresponding matrix elements.

It is common to the derivation of all three-dimensional reductions that the 
time component of the relative momentum is fixed in some covariant way, so 
that it no longer appears as an independent variable in the propagator.

Following
Blankenbecler and Sugar (BbS)~\cite{BS66},
 one possible choice for
 $g$ is
(stated in manifestly covariant form for an arbitrary frame)
\begin{eqnarray}
g_{BbS}(k,s)&=&-
\int_{4M^{2}}^{\infty}
\frac{ds'}{s'-s-i\epsilon}
\: \delta^{(+)}[\begin{array}{c} (\frac{1}{2}P'+k)^{2}-M^{2}\end{array}]
\nonumber \\ 
&&\times\delta^{(+)}[\begin{array}{c} (\frac{1}{2}P'-k)^{2}-M^{2}\end{array}]
\nonumber \\  
&&\times[\begin{array}{c} \frac{1}{2}\not\!P'+\not\!k+M\end{array}]^{(1)}
[\begin{array}{c} \frac{1}{2}\not\!P'-\not\!k+M\end{array}]^{(2)}
\label{21.7}
\end{eqnarray}
with $\delta^{(+)}$ indicating that only the positive energy root of the 
argument of the $\delta$-function is to be included; $P^2=s$ and
 $P'\equiv \frac{\sqrt{s'}}{\sqrt{s}} P$.
By construction, 
the propagator $g_{BbS}$ has the same 
imaginary part 
as ${\cal G}$ and, therefore, preserves the unitarity relation
satisfied by ${\cal T}$.
In the c.m. frame, integration yields
\begin{equation}
g_{BbS}(k,s)=\delta (k_{0})\: \bar{g}_{BbS}({\bf  k},s)
\label{21.8}
\end{equation}
with
\begin{equation} 
\bar{g}_{BbS}({\bf  k},s)=\frac{M^{2}}{E_{k}}\:
\frac{\Lambda_{+}^{(1)}({\bf  k})\:
\Lambda_{+}^{(2)}({\bf  -k})}
{\frac{1}{4}s-E^{2}_{k}+i\epsilon}
\label{21.9}
\end{equation}
where
\begin{eqnarray}
\Lambda^{(i)}_{+}({\bf  k})&=&
\left(\frac{\gamma^{0}E_{k}-{\bbox \gamma}\cdot{\bf  k}+M}{2M}\right)^{(i)}\\
                 &=&\sum_{\lambda_{i}}
              |u({\bf  k},\lambda_{i})\rangle 
               \langle \bar{u}({\bf  k},\lambda_{i})|
\label{21.11}
\end{eqnarray} 
represents
the positive-energy projection operator
for nucleon $i$ ($i$ = 1 or 2)
with $u({\bf  k})$ a positive-energy Dirac spinor of momentum
${\bf  k}$; $\bar{u}\equiv u^\dagger \gamma^0$.
$\lambda_{i}$ denotes the helicity of the respective nucleon, and 
  $E_{k}\equiv \sqrt{M^{2}+{\bf  k}^{2}}$ with $M$ the nucleon mass.
The projection operators imply that virtual anti-nucleon
contributions are suppressed. 

Using the approximation ${\cal W}\approx {\cal V}$ [cf.\ Eq.~(\ref{21.3})], 
we obtain the explicit form of 
Eq.~(\ref{21.2}) by simply
replacing  ${\cal G}$ by $g_{BbS}$ in Eq.~(\ref{21.4}). This yields
in the c.m.~frame
\begin{equation}
{\cal T}(0,{\bf  q'};0,{\bf  q}|\sqrt{s})
 =  {\cal V}(0,{\bf  q'};0,{\bf  q}) + \int d^3k\: 
{\cal V}(0,{\bf  q'};0,{\bf  k})\: \bar{g}_{BbS}({\bf  k},s)\: 
{\cal T}(0,{\bf  k};0,{\bf  q}|\sqrt{s}) .
\label{21.12}
\end{equation} 
Note that four-momentum is conserved at each vertex,
and that in the initial state the nucleons are on their mass-shell,
therefore $q=(0,{\bf  q})$.
The total c.m. energy is
\begin{equation}
\sqrt{s}=2E_{q}=2\sqrt{M^2+{\bf  q}^2}.
\label{21.13}
\end{equation}
With this we obtain, simplifying our notation,
\begin{equation}
{\cal T}({\bf  q'},{\bf  q})={\cal V}({\bf  q'},{\bf  q})+
\int d^3k\:
{\cal V}({\bf  q'},{\bf  k})\:
\frac{M^{2}}{E_{k}}\:
\frac{\Lambda_{+}^{(1)}({\bf  k})\:
\Lambda_{+}^{(2)}({\bf  -k})}
{{\bf  q}^{2}-{\bf  k}^{2}+i\epsilon}\:
{\cal T}({\bf  k},{\bf  q}).
\label{21.14}
\end{equation}
Taking matrix elements between positive-energy spinors
 yields an equation for the invariant scattering amplitude
\begin{equation}
\bar{T}({\bf  q'},{\bf  q})=\bar{V}({\bf  q'},{\bf  q})+
\int d^3k\:
\bar{V}({\bf  q'},{\bf  k})\:
\frac{M^{2}}{E_{k}}\:
\frac{1}
{{\bf  q}^{2}-{\bf  k}^{2}+i\epsilon}\:
\bar{T}({\bf  k},{\bf  q}),
\label{21.15}
\end{equation}
where helicity and isospin indices are suppressed.
\\
Defining
\begin{equation}
{T}({\bf  q'}, {\bf  q})
 = \sqrt{\frac{M}{E_{q'}}}\: \bar{T}({\bf  q'}, {\bf  q})\:
 \sqrt{\frac{M}{E_{q}}}
\label{21.16}
\end{equation}
and
\begin{equation} 
{V}({\bf  q'},{\bf  q})
 = \sqrt{\frac{M}{E_{q'}}}\:  \bar{V}({\bf  q'},{\bf  q})\:
 \sqrt{\frac{M}{E_{q}}},
\label{21.17}
\end{equation}
which has become known as ``minimal relativity''~\cite{BJK69},
we can rewrite Eq.~(\ref{21.15}) as
\begin{equation}
 {T}({\bf  q'},{\bf  q})= {V}({\bf  q'},{\bf  q})+
\int d^3k\:
{V}({\bf  q'},{\bf  k})\:
\frac{M}
{{\bf  q}^{2}-{\bf  k}^{2}+i\epsilon}\:
{T}({\bf  k},{\bf  q})
\label{21.18}
\end{equation}
which has the form of the familiar Lippmann-Schwinger equation.
The quantity $T$ has the usual (nonrelativistic) relation to
phase shifts and $NN$ observables. 
Thus, the $NN$ potential $V$
defined in Eq.~(\ref{21.17}) and used in the above Lippmann-Schwinger
equation can be applied 
in the deuteron and in conventional nuclear structure physics
in the same way as any other (nonrelativistic) potential. 
This is the great virtue of the (relativistic) BbS equation.

\subsection{R-matrix and partial wave decomposition}

In solving the scattering equation, it is more convenient to deal
with real quantities. We shall therefore introduce the
real $R$-matrix (better known as `$K$-matrix') 
defined by
\begin{equation}
T=R-i\pi R\delta(E-H_0)T
\label{22.0}
\end{equation}
The equation for the real ${R}$-matrix corresponding to the complex 
${T}$-matrix of Eq.~(\ref{21.18}) is
\begin{equation}
{R}({\bf  q'},{\bf  q})={V}({\bf  q'},{\bf  q})+
{\cal P} \int d^3k\:
{V}({\bf  q'},{\bf  k})\:
\frac{M}
{{\bf  q}^{2}-{\bf  k}^{2}}\:
{R}({\bf  k},{\bf  q})
\label{22.00}
\end{equation}
where ${\cal P}$ denotes the principal value.

Now, we need to also include the spin of the nucleons.
Relativistic scattering of particles with spin is treated 
most conveniently in the helicity formalism~\cite{JW59}.
Therefore, we will use a helicity state basis
in our further formal developments.
Our presentation will be relatively brief;
a more detailed derivation is given in Appendix C of 
Ref.~\cite{MHE87} which is based upon Refs.~\cite{JW59,EAH71}.

The helicity $\lambda_i$ of particle $i$ (with $i=1$ or 2) 
is the eigenvalue of the helicity operator 
$\frac12 \bbox{\sigma}_i \cdot {\bf p}_i/|{\bf p}_i|$ which is 
$\pm \frac12$. 

Using helicity states, the ${R}$-matrix equation reads,
 after partial wave decomposition,
\begin{eqnarray}
\langle \lambda_1' \lambda_2'|{R}^J(q',q)|\lambda_1 \lambda_2 \rangle
&=&\langle \lambda_1' \lambda_2'|{V}^J(q',q)|\lambda_1 \lambda_2 \rangle
\nonumber \\
 & & \mbox{} + \sum_{h_1,h_2} {\cal P} \int^{\infty}_0 dk\: k^2
 \frac{M}{q^2-k^2}
\langle \lambda_1' \lambda_2'|{V}^J(q',k)|h_1 h_2 \rangle \nonumber \\
 & & \mbox{} \times
\langle h_1 h_2|{R}^J(k,q)|\lambda_1 \lambda_2 \rangle 
\label{22.1}
\end{eqnarray}
where $J$ denotes the total angular momentum of the two nucleons.
Here we are changing our notation for momenta:
in the above equation and throughout the rest of Appendix A,
momenta denoted by non-bold letters are the magnitude of three-momenta, e.~g. 
$q\equiv|{\bf  q}|$, $k\equiv|{\bf  k}|$, etc.;
$h_1$ and $h_2$ are the helicities in intermediate states for nucleon 1 and 2, 
respectively.
Equation~(\ref{22.1}) is a system 
of coupled integral equations which needs to be solved to obtain the desired
matrix elements of ${R}^J$.

Ignoring anti-particles, there are $4\times 4 = 16$
 helicity amplitudes for ${R}^J$. However,
time-reversal invariance, parity conservation, and the fact that we are 
dealing with two identical fermions imply that 
only six amplitudes are independent. For these six amplitudes, we 
choose the following set:
\begin{eqnarray}
{R}^J_1(q',q)&\equiv& \langle ++|{R}^J(q',q)|++ \rangle \nonumber \\
{R}^J_2(q',q)&\equiv& \langle ++|{R}^J(q',q)|-- \rangle \nonumber \\
{R}^J_3(q',q)&\equiv& \langle +-|{R}^J(q',q)|+- \rangle \nonumber \\
{R}^J_4(q',q)&\equiv& \langle +-|{R}^J(q',q)|-+ \rangle  \\
\label{22.2}
{R}^J_5(q',q)&\equiv& \langle ++|{R}^J(q',q)|+- \rangle \nonumber \\
{R}^J_6(q',q)&\equiv& \langle +-|{R}^J(q',q)|++ \rangle \nonumber
\end{eqnarray}
where $\pm$ stands for $\pm \frac12$.
Notice that
\begin{equation}
{R}^J_5(q',q)={R}^J_6(q,q') .
\label{22.3}
\end{equation}

We have now six coupled equations.
To partially decouple this system, it is usefull to 
introduce the following linear combinations of helicity amplitudes:
\begin{eqnarray}
^0{R}^J&\equiv&{R}^J_1 - {R}^J_2  \nonumber \\
^1{R}^J&\equiv&{R}^J_3 - {R}^J_4  \nonumber \\
^{12}{R}^J&\equiv&{R}^J_1 + {R}^J_2 \label{22.4} \\
^{34}{R}^J&\equiv&{R}^J_3 + {R}^J_4 \nonumber \\
^{55}{R}^J&\equiv&2{R}^J_5 \nonumber \\
^{66}{R}^J&\equiv&2{R}^J_6 \nonumber 
\end{eqnarray}
We also introduce corresponding definitions for ${V}^J$.
Using these definitions, Eq.~(\ref{22.1}) decouples into the following 
three sub-systems of integral equations:\\
{ Spin singlet}
\begin{equation}
^0{R}^J(q',q) =\:  ^0{V}^J(q',q)+{\cal P}\int^{\infty}_0 dk\: k^2
\frac{M}{q^2-k^2}\: ^0{V}^J(q',k)\: ^0{R}^J(k,q)
\: .
\label{22.5} 
\end{equation}
{ Uncoupled spin triplet}
\begin{equation}
^1{R}^J(q',q) =\:  ^1{V}^J(q',q)+{\cal P}\int^{\infty}_0 dk\: k^2
\frac{M}{q^2-k^2}\: ^1{V}^J(q',k)\: ^1{R}^J(k,q)
\: .
\label{22.6} 
\end{equation}
{ Coupled triplet states}
\begin{eqnarray}
^{12}{R}^J(q',q)&=& ^{12}{V}^J(q',q)+{\cal P}\int^{\infty}_0 dk\: k^2
\frac{M}{q^2-k^2}[\: ^{12}{V}^J(q',k)\: ^{12}{R}^J(k,q) \nonumber \\
 & & \mbox{} +\: ^{55}{V}^J(q',k)\: ^{66}{R}^J(k,q)]  \nonumber \\
^{34}{R}^J(q',q)&=& ^{34}{V}^J(q',q)+{\cal P}\int^{\infty}_0 dk\: k^2
\frac{M}{q^2-k^2}[\: ^{34}{V}^J(q',k)\: ^{34}{R}^J(k,q) \nonumber \\
 & & \mbox{} +\: ^{66}{V}^J(q',k)\: ^{55}{R}^J(k,q)]   \nonumber \\
^{55}{R}^J(q',q)&=& ^{55}{V}^J(q',q)+{\cal P}\int^{\infty}_0 dk\: k^2
\frac{M}{q^2-k^2}[\: ^{12}{V}^J(q',k)\: ^{55}{R}^J(k,q) \nonumber \\
 & & \mbox{} +\: ^{55}{V}^J(q',k)\: ^{34}{R}^J(k,q)]  \nonumber \\
^{66}{R}^J(q',q)&=& ^{66}{V}^J(q',q)+{\cal P}\int^{\infty}_0 dk\: k^2
\frac{M}{q^2-k^2}[\: ^{34}{V}^J(q',k)\: ^{66}{R}^J(k,q) \nonumber \\
 & & \mbox{} +\: ^{66}{V}^J(q',k)\: ^{12}{R}^J(k,q)] 
\: .
\label{22.7}
\end{eqnarray}

More common in nuclear physics is the representation of 
two-nucleon states
in terms of an 
$|LSJM\rangle$ basis, 
where $S$ denotes the total spin, $L$ the total orbital 
angular momentum, and $J$ the total angular momentum with 
projection $M$. 
In this basis, we will denote the ${R}$ matrix elements by
${R}^{JS}_{L',L}\equiv \langle L'SJM|{R}|LSJM\rangle$.
  These are obtained from the helicity state matrix 
elements by the following unitary transformation:\\
{ Spin singlet}
\begin{equation}
{R}^{J0}_{J,J}\: =\: ^0{R}^J
\: .
\label{22.10}
\end{equation}
{ Uncoupled spin triplet} 
\begin{equation}
{R}^{J1}_{J,J}\: =\: ^1{R}^J
\: .
\label{22.11}
\end{equation}
{ Coupled triplet states}
\begin{eqnarray}
{R}^{J1}_{J-1,J-1} & = & \frac{1}{2J+1} \left[J\: ^{12}{R}^J 
+ (J+1)\: ^{34}{R}^J
+ \sqrt{J(J+1)}(\: ^{55}{R}^J+\: ^{66}{R}^J)\right]
\nonumber \\
{R}^{J1}_{J+1,J+1} & = & \frac{1}{2J+1} \left[(J+1)\: ^{12}{R}^J 
+ J\: ^{34}{R}^J
- \sqrt{J(J+1)}(\: ^{55}{R}^J+\: ^{66}{R}^J)\right]
\nonumber \\
{R}^{J1}_{J-1,J+1} & = & \frac{1}{2J+1} \left[\sqrt{J(J+1)}
(\: ^{12}{R}^J -\: ^{34}{R}^J)
- J\: ^{55}{R}^J + (J+1)\: ^{66}{R}^J)\right]
\nonumber  \\
{R}^{J1}_{J+1,J-1} & = & \frac{1}{2J+1} \left[\sqrt{J(J+1)}
(\: ^{12}{R}^J -\: ^{34}{R}^J)
+ (J+1)\: ^{55}{R}^J - J\: ^{66}{R}^J)\right]
\: .
\label{22.12} 
\end{eqnarray}
Similar notation and transformations apply to ${V}$.

One way to proceed is to solve the system of equations (\ref{22.7})
and then apply the transformation (\ref{22.12}).
Alternatively, one may apply the transformation (\ref{22.12})
directly in (\ref{22.7}) 
to obtain the system of four coupled integral equations
in $LSJ$ representation,
\begin{eqnarray}
{R}^{J1}_{++}(q',q)&=& {V}^{J1}_{++}(q',q)
+{\cal P}\int^{\infty}_0 dk\: k^2\: \frac{M}{q^2-k^2}
[\: {V}^{J1}_{++}(q',k)\: {R}^{J1}_{++}(k,q)  
\nonumber \\  & & \mbox{} 
+\: {V}^{J1}_{+-}(q',k)\: {R}^{J1}_{-+}(k,q)]  
\nonumber \\
{R}^{J1}_{--}(q',q)&=& {V}^{J1}_{--}(q',q)
+{\cal P}\int^{\infty}_0 dk\: k^2\: \frac{M}{q^2-k^2}
[\: {V}^{J1}_{--}(q',k)\: {R}^{J1}_{--}(k,q) 
\nonumber \\  & & \mbox{}
+\: {V}^{J1}_{-+}(q',k)\: {R}^{J1}_{+-}(k,q)]   
\nonumber \\
{R}^{J1}_{+-}(q',q)&=& {V}^{J1}_{+-}(q',q)
+{\cal P}\int^{\infty}_0 dk\: k^2\: \frac{M}{q^2-k^2}
[\: {V}^{J1}_{++}(q',k)\: {R}^{J1}_{+-}(k,q) 
\nonumber \\ & & \mbox{} 
+\: {V}^{J1}_{+-}(q',k)\: {R}^{J1}_{--}(k,q)]  
\nonumber \\
{R}^{J1}_{-+}(q',q)&=& {V}^{J1}_{-+}(q',q)
+{\cal P}\int^{\infty}_0 dk\: k^2\: \frac{M}{q^2-k^2}
[\: {V}^{J1}_{--}(q',k)\: {R}^{J1}_{-+}(k,q) 
\nonumber \\ & & \mbox{} 
+\: {V}^{J1}_{-+}(q',k)\: {R}^{J1}_{++}(k,q)]  
\nonumber \\
\mbox{} \label{22.13}
\end{eqnarray}
where we used the abbreviations  
${R}^{J1}_{++}\equiv {R}^{J1}_{J+1,J+1},\;
{R}^{J1}_{--}\equiv {R}^{J1}_{J-1,J-1},\;
{R}^{J1}_{+-}\equiv {R}^{J1}_{J+1,J-1},\;
{R}^{J1}_{-+}\equiv {R}^{J1}_{J-1,J+1}$;
and similarly for $V$.

The above integral equations can be solved numerically by the matrix
inversion method~\cite{HT70}.
The method is explained in detail in Ref~\cite{Mac93} where also
a computer code is provided.

Each two-nucleon state carries a well-defined
total isospin $T$ (which is either 0 or 1) that is fixed by
\begin{equation}
(-1)^{L+S+T}=-1\: .
\label{22.14}
\end{equation}

\subsection{Phase shifts}
Phase shifts are determined from the on-energy-shell $R$-matrix 
through:
\\
{ Spin singlet}
\begin{equation}
\tan\: ^0\delta^J(T_{lab}) = -\frac{\pi}{2} q M\: ^0{R}^J(q,q)
\label{23.1}
\end{equation}
{ Uncoupled spin triplet}
\begin{equation}
\tan\: ^1\delta^J(T_{lab}) = -\frac{\pi}{2} q M\: ^1{R}^J(q,q)
\label{23.2}
\end{equation}
For the {\it coupled states}, a unitary transformation is needed to diagonalize 
the two-by-two coupled $R$-matrix. This requires an additional parameter,
known as the `mixing 
parameter' $\epsilon_J$. Using the convention introduced by Blatt and 
Biedenharn~\cite{BB52}, the eigenphases for the coupled channels are,
in terms of the on-shell ${R}$-matrix,
\begin{eqnarray}
\tan\: \delta^J_{\mp}(T_{lab})& =& -\frac{\pi}{4} q M
\left[ {R}^J_{J-1,J-1} + {R}^J_{J+1,J+1} \pm
\frac{{R}^J_{J-1,J-1} - {R}^J_{J+1,J+1}}{\cos 2\epsilon_J}\right]
\nonumber \\
\label{23.3}  \\
\tan 2\epsilon_J(T_{lab}) & = & \frac{2{R}^J_{J+1,J-1}}
{{R}^J_{J-1,J-1} - {R}^J_{J+1,J+1}}\: . \nonumber
\end{eqnarray}
Here, all ${R}$-matrix elements carry the arguments
 $(q,q)$ where $q$ denotes the c.m.\ on-energy-shell momentum.
Based upon correct (relativistic) kinematical considerations,
the momentum $q$ and the nucleon mass $M$ to be used in the
above formulae are determined to be:
\\
Proton-proton scattering:
\begin{eqnarray}
q^2 & = & \frac12 M_p T_{lab} \: ,\\
M & = & M_p
\: .
\end{eqnarray}
Neutron-neutron scattering:
\begin{eqnarray}
q^2 & = & \frac12 M_n T_{lab} \: ,\\
M & = & M_n
\: .
\end{eqnarray}
Neutron-proton scattering:
\begin{eqnarray}
q^2 & = & \frac{M_p^2 T_{lab} (T_{lab} + 2M_n)}
               {(M_p + M_n)^2 + 2T_{lab} M_p} \: , 
\label{eq_relkin}
\\
M & = & \frac{2M_pM_n}{M_p+M_n}
\: .
\end{eqnarray}
In the above, $M_p$ denotes the proton mass, $M_n$ the
neutron mass (see Table~\ref{tab_basicpar} for their accurate
numerical values) and $T_{lab}$ is the kinetic energy of
the incident nucleon in the laboratory system.

An alternative convention for the phase parameters has been used by Stapp {\it 
et al.}~\cite{SYM57}, known as `bar' phase shifts. These are related to 
the Blatt-Biedenharn parameters by
\begin{eqnarray}
\bar{\delta}^J_+ + \bar{\delta}^J_- & = & \delta^J_+ + \delta^J_- 
\nonumber \\
\sin (\bar{\delta}^J_- - \bar{\delta}^J_+) & = & 
{\tan 2\bar{\epsilon}_J}/{\tan 2\epsilon_J}
\label{23.5} \\
\sin (\delta^J_- - \delta^J_+) & = & 
{\sin 2\bar{\epsilon}_J}/{\sin 2\epsilon_J}
\nonumber
\end{eqnarray}
In this paper, all phase shifts shown in tables or figures are in
the `bar' convention, even though we omit the bar in
our notation.

The above formulae apply to the calculation of phase shifts when only
the short-range nuclear force is taken into account (and no electromagnetic
interaction). This is, in general, appropriate for $nn$ and $np$
scattering. We also note that the above momentum space method is
exactly equivalent to calculations conducted in $r$-space
where the radial Schr\"odinger equation,
\begin{equation}
   \left[\frac{d^2}{dr^2}+q^2-\frac{L(L+1)}{r^2}-MV\right]
        \chi_L(r;q) = 0 \: , 
\label{eq_SCH}
\end{equation}
is solved for the radial wave function $\chi_L(r;q)$
which is then matched to the appropriate asymptotic form
of the wave function to obtain the phase shift.
When no long-range potential is involved, the asymptotic wave
functions are Riccati-Bessel functions~\cite{AS70}.

In $pp$ scattering, the long-range Coulomb potential must be taken into
account. The asymptotic form of the wave function then is
\begin{equation}
\chi_L(r;q) \propto
F_L (\eta',qr) + \tan \delta^C_L G_L(\eta',qr)
\label{eq_asymp}
\end{equation}
with $F_L$ and $G_L$ the regular and irregular
Coulomb functions~\cite{AS70}.
By $\delta^C$ we denote the phase shift of the nuclear plus
Coulomb interaction with respect to Coulomb wave functions; 
that is, in the notation of Ref.~\cite{Ber88}, 
$\delta^C \equiv \delta^C_{C+N}$.
The parameter $\eta'$ is the 
``relativistic'' $\eta$ defined by~\cite{Bre55,AS83,Ber88}
\begin{equation}
\eta' = \frac{\alpha}{v_{lab}} 
     = \frac{M_p}{2q} \alpha'
\: ,
\end{equation}
with 
\begin{equation}
\alpha' = \alpha \frac{E_q^2 + q^2}{M_p E_q}
\: ,
\end{equation}
and $\alpha = 1/137.035989$~\cite{PDG98}.
The total potential $V$ that appears in Eq.~(\ref{eq_SCH})
is now the sum of the nuclear potential $V_N$ and the Coulomb
potential $V_{C}$; i.~e.,
\begin{equation}
V=V_N+V_{C} \: ,
\end{equation}
where we use the ``relativistic'' Coulomb potential~\cite{AS83}
\begin{equation}
V_{C} = \frac{\alpha'}{r} \: .
\end{equation}
Since we conduct our calculations in momentum space,
we do not solve Eq.~(\ref{eq_SCH}) and, thus, do not have 
a numerical $\chi(r;q)$
available that can be matched directly to
the asymptotic form Eq.~(\ref{eq_asymp}).
However, there are ways to perform this matching
within the framework of momentum space calculations.
We follow here the method proposed by Vincent and Phatak~\cite{VP74}
in which the potential is divided into a short-range part $V_S$ 
and a long-range part $V_L$; i.~e.,
\begin{equation}
V = V_S + V_L
\end{equation}
with
\begin{eqnarray}
V_S & = & (V_N + V_{C}) \theta(R-r) \: ,\\
V_L & = &  V_{C} \theta(r-R) \: ,
\end{eqnarray}
where $R$ is to be chosen such that the short-range nuclear
potential has vanished for $r>R$ ($R\approx 10$ fm is an appropriate
choice); and $\theta$ is the usual Heaviside step function.
First, one calculates the phase shift (denoted by $\delta^S_L$)
that is produced by $V_S$ alone. 
Notice that $V_S$ is of range $R$ and consists of
the nuclear potential plus the Coulomb potential 
cut off at $r=R$. There is no problem in performing numerically the Bessel 
transformation of a cutoff Coulomb potential to produce the momentum
space version of this potential for the various partial waves. 
Since $V_S$ is of finite
range, the momentum space formalism can be used to calculate $\delta^S_L$.
The asymptotic wave function associated with $V_S$ and $\delta^S_L$ is
\begin{equation}
\chi_L^S(r;q) \propto
F_L (\eta'=0,qr) + \tan \delta^S_L G_L(\eta'=0,qr)
\end{equation}
which should match smoothly the asymptotic
function Eq.~(\ref{eq_asymp}) at $r=R$. Note that
$F_L (\eta'=0,qr)$ and $G_L(\eta'=0,qr)$ are equal to
Riccati-Bessel functions.
Matching the logarithmic derivatives yields the desired formula
for the phase shift $\delta^C_L$:
\begin{equation}
\tan \delta^C_L = 
\frac{
[F_L(0),F_L(\eta')] 
+ \tan \delta^S_L
[G_L(0),F_L(\eta')] 
}{
[G_L(\eta'),F_L(0)] 
+ \tan \delta^S_L
[G_L(\eta'),G_L(0)] 
}
\: ,
\end{equation}
where the square brackets denote the Wronskian
\begin{equation}
[F_L(0),F_L(\eta')] 
\equiv
\left[ 
F_L(0) \frac{d F_L(\eta')}{dr}
-
F_L(\eta') \frac{d F_L(0)}{dr}
\right]_{r=R}
\end{equation}
and
$F_L(0)\equiv F_L(\eta'=0,qr)$,
$F_L(\eta')\equiv F_L(\eta',qr)$;
similarly for $G_L$.

All $pp$ phase shifts shown in this paper are Coulomb phase shifts,
$\delta^C$, as defined and calculated above. However, we like to stress
that, for the calculation of observables (e.~g., to obtain
the $\chi^2$ in regard to experimental data),
we use electromagnetic phase shifts, {\it as necessary,} 
which we obtain by adding to
the Coulomb phase shifts the effects from two-photon exchange, vacuum
polarization, and magnetic moment interactions
as calculated by the Nijmegen group~\cite{Ber88,Sto95}.
This is important for $^1S_0$ below 30 MeV and negligible otherwise.

\subsection{Effective range expansion}
For low-energy $S$-wave scattering, $q\cot \delta$ can be expanded as a 
function of $q$
\begin{equation}
\frac{q}{\tan \delta}=q\cot \delta\approx -\frac{1}{a}+\frac12 r q^2
 + {\cal O}(q^4)
\label{24.1}
\end{equation}
where $a$ is called the scattering length and $r$ the effective range
(for which, in some parts of this paper, we also use the notation
$a^N$ and $r^N$).
This is appropriate for $nn$ and $np$. 

In the case of $pp$ scattering, where 
the Coulomb potential is involved, a more sophisticated 
effective range expansion must be applied~\cite{Ber88},
\begin{equation}
C_0^2(\eta')q\cot(\delta^C_{pp}) + 2q\eta'h(\eta') =
-\frac{1}{a^C_{pp}} + \frac12 r^C_{pp} q^2 + {\cal O}(q^4)
\: ,
\end{equation}
where $\delta^C_{pp}$ denotes the $^1S_0$ $pp$ phase shift
with respect to Coulomb functions
and $C^2_0$ and $h$  are the standard functions,
\begin{eqnarray}
C^2_0 (\eta') & = & \frac{2\pi\eta'}{e^{2\pi\eta'}-1} \: ,\\
h(\eta') & = & -\ln(\eta') + Re[\psi(1+i\eta')] \\
         & = & -\ln(\eta') -\gamma + \eta'^2 \sum_{n=1}^\infty
[n(n^2+\eta'^2)]^{-1} \: ,
\end{eqnarray}
where $\psi$ denotes the digamma function and
$\gamma=0.5772156649\ldots$.

This formalism takes care of the Coulomb force. However, the full
electromagnetic interaction between two protons has contributions beyond
Coulomb, e.~g., from two-photon exchange and vacuum polarization.
To include the full electromagnetic interaction
into the effective range expansion
is very involved. 
Therefore, the empirical values for the $pp$ effective range parameters
(which naturally involve the full electromagnetic interaction) have been 
corrected (in a fairly model-independent way) 
for the electromagnetic effects beyond Coulomb~\cite{SES83,Ber88}. 
This procedure yields
`empirical' values for $a_{pp}^C$ and $r_{pp}^C$ which is what
we quote in Table~\ref{tab_lep} under `Experiment'.
The existence of empirical values of this kind makes the
comparison between theory and experiment much easier; and
it justifies that a theoretican calculates predictions for just
$a_{pp}^C$ and $r_{pp}^C$ using the simple formalism outlined above.

\section{One-boson exchange potential}

\subsection{OBE amplitudes}

The Lagrangians Eqs.~(\ref{eq_pi0NN})-(\ref{eq_rhoNN})
imply the following OBE amplitudes
which we state here in terms of $i$ times the Feynman amplitude:
\begin{eqnarray}
\lefteqn{\hspace{-.6cm}\langle {\bf  q'} \lambda_{1}'\lambda_{2}'
|\bar{V}_{\pi}|
{\bf  q}\lambda_{1}\lambda_{2}\rangle} \nonumber \\
 & = &
 - \frac{g^{2}_{\pi}
 }{(2\pi)^{3}}
  \bar{u}({\bf  q'},\lambda_{1}') i \gamma^{5}
 u({\bf  q},\lambda_{1}) 
 \bar{u}({\bf  -q'},\lambda_{2}') i \gamma^{5}
 u({\bf  -q},\lambda_{2}) 
 /[({\bf  q'-q})^{2}+m_{\pi}^{2}] 
\: ,
\label{31.6}\\
\nonumber \\
\lefteqn{\hspace{-.6cm}\langle {\bf  q'} \lambda_{1}'\lambda_{2}'
|\bar{V}_{\sigma}|
{\bf  q}\lambda_{1}\lambda_{2}\rangle}\nonumber \\
 & = & -\frac{g^{2}_{\sigma}}{(2\pi)^3}
\bar{u}({\bf  q'},\lambda_{1}')    u({\bf  q},\lambda_{1})
\bar{u}({\bf  -q'},\lambda_{2}')   u({\bf  -q},\lambda_{2})
  /[({\bf  q'-q})^{2}+m_{\sigma}^{2}] 
\: ,
\label{31.8}\\
\nonumber \\
\lefteqn{\hspace{-.6cm}\langle {\bf  q'} \lambda_{1}'\lambda_{2}'
|\bar{V}_{\omega}|
{\bf  q}\lambda_{1}\lambda_{2}\rangle}\nonumber \\
 &  = &
\frac{g^2_{\omega}}{(2\pi)^3} 
\{ \bar{u}({\bf  q'},\lambda_{1}')  
   \gamma_{\mu} u({\bf  q},\lambda_{1}) \}
\{ \bar{u}({\bf  -q'},\lambda_{2}')
  \gamma^{\mu} u({\bf  -q},\lambda_{2}) \}
  /[({\bf  q'-q})^{2}+m_{\omega}^{2}]
\: ,
\label{31.8a}\\
\nonumber \\
\lefteqn{\hspace{-.6cm}\langle {\bf  q'} \lambda_{1}'\lambda_{2}'
|\bar{V}_{\rho}|
{\bf  q}\lambda_{1}\lambda_{2}\rangle}\nonumber \\
 &  = &
 \frac{\bbox{\tau}_1 \cdot \bbox{\tau}_2}{(2\pi)^3} 
\{g_{\rho}
\bar{u}({\bf  q'},\lambda_{1}')  \gamma_{\mu} u({\bf  q},\lambda_{1})
+\frac{f_{\rho}}{2M_p}
\bar{u}({\bf  q'},\lambda_{1}')
\sigma_{\mu\nu}i(q'-q)^{\nu}
 u({\bf  q},\lambda_{1})\}
\nonumber \\ & & \times
\{g_{\rho}\bar{u}({\bf  -q'},\lambda_{2}')
  \gamma^{\mu} u({\bf  -q},\lambda_{2})
-\frac{f_{\rho}}{2M_p}\bar{u}({\bf  -q'},\lambda_{2}')
\sigma^{\mu\nu}i(q'-q)_{\nu}
   u({\bf  -q},\lambda_{2})\}
\nonumber \\ 
& &  /[({\bf  q'-q})^{2}+m_{\rho}^{2}]
\nonumber \\ & = &
 \frac{\bbox{\tau}_1 \cdot \bbox{\tau}_2}{(2\pi)^3} 
 \{(g_{\rho}+f_{\rho})
\bar{u}({\bf  q'},\lambda_{1}')  \gamma_{\mu} u({\bf  q},\lambda_{1})
\nonumber \\ & &
-\frac{f_{\rho}}{2M_p}
\bar{u}({\bf  q'},\lambda_{1}')
[(q'+q)_{\mu}+(E'-E)(g_{\mu 0}-\gamma_{\mu}\gamma_{0})]
 u({\bf  q},\lambda_{1})\}
\nonumber \\ & & \times
\{(g_{\rho}+f_{\rho})\bar{u}({\bf  -q'},\lambda_{2}')
  \gamma^{\mu} u({\bf  -q},\lambda_{2})
\nonumber \\ & &
-\frac{f_{\rho}}{2M_p}\bar{u}({\bf  -q'},\lambda_{2}')
[\overline{(q'+q)}^{\mu}+(E'-E)(g^{\mu 0}-\gamma^{\mu}\gamma^{0})]
   u({\bf  -q},\lambda_{2})\}
\nonumber \\  &  &  /[({\bf  q'-q})^{2}+m_{\rho}^{2}] 
\: ,
\label{31.9}
\end{eqnarray}
\normalsize
where for the pion we have suppressed isospin factors and charge-dependence
which will be included later.
Working in the two-nucleon c.m.\ frame,
the momenta of the two incoming (outgoing) nucleons are
${\bf  q}$ and $-{\bf  q}$ (${\bf  q'}$ and $-{\bf  q'}$). 
$E\equiv\sqrt{M^{2}+{\bf  q}^{2}}$, $E'\equiv\sqrt{M^{2}+{\bf  q'}^{2}}$,
and $M$ is the nucleon mass.
Using the BbS equation~\cite{BS66}, the four-momentum transfer between
the two nucleons is $(q'-q)^\mu=(0,{\bf  q'-q})$.
The Gordon identity~\cite{BD64} has been applied in the evaluation of
the tensor coupling of the $\rho$; 
$(q'+q)^\mu \equiv (E'+E,{\bf q}' + {\bf q})$
and
$\overline{(q'+q)}^\mu \equiv (E'+E,-{\bf q}' - {\bf q})$.
The propagator for vector bosons is
\begin{equation}
i\frac{-g_{\mu\nu}+
(q'-q)_{\mu}(q'-q)_{\nu}
/m_{v}^{2}}{-({{\bf  q'-q})^{2}-m_{v}^{2}}}
\label{31.10}
\end{equation}
where we drop the $(q'-q)_{\mu}(q'-q)_{\nu}$-term
 which vanishes on-shell, anyhow, since the nucleon current
is conserved.
The off-shell effect of this term was examined
in Ref.~\cite{HM1} and found to be unimportant.

The Dirac spinors in helicity representation are given by
\begin{eqnarray}
u({\bf  q},\lambda_1)&=&\sqrt{\frac{E+M}{2M}}
\left( \begin{array}{c}
       1\\
       \frac{2\lambda_1 |{\bf q}|}{E+M}
       \end{array} \right)
|\lambda_1\rangle
\: ,
\label{31.11} \\
u(-{\bf  q},\lambda_2)&=&\sqrt{\frac{E+M}{2M}}
\left( \begin{array}{c}
       1\\
       \frac{2\lambda_2 |{\bf q}|}{E+M}
       \end{array} \right)
|\lambda_2\rangle
\: ,
\end{eqnarray}
which are normalized such that
\begin{equation}
\bar{u}({\bf  q},\lambda) u({\bf  q},\lambda)=1.
\, ,
\label{31.12}
\end{equation}
with $\bar{u}=u^{\dagger}\gamma^{0}$.

At each meson-nucleon vertex, a form factor is applied which has the 
analytical form
\begin{equation}
{\cal F}_\alpha[({\bf  q}'-{\bf  q})^2]=
\frac{\Lambda^2_\alpha-m^2_\alpha}
{\Lambda^2_\alpha+({\bf  q}'-{\bf  q})^2} 
\label{31.13}
\end{equation}
with $m_\alpha$ the mass of the meson involved and  
$\Lambda_\alpha$ the so-called cutoff mass.
Thus, to obtain the final OBE potential $V$, 
the amplitudes Eqs.~(\ref{31.6})-(\ref{31.9}) are to be multiplied
by ${\cal F}_\alpha^2$ and certain square-root factors 
[cf.\ Eq.~(\ref{eq_OBEP})].

\subsection{Partial wave decomposition}

The potential is decomposed into partial waves according to
\begin{eqnarray}
\langle \lambda_1' \lambda_2' |V^J(q',q)| \lambda_1 \lambda_2 \rangle =
2\pi \int^{+1}_{-1} d(\cos \theta)\: d^J_{\lambda_1-\lambda_2, 
\lambda_1'-\lambda_2'}(\theta)
\langle {\bf  q}'\lambda_1'\lambda_2'|V|{\bf  q}\lambda_1\lambda_2\rangle
\nonumber \\ \label{32.1}
\end{eqnarray}
where $\theta$ is the angle between ${\bf  q}$ and ${\bf  q}'$ and
$d^J_{m,m'}(\theta)$ are the conventional reduced rotation matrices
which can be expressed in terms of Legendre 
polynominals $P_J(\cos \theta)$. 
The following types of integrals occur:
\begin{eqnarray}
\nonumber \\ 
I^{(0)}_J & \equiv &
\int_{-1}^{+1} dt\: 
\frac{P_J(t)}{({\bf  q}'-{\bf  q})^2+m^2_\alpha}
 = \frac{Q_J(z_\alpha)}{q'q}
\; ,
\label{32.21} \\
\nonumber \\ 
I^{(1)}_J & \equiv &
\int_{-1}^{+1} dt\: \frac{ t P_J(t)}{({\bf  q}'-{\bf  q})^2+m^2_\alpha}
 = \frac{Q_J^{(1)}(z_\alpha)}{q'q}
\; ,
\label{32.22} \\
\nonumber \\
I^{(2)}_J & \equiv & \frac{1}{J+1}
\int_{-1}^{+1} dt\: \frac{J t P_J(t)+P_{J-1}(t)}{({\bf  q}'-{\bf  q})^2+m^2_\alpha}
 = \frac{Q_J^{(2)}(z_\alpha)}{q'q}
\; ,
\label{32.23} \\
\nonumber \\
I^{(3)}_J & \equiv & \sqrt{\frac{J}{J+1}}
\int_{-1}^{+1} dt\: \frac{ t P_J(t)-P_{J-1}(t)}{({\bf  q}'-{\bf  q})^2+m^2_\alpha}
 = \frac{Q_J^{(3)}(z_\alpha)}{q'q}
\; ,
\label{32.24} \\
\nonumber \\ 
I^{(4)}_J & \equiv &
\int_{-1}^{+1} dt\: \frac{ t^2 P_J(t)}{({\bf  q}'-{\bf  q})^2+m^2_\alpha}
 = \frac{Q_J^{(4)}(z_\alpha)}{q'q}
\; ,
\label{32.25} \\
\nonumber \\
I^{(5)}_J & \equiv & \frac{1}{J+1}
\int_{-1}^{+1} dt\: \frac{J t^2 P_J(t) + t P_{J-1}(t)}
{({\bf  q}'-{\bf  q})^2+m^2_\alpha}
 = \frac{Q_J^{(5)}(z_\alpha)}{q'q}
\; ,
\label{32.26} \\
\nonumber \\
I^{(6)}_J & \equiv & \sqrt{\frac{J}{J+1}}
\int_{-1}^{+1} dt\: \frac{ t^2 P_J(t) - t P_{J-1}(t)}
{({\bf  q}'-{\bf  q})^2+m^2_\alpha}
 = \frac{Q_J^{(6)}(z_\alpha)}{q'q}
\; ,
\label{32.27} \\
\nonumber
\end{eqnarray}
with $t\equiv\cos \theta$ and
$z_\alpha \equiv (q'^2 + q^2 + m_\alpha^2)/2q'q$ 
where our notation for momenta is
$q'\equiv |{\bf q}'|$, and $q\equiv |{\bf q}|$
which we will use throughout the remainder of the appendices.

The $Q_J(z)$ are the Legendre functions of the second 
kind~\cite{AS70}; e.~g., $Q_0(z)=\frac12 \ln [(z+1)/(z-1)]$.
The combinations needed above are defined by:
\begin{eqnarray}
Q^{(1)}_J(z) & \equiv & zQ_J-\delta_{J0}
\; ,
\label{eq_QJ1} \\
Q^{(2)}_J(z) & \equiv & \frac{1}{J+1} ( JzQ_J+Q_{J-1} )
\; ,
\label{eq_QJ2} \\
Q^{(3)}_J(z) & \equiv & \sqrt{\frac{J}{J+1}} ( zQ_J-Q_{J-1} )
\; ,
\label{eq_QJ3} \\
Q^{(4)}_J(z) & \equiv & zQ_J^{(1)}-\frac13\delta_{J1}
\; ,
\label{eq_QJ4} \\
Q^{(5)}_J(z) & \equiv & zQ_J^{(2)}-\frac23\delta_{J1}
\; ,
\label{eq_QJ5} \\
Q^{(6)}_J(z) & \equiv & zQ_J^{(3)}+\frac13\sqrt{2}\delta_{J1}
\; .
\label{eq_QJ6} \\
\end{eqnarray}
The integrals
Eqs.~(\ref{32.21})-(\ref{32.27}) can be evaluated either numerically
or analytically by using the Legendre functions of the second kind.
The latter method is better if the correct threshold behavior
of $V^J(q',q)$ for $q',q \rightarrow 0$ is important.

The above expressions still ignore the cutoff
which is included by replacing
\begin{equation}
\frac{1}{({\bf  q}'-{\bf  q})^2+m^2_\alpha}
\longrightarrow
\frac{{\cal F}^2_\alpha[({\bf  q}'-{\bf  q})^2]}
{({\bf  q}'-{\bf  q})^2+m^2_\alpha}
\end{equation}
in Eqs.~(\ref{32.21})-(\ref{32.27}).
If the Legendre functions of the second kind are used,
then the product of propagator and cutoff must be decomposed according to
\begin{eqnarray}
\lefteqn{ \frac{{\cal F}^2_\alpha[({\bf  q}'-{\bf  q})^2]}
{({\bf  q}'-{\bf  q})^2+m^2_\alpha}  } \nonumber \\
 & = &
\frac{1}{({\bf  q}'-{\bf  q})^2+m^2_\alpha}
- 
\left(
\frac{\Lambda^2_{\alpha 2} - m^2_\alpha}
{\Lambda^2_{\alpha 2} - \Lambda^2_{\alpha 1}}
\right)
\frac{1}{({\bf  q}'-{\bf  q})^2+\Lambda^2_{\alpha 1}}
+
\left(
\frac{\Lambda^2_{\alpha 1} - m^2_\alpha}
{\Lambda^2_{\alpha 2} - \Lambda^2_{\alpha 1}}
\right)
\frac{1}{({\bf  q}'-{\bf  q})^2+\Lambda^2_{\alpha 2}}
\end{eqnarray}
where
$\Lambda_{\alpha 1/2} \equiv \Lambda_\alpha \pm \epsilon$
with $\epsilon \rightarrow 0$; i.~e., $\epsilon \ll \Lambda_\alpha$, 
e.~g., $\epsilon \approx 1$ MeV.
To give an example, $I^{(0)}_J$ with cutoff is given by
\begin{eqnarray}
 I^{(0)}_J  & = & 
\int_{-1}^{+1} dt\: 
\frac{P_J(t)
{\cal F}^2_\alpha[({\bf  q}'-{\bf  q})^2] 
}
{({\bf  q}'-{\bf  q})^2+m^2_\alpha}
\\
 & = & 
\frac{Q_J(m_\alpha)}{q'q}
- 
\left(
\frac{\Lambda^2_{\alpha 2} - m^2_\alpha}
{\Lambda^2_{\alpha 2} - \Lambda^2_{\alpha 1}}
\right)
\frac{Q_J(\Lambda_{\alpha 1})}{q'q}
+
\left(
\frac{\Lambda^2_{\alpha 1} - m^2_\alpha}
{\Lambda^2_{\alpha 2} - \Lambda^2_{\alpha 1}}
\right)
\frac{Q_J(\Lambda_{\alpha 2})}{q'q}
\; ,
\end{eqnarray}
and similarly for the other $I^{(i)}_J$.
Notice that the $I^{(i)}_J$ are functions of
$q'$, $q$, $m_\alpha$, and $\Lambda_\alpha$
even though our notation does not indicate this.

\subsection{Final potential expressions}
Here, we will present the final potential expressions in partial wave
decomposition.  More details concerning 
their derivation can be found in Appendix E of Ref.~\cite{MHE87}.
First, we state the potentials in terms of the combinations of helicity
states defined in Eq.~(\ref{22.4}).
\\ \\
One-pion-exchange:
\begin{eqnarray}
^0V^J_{\pi} & = &
C_{\pi}\: ( F^{(0)}_{\pi}\: I^{(0)}_J + F^{(1)}_{\pi}\: I^{(1)}_J )
\nonumber \\
^1V^J_{\pi} & = &
C_{\pi}\: ( -F^{(0)}_{\pi}\: I^{(0)}_J - F^{(1)}_{\pi}\: I^{(2)}_J )
\nonumber \\
^{12}V^J_{\pi} & = &
C_{\pi}\: ( F^{(1)}_{\pi}\: I^{(0)}_J + F^{(0)}_{\pi}\: I^{(1)}_J )
\label{32.3} \\
^{34}V^J_{\pi} & = &
C_{\pi}\: ( -F^{(1)}_{\pi}\: I^{(0)}_J - F^{(0)}_{\pi}\: I^{(2)}_J )
\nonumber \\
^{55}V^J_{\pi} & = &
C_{\pi}\:  F^{(2)}_{\pi}\: I^{(3)}_J
\nonumber \\
^{66}V^J_{\pi} & = &
-C_{\pi}\:  F^{(2)}_{\pi}\: I^{(3)}_J
\nonumber
\end{eqnarray}
with
\begin{equation}
C_{\pi}= \frac{g^2_{\pi}}{4\pi}\: \frac{
1
}{2\pi M^2}
\sqrt{\frac{M}{E'}}
\sqrt{\frac{M}{E}}
\label{32.4}
\end{equation}
and
\begin{eqnarray}
F^{(0)}_{\pi} & = & E'E-M^2 \nonumber \\
F^{(1)}_{\pi} & = & -q'q \label{32.5} \\
F^{(2)}_{\pi} & = & -M(E'-E) . \nonumber
\end{eqnarray}
\\ \\
One-sigma-exchange:
\begin{eqnarray}
^0V^J_{\sigma} & = &
C_{\sigma}\: ( F^{(0)}_{\sigma}\: I^{(0)}_J + F^{(1)}_{\sigma}\: I^{(1)}_J )
\nonumber \\
^1V^J_{\sigma} & = &
C_{\sigma}\: ( F^{(0)}_{\sigma}\: I^{(0)}_J + F^{(1)}_{\sigma}\: I^{(2)}_J )
\nonumber \\
^{12}V^J_{\sigma} & = &
C_{\sigma}\: ( F^{(1)}_{\sigma}\: I^{(0)}_J + F^{(0)}_{\sigma}\: I^{(1)}_J )
\label{32.9} \\
^{34}V^J_{\sigma} & = &
C_{\sigma}\: ( F^{(1)}_{\sigma}\: I^{(0)}_J + F^{(0)}_{\sigma}\: I^{(2)}_J )
\nonumber \\
^{55}V^J_{\sigma} & = &
C_{\sigma}\:  F^{(2)}_{\sigma}\: I^{(3)}_J
\nonumber \\
^{66}V^J_{\sigma} & = &
C_{\sigma}\:  F^{(2)}_{\sigma}\: I^{(3)}_J
\nonumber
\end{eqnarray}
with
\begin{equation}
C_{\sigma}= \frac{g^2_{\sigma}}{4\pi}\: \frac{1}{2\pi M^2}
\sqrt{\frac{M}{E'}}
\sqrt{\frac{M}{E}}
\label{32.10}
\end{equation}
and
\begin{eqnarray}
F^{(0)}_{\sigma} & = & - (E'E+M^2) \nonumber \\
F^{(1)}_{\sigma} & = & q'q \label{32.11} \\
F^{(2)}_{\sigma} & = & M(E'+E) .  \nonumber
\end{eqnarray}
\\ \\
One-omega-exchange:
\begin{eqnarray}
^0V^J_{\omega} & = &
C_{\omega}\: (2E'E-M^2)\: I^{(0)}_J 
\nonumber \\
^1V^J_{\omega} & = &
C_{\omega}\: ( E'E\: I^{(0)}_J + q'q\: I^{(2)}_J )
\nonumber \\
^{12}V^J_{\omega} & = &
C_{\omega}\: ( 2q'q\: I^{(0)}_J + M^2\: I^{(1)}_J )
\label{32.12} \\
^{34}V^J_{\omega} & = &
C_{\omega}\: ( q'q\: I^{(0)}_J + E'E\: I^{(2)}_J )
\nonumber \\
^{55}V^J_{\omega} & = &
-C_{\omega}\:  ME\: I^{(3)}_J
\nonumber \\
^{66}V^J_{\omega} & = &
-C_{\omega}\:  ME'\: I^{(3)}_J
\nonumber
\end{eqnarray}
with
\begin{equation}
C_{\omega}= \frac{g^2_{v}}{4\pi}\: \frac{1}{\pi M^2}
\sqrt{\frac{M}{E'}}
\sqrt{\frac{M}{E}}
\: .
\label{32.13a}
\end{equation}
\\ \\
The one-rho-exchange potential is the sum of three
terms,
\begin{equation}
V_\rho = V_{vv} + V_{tt} + V_{vt}
\, .
\end{equation}
Vector-vector coupling
\begin{eqnarray}
^0V^J_{vv} & = &
C_{vv}\: (2E'E-M^2)\: I^{(0)}_J 
\nonumber \\
^1V^J_{vv} & = &
C_{vv}\: ( E'E\: I^{(0)}_J + q'q\: I^{(2)}_J )
\nonumber \\
^{12}V^J_{vv} & = &
C_{vv}\: ( 2q'q\: I^{(0)}_J + M^2\: I^{(1)}_J )
\label{32.12a} \\
^{34}V^J_{vv} & = &
C_{vv}\: ( q'q\: I^{(0)}_J + E'E\: I^{(2)}_J )
\nonumber \\
^{55}V^J_{vv} & = &
-C_{vv}\:  ME\: I^{(3)}_J
\nonumber \\
^{66}V^J_{vv} & = &
-C_{vv}\:  ME'\: I^{(3)}_J
\nonumber
\end{eqnarray}
with
\begin{equation}
C_{vv}= \frac{g^2_{\rho}}{4\pi}\: 
\frac{
\bbox{\tau}_1 \cdot \bbox{\tau}_2
}{\pi M^2}
\sqrt{\frac{M}{E'}}
\sqrt{\frac{M}{E}}
\: .
\label{32.13}
\end{equation}
Tensor-tensor coupling
\begin{eqnarray}
^0V^J_{tt} & = &
C_{tt}\: \{ (q'^2+q^2)(3E'E+M^2)\: I^{(0)}_J \nonumber \\
 & & + [q'^2+q^2-2(3E'E+M^2)]q'q\: I^{(1)}_J - 2q'^2q^2\: I^{(4)}_J \}
\nonumber \\
^1V^J_{tt} & = &
C_{tt}\: \{ [4q'^2q^2+(q'^2+q^2)(E'E-M^2)]\: I^{(0)}_J \nonumber \\
 & & + 2(E'E+M^2)q'q\: I^{(1)}_J \nonumber \\
 & & - (q'^2+q^2+4E'E)q'q\: I^{(2)}_J - 2q'^2q^2\: I^{(5)}_J \}
\label{32.14} \\
^{12}V^J_{tt} & = &
C_{tt}\: \{ [4M^2-3(q'^2+q^2)]q'q\: I^{(0)}_J \nonumber \\
 & & + [6q'^2q^2-(q'^2+q^2)(E'E+3M^2)]\: I^{(1)}_J +
  2(E'E+M^2)q'q\: I^{(4)}_J \}
\nonumber \\
^{34}V^J_{tt} & = &
C_{tt}\: \{ -(q'^2+q^2+4E'E)q'q\: I^{(0)}_J - 2q'^2q^2\: I^{(1)}_J
\nonumber \\
 & & +[4q'^2q^2+(q'^2+q^2)(E'E-M^2)]\: I^{(2)}_J
     +2(E'E+M^2)q'q\: I^{(5)}_J \}
\nonumber \\
^{55}V^J_{tt} & = &
C_{tt}\:  M \{ [ E'(q'^2+q^2) + E (3q'^2 - q^2)]\: I^{(3)}_J
 - 2(E'+E)q'q\: I^{(6)}_J \}
\nonumber \\
^{66}V^J_{tt} & = &
C_{tt}\:  M \{ [ E(q'^2+q^2) + E'(3q^2 - q'^2)]\: I^{(3)}_J
 - 2(E'+E)q'q\: I^{(6)}_J \}
\nonumber
\end{eqnarray}
with
\begin{equation}
C_{tt}= \frac{f^2_{\rho}}{4\pi M_p^2}\: 
\frac{
\bbox{\tau}_1 \cdot \bbox{\tau}_2
}{8\pi M^2}
\sqrt{\frac{M}{E'}}
\sqrt{\frac{M}{E}}
\: .
\label{32.15}
\end{equation}
Vector-tensor coupling
\begin{eqnarray}
^0V^J_{vt} & = &
C_{vt}\: M [ (q'^2+q^2)\: I^{(0)}_J - 2q'q\: I^{(1)}_J ]
\nonumber \\
^1V^J_{vt} & = &
C_{vt}\: M [ -(q'^2+q^2)\: I^{(0)}_J + 2q'q\: I^{(2)}_J ]
\nonumber \\
^{12}V^J_{vt} & = &
C_{vt}\: M [ 6q'q\: I^{(0)}_J - 3(q'^2+q^2)\: I^{(1)}_J ]
\label{32.16} \\
^{34}V^J_{vt} & = &
C_{vt}\: M [ 2q'q\: I^{(0)}_J - (q'^2+q^2)\: I^{(2)}_J ]
\nonumber \\
^{55}V^J_{vt} & = &
C_{vt}\:  (E'q^2+3Eq'^2)\: I^{(3)}_J
\nonumber \\
^{66}V^J_{vt} & = &
C_{vt}\:  (Eq'^2+3E'q^2)\: I^{(3)}_J
\nonumber
\end{eqnarray}
with
\begin{equation}
C_{vt}= \frac{g_{\rho}f_{\rho}}{4\pi M_p}\: \frac{
\bbox{\tau}_1 \cdot \bbox{\tau}_2
}{2\pi M^2}
\sqrt{\frac{M}{E'}}
\sqrt{\frac{M}{E}}
\: .
\label{32.17}
\end{equation}
Note that in the $\rho$ potential, $M_p$ is a scaling mass associated
with the tensor-coupling constant $f_\rho$. For this scaling
mass, the same is to be used in
$pp$, $np$, and $nn$ scattering.

More common in nuclear physics is the representation of 
two-nucleon states
in terms of an 
$|LSJM\rangle$ basis, 
where $S$ denotes the total spin, $L$ the total orbital 
angular momentum, and $J$ the total angular momentum with 
projection $M$. 
In this basis, we denote the potential by
${V}^{JS}_{L',L}\equiv \langle L'SJM|{V}|LSJM\rangle$.
  These are obtained from the above helicity state matrix 
elements by the following unitary transformation:\\
{ Spin singlet}
\begin{equation}
{V}^{J0}_{J,J}\: =\: ^0{V}^J
\: .
\label{22.10V}
\end{equation}
{ Uncoupled spin triplet} 
\begin{equation}
{V}^{J1}_{J,J}\: =\: ^1{V}^J
\: .
\label{22.11V}
\end{equation}
{ Coupled triplet states}
\begin{eqnarray}
{V}^{J1}_{J-1,J-1} & = & \frac{1}{2J+1} \left[J\: ^{12}{V}^J 
+ (J+1)\: ^{34}{V}^J
+ \sqrt{J(J+1)}(\: ^{55}{V}^J+\: ^{66}{V}^J)\right]
\nonumber \\
{V}^{J1}_{J+1,J+1} & = & \frac{1}{2J+1} \left[(J+1)\: ^{12}{V}^J 
+ J\: ^{34}{V}^J
- \sqrt{J(J+1)}(\: ^{55}{V}^J+\: ^{66}{V}^J)\right]
\nonumber \\
{V}^{J1}_{J-1,J+1} & = & \frac{1}{2J+1} \left[\sqrt{J(J+1)}
(\: ^{12}{V}^J -\: ^{34}{V}^J)
- J\: ^{55}{V}^J + (J+1)\: ^{66}{V}^J)\right]
\nonumber  \\
{V}^{J1}_{J+1,J-1} & = & \frac{1}{2J+1} \left[\sqrt{J(J+1)}
(\: ^{12}{V}^J -\: ^{34}{V}^J)
+ (J+1)\: ^{55}{V}^J - J\: ^{66}{V}^J)\right]
\: .
\label{22.12V} 
\end{eqnarray}

The final charge-dependent potentials are
\begin{equation}
V (N_1N_2) = V^{OPE} (N_1N_2) \
+ \sum_{\alpha = \rho, \omega, \sigma_1, \sigma_2} 
    V_\alpha [M(N_1N_2)]
\label{eq_pot}
\end{equation}
with $N_1 N_2$ either $pp$, $nn$, or $np$.
The nucleon mass referred to by $M(N_1N_2)$ in the above equation
is fixed as follows
\begin{eqnarray}
M(pp) & = & M_p \\
M(nn) & = & M_n \\
M(np) & = & \check{M} \equiv \sqrt{M_pM_n} = 938.91875 \mbox{ MeV}
\label{eq_mav}
\: ,
\end{eqnarray}
with the precise values for $M_p$ and $M_n$ 
given in Table~\ref{tab_basicpar}.
The charge-dependent OPE potentials are given by
\begin{eqnarray}
V^{OPE}(pp) & = & V_{\pi}[g_{\pi}(M_p), m_{\pi^0}, M_p] 
\label{eq_opepp}
\\
V^{OPE}(nn) & = & V_{\pi}[g_{\pi}(M_n), m_{\pi^0}, M_n] \\
V^{OPE}(np, T=1) & = & 
- V_{\pi}[g_{\pi}(\check{M}), m_{\pi^0}, \check{M}]
+2V_{\pi}[g_{\pi}(\check{M}), m_{\pi^\pm}, \check{M}] 
\label{eq_openp1}
\\
V^{OPE}(np, T=0) & = & 
- V_{\pi}[g_{\pi}(\check{M}), m_{\pi^0}, \check{M}]
-2V_{\pi}[g_{\pi}(\check{M}), m_{\pi^\pm}, \check{M}]
\: ,
\end{eqnarray}
with $m_{\pi^0}$ and $m_{\pi^\pm}$
as given in Table~\ref{tab_basicpar}.
Most modern determinations~\cite{STS93} of the $\pi NN$ coupling constant
yield a value for the so-called pseudovector coupling constant,
$f_\pi$~\cite{Dum83}. Assuming that $f_\pi$ is fundamentally constant,
then $g_\pi$ has a small charge dependence, since
the two coupling constants are related by
\begin{equation}
\frac{g^2_\pi(M)}{4\pi} 
= \frac{4M^2}{m_{\pi^\pm}^2}
\frac{f^2_\pi}{4\pi} 
\: ,
\label{eq_fg}
\end{equation}
with $M$ the mean of the masses of the two nucleons involved in
the $\pi NN$ vertex.
We take this very small effect into account by using
in our $V^{OPE}$ the $\pi NN$ coupling constant
\begin{equation}
\frac{g^2_\pi(M)}{4\pi} 
\equiv \frac{M^2}{M_p^2}
\frac{\bar{g}^2_\pi}{4\pi} 
\: ,
\end{equation}
with
\begin{equation}
\frac{\bar{g}^2_\pi}{4\pi} 
= 13.6
\: .
\end{equation}
Defining,
\begin{equation}
\frac{\bar{g}^2_\pi}{4\pi} 
= \frac{4M_p^2}{m_{\pi^\pm}^2}
\frac{f^2_\pi}{4\pi} 
\: ,
\end{equation}
recovers Eq.~(\ref{eq_fg}).

Since we use units such that $\hbar=c=1$,
energies, masses and momenta are in units of MeV.
The potential is in units of MeV$^{-2}$.
 The conversion factor is $\hbar c 
=197.327053$ MeV fm.  If the user wants to relate our units 
and conventions to the ones used by other practitioners, he/she should 
compare our Eq.~(\ref{22.5}) and our phase shift relation Eq.~(\ref{23.1}) 
with the corresponding equations used by others.
A FORTRAN77 computer code for the CD-Bonn potential is available from the
author.

\section{Potential parameters}
For the `basic' mesons, $\pi$, $\omega$, and $\rho$, we use, in
general, the parameters shown in Table~\ref{tab_basicpar}.
Note that (except for the cutoff masses) these parameters
are determined from empirical or semi-empirical
sources and, therefore, they are not free parameters of our model.
The intermediate range attraction is described by
two scalar isoscalar bosons, $\sigma_1$ and $\sigma_2$,
that are also used for the fine-tuning of individual partial
waves.  The $\sigma$ parameters are given 
in Table~\ref{tab_par1} for the $pp$ $T=1$ potential 
and in Table~\ref{tab_par2} for the $T=0$ $np$ potential. 
For partial waves with $J\geq 6$, we take 
$g^2_{\sigma_1}/4\pi = 2.3$ 
and $m_{\sigma_1} = 452$ MeV.
The cutoff mass for the two $\sigma$ is 
$\Lambda_{\sigma_1}
=\Lambda_{\sigma_2} = 2.5$ GeV, 
in all partial waves.
In two cases, we vary the cutoff
parameter of one of the `basic' mesons:
in $^1P_1$ we apply $\Lambda_\omega \rightarrow \infty$ (i.~e., the
$\omega$ cutoff is omitted), and in $^3P_2/^3F_2$ we use
$\Lambda_\pi = 3.0$ GeV; otherwise, the same cutoff masses
(namely, the ones shown in Table~\ref{tab_basicpar} and
$\Lambda_{\sigma_1}
=\Lambda_{\sigma_2} = 2.5$ GeV) are used in all partial waves.

The $nn$ $T=1$ potential is constructed by starting from the
$pp$ $T=1$ potential, replacing the proton mass by the neutron mass
and adjusting the coupling constants of the two $\sigma$ such
that the CSB phase shift differences listed in the last column
(`Total') of Table~\ref{tab_CSB} are reproduced. Thus, the 
$\sigma$ coupling constants of the $nn$ potential (which are
given in Table~\ref{tab_par3}) are not free parameters.
The procedure for the $T=1$ $np$ potential is similar.
We start from the $pp$ $T=1$ potential, replace the proton mass
by the average mass given in Eq.~(\ref{eq_mav}),
apply the appropriate OPE potential [i.~e., we replace
Eq.~(\ref{eq_opepp}) by (\ref{eq_openp1})],
and then adjust the $\sigma$ coupling constants such that
the CIB phase shift differences listed in column `Total'
of Table~\ref{tab_CIB} are reproduced which, again, does not
generate any free parameters. The exception is the $^1S_0$
state where the $\sigma$ parameters are used to minimized
the $\chi^2$ in regard to the $np$ data. 
The charge-dependence caused by the Bonn Full Model
produces also a small charge-dependent tensor force of
$2\pi$ range that can be simulated with the help of
the $\rho$ coupling. 
A noticeable effect occurs only in the coupled $^3P_2/^3F_2$
states where we use
$g^2_\rho/4\pi=0.844$ for $nn$
 and $g^2_\rho/4\pi=0.862$ for $np$
(in all other states $g^2_\rho/4\pi=0.84$).
Again, these choices are made to reproduce the CSB and CIB
predicted by other sources and, thus, does not introduce
new parameters.

The free (`fit') parameters of our model are the ones
given in Table~\ref{tab_par1} and \ref{tab_par2}
plus two parameters for $^1S_0$ $np$
and the cutoff masses which adds up to a total
of 43 free parameters.

\section{Deuteron calculations}
In momentum space, the deuteron wave function is given by
\begin{equation}
  \Psi_d^M({\bf k}) 
= \left[
  \psi_0(k) {\cal Y}^{1M}_{01}({\bf \hat{k}}) +
  \psi_2(k) {\cal Y}^{1M}_{21}({\bf \hat{k}}) \right]
  \zeta_0^0,                                           \label{DWQ}
\end{equation}
where 
${\cal Y}^{JM}_{LS}({\bf \hat{k}})$
are the normalized eigenfunctions of the two-nucleon
orbital angular momentum $L$, spin $S$, and total angular
momentum $J$ with projection $M$;
$\zeta_T^{M_T}$ denotes the normalized eigenstates of the total
isospin $T$ with projection $M_T$ of the two nucleons.
The normalization is
\begin{equation}
\langle \Psi_d^M | \Psi_d^M \rangle
 =
  \int_0^\infty dk k^2 \left[ \psi_0^2(k) + \psi_2^2(k) \right]=1.
\end{equation}

The wave functions are obtained by solving
the bound state equation which is the homogenous version
of the scattering equation~(\ref{21.18}):
\begin{equation}
\psi({\bf k}) = \frac{M}{-\gamma^2 - k^2} \int d^3k' V({\bf k},
{\bf k}') \psi({\bf k}')
\: .
\label{eq_bound}
\end{equation}
Note that the deuteron is a pole in the $S$ matrix at $q=i\gamma$.
Since we use relativistic kinematics in $np$ scattering [cf.\
Eq.~(\ref{eq_relkin})],
consistency requires that we determine $\gamma$ based upon
relativistic kinematics which is,
\begin{equation}
   M_d \equiv M_p + M_n - B_d = \sqrt{M_p^2-\gamma^2} +
            \sqrt{M_n^2-\gamma^2}  
\, ,
\label{DGAMM}
\end{equation}
where $M_d$ denotes the deuteron rest mass
and $B_d$ the binding energy.
The formal solution of Eq.~(\ref{DGAMM}) is
\begin{equation}
   \gamma^2 = \left[4M_p^2 M_n^2 - (M_d^2-M_p^2-M_n^2)^2\right]/4M_d^2,
\label{DGAMM2}
\end{equation}
and, using $B_d=2.224575$ MeV and $\hbar c = 197.327053$ MeV fm,
the accurate numerical value for $\gamma$ comes out to be
\begin{equation}
   \gamma = 0.2315380 \mbox{ fm}^{-1}.  \label{DGAMMA}
\end{equation}
To obtain more insight into $\gamma^2$,
we rewrite Eq.~(\ref{DGAMM2}) in factorized form,
\begin{eqnarray}
4M_d^2\gamma^2 &=& \left[ (M_n+M_p)^2 - M_d^2 \right]
                 \left[ M_d^2 - (M_n-M_p)^2 \right] \nonumber\\
               &=& B_d (4\overline{M}-B_d)(M_d^2-\delta M^2)
\end{eqnarray}
where we introduce the average nucleon mass,
\begin{equation}
\overline{M} \equiv \frac{M_p + M_n}{2} = 938.91897 \mbox{ MeV,}
\end{equation}
and the nucleon mass difference
$\delta M \equiv M_n-M_p=1.29332$ MeV, and used
$M_d=2\overline{M}-B_d$.
From this we get
\begin{equation}
\gamma^2 = \overline{M}B_d \left( 1 - \frac{B_d}{4\overline{M}} \right)
                \left( 1 - \frac{\delta M^2}{M_d^2} \right) \; ,
\end {equation}
and, in terms of twice the reduced nucleon mass, 
$\widehat{M}$, which is defined by
\begin{equation}
   \widehat{M} \equiv \frac{2 M_p M_n}{M_p + M_n}
 = \overline{M}
\left( 1 -\frac{\delta M^2}{4\overline{M}^2} \right) 
 = 938.91852 \mbox{ MeV}
\; ,
\end{equation}
we finally obtain
\begin{eqnarray}
\gamma^2 &=& \widehat{M} B_d \left( 1 - \frac{B_d}{4\overline{M}} \right)
           \frac{1-\frac{\delta M^2}{M_d^2}}
                {1-\frac{\delta M^2}{4\overline{M}^2}} \nonumber\\
 & \approx & \widehat{M} B_d \left( 1 - \frac{B_d}{4\overline{M}} \right) \; .
\label{DGAMMAPPR}
\end{eqnarray}
The approximation involved in Eq.~(\ref{DGAMMAPPR}) is
good to one part in $10^9$.
Therefore, this equation reproduces the exact value for $\gamma$
to all digits given in Eq.~(\ref{DGAMMA}).
One can now identify the term $\widehat{M}B_d$ as the non-relativistic
approximation to $\gamma^2$ and the factor $(1-B_d/4\overline{M})$ as the
essential relativistic correction. 

Partial wave decomposition of Eq.~(\ref{eq_bound})
yields for the coupled $^3S_1$ and $^3D_1$ states,
\begin{eqnarray}
  \psi_0(k) & = & -\frac{\widehat{M}}{\gamma^2+k^2} \int_0^\infty dk' k'^2
  \left[ V_{00}(k,k')\psi_0(k') + V_{02}(k,k')\psi_2(k') \right],
                                                            \nonumber\\
  \psi_2(k) & = & -\frac{\widehat{M}}{\gamma^2+k^2} \int_0^\infty dk' k'^2
  \left[ V_{20}(k,k')\psi_0(k') + V_{22}(k,k')\psi_2(k') \right]
\: ,
                                                            \label{DEQ}
\end{eqnarray}
from which $\psi_0$ and $\psi_2$ are obtained.
Considering a finite set of discrete arguments for the functions on the
l.h.s.\ and using the same set of momenta to discretize the integrals
on the r.h.s.\ produces a matrix equation that is solved easily by the
matrix-inversion method~\cite{HT70}.

The momentum-space wave funtions can be
Fourier transformed into 
the configuration-space wave functions $u$ and $w$ by
\begin{equation}
  \frac{u_L(r)}{r} = \sqrt{\frac{2}{\pi}}
  \int_0^\infty dk k^2 j_L(kr) \psi_L(k) \>\>,         \label{DFTW}
\end{equation}
with $u_0(r)\equiv u(r)$, $u_2(r)\equiv w(r)$,
and $j_L$ the spherical Bessel functions.
The normalization is
\begin{equation}
   \int_0^\infty dr \left[ u^2(r) + w^2(r) \right] = 1.
\end{equation}
The asymptotic behavior of the wave functions for large values of $r$
are
\begin{eqnarray}
  u(r) & \sim & A_S {\rm e}^{-\gamma r},  \nonumber\\
  w(r) & \sim & A_D {\rm e}^{-\gamma r}
      \left[ 1 + \frac{3}{(\gamma r)} + \frac{3}{(\gamma r)^2} \right],
                                    \label{DAW}
\end{eqnarray}
where $A_S$ and $A_D$ are known as the asymptotic $S$- and $D$-state
normalizations, respectively. In addition, one defines the
``$D/S$-state ratio'' $\eta\equiv A_D/A_S$.
Other deuteron parameters of interest are the quadrupole moment
\begin{equation}
  Q_d = \frac{1}{20} \int_0^\infty dr r^2 w(r)
        \left[ \sqrt{8} u(r) - w(r) \right],
\end{equation}
the root-mean-square or matter radius
\begin{equation}
  r_d = \frac12 \left\{\int_0^\infty dr r^2 \left[u^2(r) + w^2(r)
                \right]\right\}^{1/2},
\end{equation}
and the $D$-state probability
\begin{equation}
   P_D = \int_0^\infty dr w^2(r).
\end{equation}
The predictions by the CD-Bonn potential for the properties
of the deuteron are given in Table~\ref{tab_deu}; numerical values for
the wave functions are listed in Table~\ref{tab_dwaves}
and plots are shown in
Figs.~\ref{fig_dwaves} and \ref{fig_dwaves2}.

In some applications, it is convenient to have the deuteron
wave functions in analytic form. Therefore, we present here
a simple parametrization of the deuteron functions (that was first introduced
in Ref.~\cite{Lac81}).
The ansatz for the analytic version of the $r$-space wave functions is
\begin{eqnarray}
u_{a}(r) & = & \sum^{n}_{j=1} C_{j} \exp (-m_{j}r)\\
w_{a}(r) & = & \sum^{n}_{j=1} D_{j} \exp (-m_{j}r)
\left[1+\frac{3}{m_{j}r}+\frac{3}{(m_{j}r)^{2}}\right] \, .
\end{eqnarray}
The corresponding momentum space wave functions are
\begin{eqnarray}
\psi^{a}_{0}(q) & = & (2/\pi)^{1/2}
\sum_{j=1}^{n} \frac{C_{j}}{q^{2}+m^{2}_{j}}\\
\psi^{a}_{2}(q) & = & (2/\pi)^{1/2}
\sum_{j=1}^{n} \frac{D_{j}}{q^{2}+m^{2}_{j}}
\: .
\end{eqnarray}
The boundary conditions $u_{a}(r) \rightarrow r$ and
$w_{a}(r)\rightarrow r^{3}$ as $r\rightarrow 0$ lead to one constraint
for the $C_{j}$ and three constraints for the $D_{j}$~\cite{Lac81}, namely
\begin{eqnarray}
C_{n} & = & -\sum_{j=1}^{n-1} C_{j}
\label{eq_dbound1}
\\
D_{n-2} & = & \frac{m^{2}_{n-2}}{(m^{2}_{n}-m^{2}_{n-2})
(m^{2}_{n-1}-m^{2}_{n-2})} \left[ -m^{2}_{n-1}m^{2}_{n}\sum_{j=1}^{n-3}
\frac{D_{j}}{m^{2}_{j}}
\right.
\nonumber \\
 & & 
\left.
+(m^{2}_{n-1}+m^{2}_{n})\sum_{j=1}^{n-3}D_{j}
-\sum_{j=1}^{n-3}D_{j}m^{2}_{j} \right]
\label{eq_dbound2}
\end{eqnarray}
and two other relations obtained by circular permutation of
$n-2, n-1, n$. The masses are
\begin{equation}
m_{j}=\gamma+(j-1)m_{0}
\end{equation}
with $m_{0}=0.9$ fm$^{-1}$ and $\gamma$ given in Eq.~(\ref{DGAMMA}).
The parameters are given in Table~\ref{tab_dwpar}.
The constraints Eqs.~(\ref{eq_dbound1}) and (\ref{eq_dbound2})
must be enforced by double precision (i.~e., to about 15 decimal
digits), otherwise the wave function is not reproduced correctly
for $r\leq 0.5$ fm. This applies, particularly, to the $D$ wave.
The accuracy of the parametrization is characterized by
\begin{equation}
\left\{\int_0^\infty dr \left[ u(r) - u_a(r) \right]^2\right\}^{1/2}
= 2.2 \times 10^{-4}
\end{equation}
and
\begin{equation}
\left\{\int_0^\infty dr \left[ w(r) - w_a(r) \right]^2\right\}^{1/2}
= 1.1 \times 10^{-4} \: .
\end{equation}
Data files for the deuteron wave functions in $r$-space as well as in
momentum space can be obtained from the author upon request.

\pagebreak

\begin{table}
\caption{Basic constants and parameters adopted for the CD-Bonn potential.}
\begin{tabular}{ldddd}
  Particle
 & Mass (MeV)
 & $g^2/4\pi$
 & $f/g$
 & $\Lambda$ (GeV)
\\
\hline 
$\pi^\pm$  &  139.56995  &  13.6  &  --   &  1.72  \\
$\pi^0$    &  134.9764   &  13.6  &  --   &  1.72  \\
$\rho^\pm, \rho^0$& 769.9 & 0.84  &  6.1  &  1.31  \\
$\omega$   &  781.94     &  20.0  &  0.0  &  1.5   \\
Proton ($p$)& 938.27231    \\
Neutron ($n$)&939.56563    \\
\end{tabular}
\label{tab_basicpar}
\end{table}


\vspace*{6cm}

\begin{table}
\caption{Differences between the $pp$ and $nn$ $^1S_0$ effective 
range parameters [as defined in 
Eqs.~(\ref{eq_CSBlep1})
and (\ref{eq_CSBlep2})]
due to the impact of
nucleon mass splitting on the kinetic energy (Kin.\ en.),
one-boson exchange diagrams (OBE), and two-boson exchanges (TBE).
Total denotes the sum of the three contributions and empirical
information is given in the last column.}
\begin{tabular}{lrrrrr}
   
 & Kin.\ en.\
 & OBE
 & TBE
 & Total
 & Empirical
\\
\hline 
$\Delta a_{CSB}$ (fm)&  0.263 & --0.030 &   1.275 &   1.508 & 1.6 $\pm$ 0.6 \\
$\Delta r_{CSB}$ (fm)&  0.004 &   0.000 &   0.022 &   0.026 & 0.10 $\pm$ 0.12\\
\end{tabular}
\label{tab_CSBlep}
\end{table}

\pagebreak

\begin{table}
\caption{
Difference $\delta_{nn} - \delta_{pp}$ 
(in degrees) due to the impact of
nucleon mass splitting on kinematics~\protect\cite{foot1},
one-boson exchange diagrams (OBE), and two-boson exchanges (TBE).
Total is the sum of all.}
\begin{tabular}{lrrrr}
$T_{lab}$ (MeV)
 & Kinematics
 & OBE
 & TBE
 & Total
\\
 \hline 
\hline                                                                          
\\ \multicolumn{5}{c}{$^1S_0$} \\                                               
0.38254&  0.404 & --0.045 &   1.795 &   2.154 \\
    1 &   0.324 & --0.036 &   1.440 &   1.728 \\
    5 &   0.165 & --0.018 &   0.785 &   0.932 \\
   10 &   0.114 & --0.013 &   0.591 &   0.692 \\
   25 &   0.062 & --0.006 &   0.408 &   0.464 \\
   50 &   0.031 & --0.001 &   0.310 &   0.340 \\
  100 &   0.003 &   0.005 &   0.239 &   0.247 \\
  150 & --0.013 &   0.010 &   0.206 &   0.203 \\
  200 & --0.023 &   0.014 &   0.185 &   0.176 \\
  300 & --0.039 &   0.021 &   0.160 &   0.142 \\
\hline                                                                          
\\ \multicolumn{5}{c}{$^3P_0$} \\                                               
    5 &   0.006 &   0.001 &   0.001 &   0.008 \\
   10 &   0.013 &   0.003 &   0.002 &   0.018 \\
   25 &   0.022 &   0.010 &   0.008 &   0.040 \\
   50 &   0.021 &   0.021 &   0.014 &   0.056 \\
  100 &   0.004 &   0.036 &   0.020 &   0.060 \\
  150 & --0.011 &   0.045 &   0.024 &   0.058 \\
  200 & --0.022 &   0.052 &   0.024 &   0.054 \\
  300 & --0.040 &   0.063 &   0.025 &   0.048 \\
\hline                                                                          
\\ \multicolumn{5}{c}{$^3P_1$} \\                                               
    5 & --0.003 &   0.000 &   0.002 & --0.001 \\
   10 & --0.006 &   0.000 &   0.004 & --0.002 \\
   25 & --0.011 &   0.001 &   0.012 &   0.002 \\
   50 & --0.017 &   0.002 &   0.027 &   0.012 \\
  100 & --0.026 &   0.006 &   0.049 &   0.029 \\
  150 & --0.033 &   0.009 &   0.065 &   0.041 \\
  200 & --0.039 &   0.011 &   0.076 &   0.048 \\
  300 & --0.050 &   0.016 &   0.090 &   0.056 \\
\hline                                                                          
\\ \multicolumn{5}{c}{$^1D_2$} \\                                               
   10 &   0.001 &   0.000 &   0.000 &   0.001 \\
   25 &   0.002 &   0.000 &   0.002 &   0.004 \\
   50 &   0.005 &   0.000 &   0.006 &   0.011 \\
  100 &   0.011 &   0.002 &   0.019 &   0.032 \\
  150 &   0.016 &   0.005 &   0.033 &   0.054 \\
  200 &   0.019 &   0.010 &   0.046 &   0.075 \\
  300 &   0.022 &   0.022 &   0.068 &   0.112 \\
\hline                                                                          
\\ \multicolumn{5}{c}{$^3P_2$} \\                                               
    5 &   0.001 &   0.000 &   0.001 &   0.002 \\
   10 &   0.003 &   0.000 &   0.004 &   0.007 \\
   25 &   0.010 &   0.001 &   0.013 &   0.024 \\
   50 &   0.021 &   0.002 &   0.031 &   0.054 \\
  100 &   0.032 &   0.006 &   0.062 &   0.100 \\
  150 &   0.036 &   0.010 &   0.081 &   0.127 \\
  200 &   0.035 &   0.015 &   0.093 &   0.143 \\
  300 &   0.032 &   0.023 &   0.105 &   0.160 \\
\end{tabular}
\label{tab_CSB}
\end{table}

\pagebreak

\begin{table}
\caption{Differences between the $pp$ and $np$ $^1S_0$ effective 
range parameters [as defined in 
Eqs.~(\ref{eq_CIBlep1})
and (\ref{eq_CIBlep2})]
produced by various CIB mechanisms and phenomenology.
Total is the sum of all contributions listed left of column `Total'.
$\Delta M$ denotes all effects caused by nucleon mass splitting.
Empirical information is given in the last column.}
\begin{tabular}{lrrrrrrr}
                   
 & $\Delta M$
 & OPE
 & TBE 
 & $\pi\gamma$
 & phenom.
 & Total 
 & Empirical
\\
\hline                                                                          
$\Delta a_{CIB}$ (fm)& 0.754 & 3.035 & 1.339 & --0.405 & 1.555 & 6.278 
 & $6.44 \pm 0.40$ \\
$\Delta r_{CIB}$ (fm)& 0.013 & 0.092 & 0.016 & --0.004 & 0.057 & 0.174 
 & $0.08 \pm 0.06$ \\
\end{tabular}
\label{tab_CIBlep}
\end{table}


\vspace*{6cm}

\begin{table}
\caption{
Difference 
$\delta_{np} - \delta_{pp}$ 
(in degrees) in the $^1S_0$ state
as produced by various CIB mechanisms and phenomenology.
$\Delta M$ stands for all 
effects caused by nucleon mass splitting~\protect\cite{foot3}.
Total is the sum of all contributions listed left of column `Total'.
`All' denotes the sum of Total and Coulomb, where Coulomb is the difference
$\delta_{pp} - \delta_{pp}^C$. 
}
\begin{tabular}{lrrrrrrrr}
 $T_{lab}$ (MeV) 
 & $\Delta M$
 & OPE
 & TBE 
 & $\pi\gamma$
 & phenom.
 & Total 
 & Coulomb
 & All
\\
 \hline 
\hline                                                                          
0.38254& 1.077 & 3.541 & 1.655 & --0.412 & 1.953 & 7.814 &32.085 &39.894 \\
    1 & 0.859 & 2.851 & 1.260 &-0.305 & 1.521 & 6.186 &23.114 &29.300 \\
    5 & 0.468 & 1.650 & 0.654 &-0.152 & 0.982 & 3.602 & 5.219 & 8.821 \\
   10 & 0.350 & 1.271 & 0.482 &-0.106 & 0.909 & 2.906 & 1.896 & 4.802 \\
   25 & 0.240 & 0.875 & 0.320 &-0.058 & 0.970 & 2.347 &-0.044 & 2.304 \\
   50 & 0.182 & 0.656 & 0.233 &-0.028 & 1.142 & 2.185 &-0.589 & 1.597 \\
  100 & 0.139 & 0.513 & 0.165 &-0.002 & 1.433 & 2.248 &-0.772 & 1.476 \\
  150 & 0.119 & 0.469 & 0.130 & 0.012 & 1.656 & 2.386 &-0.796 & 1.590 \\
  200 & 0.108 & 0.457 & 0.103 & 0.021 & 1.839 & 2.528 &-0.796 & 1.733 \\
  300 & 0.094 & 0.477 & 0.058 & 0.034 & 2.124 & 2.787 &-0.782 & 2.005 \\
\end{tabular}
\label{tab_CIB1S0}
\end{table}

\pagebreak

\begin{table}
\caption{
Difference 
$\delta_{np} - \delta_{pp}$ 
(in degrees) for partial waves with $L>0$
as produced by various CIB mechanisms. 
Notation as in Table~\ref{tab_CIB1S0}.}
\begin{tabular}{lrrrrrrr}
 $T_{lab}$ (MeV)
 & $\Delta M$
 & OPE
 & TBE 
 & $\pi\gamma$
 & Total 
 & Coulomb
 & All
\\
 \hline 
\hline                                                                          
\\ \multicolumn{8}{c}{$^3P_0$} \\                                               
    1 & 0.000 &-0.030 & 0.000 & 0.000 &-0.030 & 0.073 & 0.043 \\
    5 & 0.000 &-0.230 &-0.003 & 0.000 &-0.233 & 0.262 & 0.029 \\
   10 & 0.000 &-0.448 &-0.009 & 0.000 &-0.457 & 0.353 &-0.104 \\
   25 & 0.012 &-0.770 &-0.027 &-0.017 &-0.802 & 0.320 &-0.481 \\
   50 & 0.032 &-0.846 &-0.050 &-0.050 &-0.914 & 0.111 &-0.803 \\
  100 & 0.050 &-0.742 &-0.074 &-0.087 &-0.853 &-0.142 &-0.996 \\
  150 & 0.050 &-0.649 &-0.083 &-0.104 &-0.786 &-0.255 &-1.041 \\
  200 & 0.047 &-0.586 &-0.088 &-0.113 &-0.740 &-0.314 &-1.054 \\
  300 & 0.045 &-0.513 &-0.096 &-0.125 &-0.689 &-0.369 &-1.058 \\
\hline                                                                          
\\ \multicolumn{8}{c}{$^3P_1$} \\                                               
    1 & 0.000 & 0.016 & 0.000 & 0.000 & 0.016 &-0.043 &-0.026 \\
    5 & 0.002 & 0.110 & 0.001 &-0.002 & 0.111 &-0.140 &-0.028 \\
   10 & 0.004 & 0.193 & 0.003 &-0.002 & 0.198 &-0.187 & 0.011 \\
   25 & 0.006 & 0.298 & 0.008 & 0.003 & 0.315 &-0.224 & 0.091 \\
   50 & 0.008 & 0.330 & 0.018 & 0.016 & 0.372 &-0.240 & 0.133 \\
  100 & 0.016 & 0.307 & 0.038 & 0.038 & 0.399 &-0.265 & 0.133 \\
  150 & 0.022 & 0.274 & 0.055 & 0.054 & 0.405 &-0.287 & 0.118 \\
  200 & 0.028 & 0.246 & 0.069 & 0.064 & 0.407 &-0.303 & 0.103 \\
  300 & 0.033 & 0.202 & 0.099 & 0.077 & 0.411 &-0.325 & 0.085 \\
\hline                                                                          
\\ \multicolumn{8}{c}{$^1D_2$} \\                                               
    5 & 0.000 &-0.009 & 0.000 & 0.000 &-0.009 & 0.007 &-0.002 \\
   10 & 0.000 &-0.024 & 0.000 & 0.000 &-0.024 & 0.015 &-0.009 \\
   25 & 0.000 &-0.049 & 0.001 & 0.001 &-0.047 & 0.031 &-0.016 \\
   50 & 0.002 &-0.043 & 0.005 &-0.002 &-0.038 & 0.049 & 0.011 \\
  100 & 0.014 & 0.003 & 0.013 &-0.011 & 0.019 & 0.071 & 0.090 \\
  150 & 0.024 & 0.041 & 0.023 &-0.018 & 0.070 & 0.081 & 0.151 \\
  200 & 0.034 & 0.068 & 0.030 &-0.025 & 0.107 & 0.083 & 0.190 \\
  300 & 0.045 & 0.095 & 0.042 &-0.033 & 0.149 & 0.073 & 0.222 \\
\hline                                                                          
\\ \multicolumn{8}{c}{$^3P_2$} \\                                               
    5 & 0.000 &-0.009 &-0.001 & 0.000 &-0.010 & 0.049 & 0.040 \\
   10 & 0.001 &-0.028 &-0.002 & 0.000 &-0.029 & 0.094 & 0.065 \\
   25 & 0.004 &-0.090 &-0.005 &-0.001 &-0.092 & 0.188 & 0.097 \\
   50 & 0.017 &-0.162 &-0.011 &-0.006 &-0.162 & 0.257 & 0.095 \\
  100 & 0.043 &-0.211 &-0.024 &-0.020 &-0.212 & 0.260 & 0.048 \\
  150 & 0.058 &-0.210 &-0.032 &-0.030 &-0.214 & 0.221 & 0.007 \\
  200 & 0.065 &-0.196 &-0.035 &-0.037 &-0.203 & 0.184 &-0.019 \\
  300 & 0.072 &-0.169 &-0.034 &-0.044 &-0.175 & 0.130 &-0.044 \\
\hline                                                                          
\\ \multicolumn{8}{c}{$^3F_2$} \\                                               
   10 & 0.000 &-0.004 & 0.000 & 0.000 &-0.004 & 0.001 &-0.002 \\
   25 & 0.000 &-0.019 & 0.000 & 0.000 &-0.019 & 0.004 &-0.015 \\
   50 & 0.000 &-0.043 & 0.000 & 0.001 &-0.042 & 0.007 &-0.036 \\
  100 & 0.000 &-0.068 & 0.000 & 0.002 &-0.066 & 0.008 &-0.058 \\
  150 & 0.003 &-0.081 &-0.001 & 0.002 &-0.077 & 0.007 &-0.070 \\
  200 & 0.007 &-0.090 &-0.001 & 0.002 &-0.082 & 0.003 &-0.079 \\
  300 & 0.008 &-0.099 &-0.001 & 0.002 &-0.090 &-0.009 &-0.098 \\
\hline                                                                          
\\ \multicolumn{8}{c}{$\epsilon_2$} \\                                               
    5 & 0.000 & 0.011 & 0.000 & 0.000 & 0.011 &-0.008 & 0.004 \\
   10 & 0.001 & 0.034 & 0.000 &-0.001 & 0.034 &-0.016 & 0.018 \\
   25 & 0.002 & 0.086 & 0.000 &-0.002 & 0.086 &-0.028 & 0.058 \\
   50 &-0.001 & 0.111 & 0.003 & 0.001 & 0.114 &-0.025 & 0.089 \\
  100 &-0.004 & 0.087 & 0.007 & 0.010 & 0.100 &-0.003 & 0.097 \\
  150 &-0.004 & 0.051 & 0.010 & 0.018 & 0.075 & 0.017 & 0.092 \\
  200 &-0.001 & 0.020 & 0.012 & 0.024 & 0.055 & 0.032 & 0.087 \\
  300 & 0.008 &-0.020 & 0.014 & 0.032 & 0.034 & 0.047 & 0.080 \\
\end{tabular}
\label{tab_CIB}
\end{table}

\pagebreak

\begin{table}
\caption{$pp$ phase shifts in degrees.}
\begin{tabular}{lrrrrrrrrrr}
 $T_{lab}$ (MeV)
 & $^1S_0$
 & $^3P_0$
 & $^3P_1$
 & $^1D_2$
 & $^3P_2$
 & $^3F_2$
 & $\epsilon_2$
 & $^3F_3$
 & $^1G_4$
 & $^3F_4$
\\
 \hline 
    1 &  32.79 &   0.13 &  -0.08 &   0.00 &   0.01 &   0.00 &   0.00 &   0.00 &   0.00 &   0.00 \\
    5 &  54.85 &   1.58 &  -0.90 &   0.04 &   0.22 &   0.00 &  -0.05 &   0.00 &   0.00 &   0.00 \\
   10 &  55.20 &   3.72 &  -2.05 &   0.17 &   0.66 &   0.01 &  -0.20 &  -0.03 &   0.00 &   0.00 \\
   25 &  48.63 &   8.58 &  -4.90 &   0.70 &   2.50 &   0.10 &  -0.81 &  -0.23 &   0.04 &   0.02 \\
   50 &  38.86 &  11.54 &  -8.31 &   1.71 &   5.84 &   0.33 &  -1.73 &  -0.70 &   0.15 &   0.12 \\
  100 &  24.91 &   9.57 & -13.37 &   3.77 &  10.97 &   0.78 &  -2.72 &  -1.53 &   0.42 &   0.50 \\
  150 &  14.73 &   4.76 & -17.62 &   5.67 &  13.98 &   1.10 &  -2.99 &  -2.12 &   0.69 &   1.04 \\
  200 &   6.58 &  -0.49 & -21.49 &   7.26 &  15.68 &   1.27 &  -2.88 &  -2.48 &   0.97 &   1.63 \\
  250 &  -0.29 &  -5.62 & -25.05 &   8.55 &  16.63 &   1.26 &  -2.59 &  -2.68 &   1.26 &   2.19 \\
  300 &  -6.26 & -10.48 & -28.36 &   9.54 &  17.12 &   1.08 &  -2.21 &  -2.75 &   1.55 &   2.69 \\
  350 & -11.56 & -15.04 & -31.45 &  10.27 &  17.33 &   0.73 &  -1.80 &  -2.72 &   1.83 &   3.11 \\
\end{tabular}
\label{tab_phpp}
\end{table}

\vspace*{5cm}

\begin{table}
\caption{$nn$ phase shifts in degrees.}
\begin{tabular}{lrrrrrrrrrr}
 $T_{lab}$ (MeV)
 & $^1S_0$
 & $^3P_0$
 & $^3P_1$
 & $^1D_2$
 & $^3P_2$
 & $^3F_2$
 & $\epsilon_2$
 & $^3F_3$
 & $^1G_4$
 & $^3F_4$
\\
 \hline 
    1 &  57.63 &   0.21 &  -0.12 &   0.00 &   0.02 &   0.00 &   0.00 &   0.00 &   0.00 &   0.00 \\
    5 &  61.00 &   1.85 &  -1.04 &   0.05 &   0.27 &   0.00 &  -0.06 &  -0.01 &   0.00 &   0.00 \\
   10 &  57.79 &   4.10 &  -2.24 &   0.18 &   0.76 &   0.01 &  -0.22 &  -0.04 &   0.00 &   0.00 \\
   25 &  49.05 &   8.94 &  -5.13 &   0.74 &   2.71 &   0.11 &  -0.85 &  -0.24 &   0.04 &   0.02 \\
   50 &  38.61 &  11.71 &  -8.54 &   1.77 &   6.15 &   0.34 &  -1.76 &  -0.71 &   0.16 &   0.12 \\
  100 &  24.38 &   9.49 & -13.60 &   3.88 &  11.33 &   0.79 &  -2.73 &  -1.55 &   0.42 &   0.52 \\
  150 &  14.14 &   4.56 & -17.87 &   5.80 &  14.32 &   1.11 &  -2.97 &  -2.13 &   0.70 &   1.06 \\
  200 &   5.96 &  -0.75 & -21.74 &   7.42 &  16.01 &   1.28 &  -2.85 &  -2.49 &   0.98 &   1.66 \\
  250 &  -0.92 &  -5.92 & -25.31 &   8.72 &  16.94 &   1.27 &  -2.54 &  -2.68 &   1.28 &   2.23 \\
  300 &  -6.90 & -10.80 & -28.63 &   9.72 &  17.42 &   1.09 &  -2.15 &  -2.74 &   1.57 &   2.73 \\
  350 & -12.21 & -15.38 & -31.72 &  10.46 &  17.60 &   0.73 &  -1.74 &  -2.70 &   1.86 &   3.15 \\
\end{tabular}
\label{tab_phnn}
\end{table}

\pagebreak

\begin{table}
\caption{$T=1$ $np$ phase shifts in degrees.}
\begin{tabular}{lrrrrrrrrrr}
 $T_{lab}$ (MeV)
 & $^1S_0$
 & $^3P_0$
 & $^3P_1$
 & $^1D_2$
 & $^3P_2$
 & $^3F_2$
 & $\epsilon_2$
 & $^3F_3$
 & $^1G_4$
 & $^3F_4$
\\
 \hline 
    1 &  62.09 &   0.18 &  -0.11 &   0.00 &   0.02 &   0.00 &   0.00 &   0.00 &   0.00 &   0.00 \\
    5 &  63.67 &   1.61 &  -0.93 &   0.04 &   0.26 &   0.00 &  -0.05 &   0.00 &   0.00 &   0.00 \\
   10 &  60.01 &   3.62 &  -2.04 &   0.16 &   0.72 &   0.01 &  -0.18 &  -0.03 &   0.00 &   0.00 \\
   25 &  50.93 &   8.10 &  -4.81 &   0.69 &   2.60 &   0.09 &  -0.76 &  -0.20 &   0.03 &   0.02 \\
   50 &  40.45 &  10.74 &  -8.18 &   1.73 &   5.93 &   0.30 &  -1.64 &  -0.62 &   0.13 &   0.11 \\
  100 &  26.38 &   8.57 & -13.23 &   3.86 &  11.01 &   0.72 &  -2.63 &  -1.42 &   0.39 &   0.48 \\
  150 &  16.32 &   3.72 & -17.51 &   5.82 &  13.98 &   1.03 &  -2.90 &  -1.98 &   0.67 &   1.01 \\
  200 &   8.31 &  -1.55 & -21.38 &   7.45 &  15.66 &   1.19 &  -2.79 &  -2.33 &   0.96 &   1.59 \\
  250 &   1.59 &  -6.68 & -24.96 &   8.76 &  16.59 &   1.17 &  -2.50 &  -2.51 &   1.26 &   2.15 \\
  300 &  -4.25 & -11.54 & -28.27 &   9.76 &  17.08 &   0.98 &  -2.13 &  -2.57 &   1.56 &   2.65 \\
  350 &  -9.44 & -16.10 & -31.37 &  10.49 &  17.28 &   0.62 &  -1.72 &  -2.53 &   1.85 &   3.06 \\
\end{tabular}
\label{tab_phnp1}
\end{table}

\vspace*{5cm}

\begin{table}
\caption{$T=0$ $np$ phase shifts in degrees.}
\begin{tabular}{lrrrrrrrrrr}
 $T_{lab}$ (MeV)
 & $^1P_1$
 & $^3S_1$
 & $^3D_1$
 & $\epsilon_1$
 & $^3D_2$
 & $^1F_3$
 & $^3D_3$
 & $^3G_3$
 & $\epsilon_3$
 & $^3G_4$
\\
 \hline 
    1 &  -0.19 & 147.75 &  -0.01 &   0.11 &   0.01 &   0.00 &   0.00 &   0.00 &   0.00 &   0.00 \\
    5 &  -1.49 & 118.18 &  -0.18 &   0.68 &   0.22 &  -0.01 &   0.00 &   0.00 &   0.01 &   0.00 \\
   10 &  -3.05 & 102.62 &  -0.68 &   1.17 &   0.85 &  -0.07 &   0.01 &   0.00 &   0.08 &   0.01 \\
   25 &  -6.35 &  80.63 &  -2.80 &   1.81 &   3.72 &  -0.42 &   0.05 &  -0.05 &   0.55 &   0.17 \\
   50 &  -9.73 &  62.73 &  -6.44 &   2.13 &   8.97 &  -1.11 &   0.33 &  -0.26 &   1.61 &   0.72 \\
  100 & -14.43 &  43.06 & -12.25 &   2.45 &  17.22 &  -2.15 &   1.45 &  -0.94 &   3.49 &   2.17 \\
  150 & -18.33 &  30.47 & -16.50 &   2.79 &  22.09 &  -2.87 &   2.70 &  -1.76 &   4.83 &   3.64 \\
  200 & -21.77 &  20.95 & -19.68 &   3.18 &  24.51 &  -3.48 &   3.70 &  -2.60 &   5.76 &   4.99 \\
  250 & -24.84 &  13.21 & -22.12 &   3.60 &  25.36 &  -4.08 &   4.31 &  -3.39 &   6.40 &   6.18 \\
  300 & -27.57 &   6.65 & -24.03 &   4.00 &  25.21 &  -4.73 &   4.54 &  -4.09 &   6.83 &   7.21 \\
  350 & -30.00 &   0.92 & -25.53 &   4.38 &  24.44 &  -5.45 &   4.44 &  -4.71 &   7.14 &   8.07 \\
\end{tabular}
\label{tab_phnp0}
\end{table}

\pagebreak

\begin{table}
\caption{Scattering lengths ($a$) and effective ranges ($r$) in units of fm.}
\begin{tabular}{rddl}
                   
 & CD-Bonn 
 & Experiment
 & Reference(s)
\\
\hline 
\\
\multicolumn{4}{c}{\boldmath $^1S_0$} \\
$a_{pp}^C$ & --7.8154  &--7.8149 $\pm$ 0.0029 & \cite{SES83}\\
$r_{pp}^C$ &  2.773   & 2.769 $\pm$ 0.014  & \cite{SES83}\\
$a_{pp}^N$ &--17.4602  &                   &             \\
$r_{pp}^N$ &  2.845   &                   &             \\
$a_{nn}^N$ &--18.9680  &--18.9 $\pm$ 0.4    & \cite{How98,Gon99}\\
$r_{nn}^N$ &  2.819   & 2.75 $\pm$ 0.11   & \cite{MNS90}\\
$a_{np}  $ &--23.7380  &--23.740 $\pm$ 0.020  & \cite{KMS84}\\
$r_{np}  $ &  2.671   &( 2.77  $\pm$ 0.05 )  & \cite{KMS84}\\
\\
\multicolumn{4}{c}{\boldmath $^3S_1$} \\
$a_t$ &  5.4196  & 5.419 $\pm$ 0.007  & \cite{KMS84}\\
$r_t$ &  1.751   & 1.753 $\pm$ 0.008  & \cite{KMS84}\\
\end{tabular}
\label{tab_lep}
\end{table}

\pagebreak

\begin{table}
\caption{After-1992 $pp$ data below 350 MeV included in the 1999 $pp$ database.
Error refers to the normalization error. This table contains 1113 observables
and 32 normalizations resulting in a total of 1145 data.}
\begin{tabular}{lcccr}
                   
   $T_{lab}$ (MeV)
 & \# observable
 & Error (\%)
 & Institution(s)
 & Reference

\\
\hline                                                                          
25.68 & 8 $D$ & 1.3 & Erlangen, Z\"urich, PSI & \cite{Kr94} \\
25.68 & 6 $R$ & 1.3 & Erlangen, Z\"urich, PSI & \cite{Kr94} \\
25.68 & 2 $A$ & 1.3 & Erlangen, Z\"urich, PSI & \cite{Kr94} \\
197.4 & 41 $P$ & 1.3 &  Wisconsin, IUCF & \cite{Rat98} \\
197.4 & 41 $A_{xx}$ & 2.5 &  Wisconsin, IUCF & \cite{Rat98} \\
197.4 & 41 $A_{yy}$ & 2.5 &  Wisconsin, IUCF & \cite{Rat98} \\
197.4 & 41 $A_{xz}$ & 2.5 &  Wisconsin, IUCF & \cite{Rat98} \\
197.4 & 39 $A_{zz}$ & 2.0 &  Wisconsin, IUCF & \cite{Lo00} \\
197.8 & 14 $P$ & 1.3 &  Wisconsin, IUCF & \cite{Ha97} \\
197.8 & 14 $A_{xx}$ & 2.4 &  Wisconsin, IUCF & \cite{Ha97} \\
197.8 & 14 $A_{yy}$ & 2.4 &  Wisconsin, IUCF & \cite{Ha97} \\
197.8 & 14 $A_{xz}$ & 2.4 &  Wisconsin, IUCF & \cite{Ha97} \\
197.8 & 10 $D$ & None &  IUCF & \cite{Wi99} \\
197.8 &  5 $R$ & None &  IUCF & \cite{Wi99} \\
197.8 &  5 $R'$ & None &  IUCF & \cite{Wi99} \\
197.8 &  5 $A$ & None &  IUCF & \cite{Wi99} \\
197.8 &  5 $A'$ & None &  IUCF & \cite{Wi99} \\
250.0 & 41 $P$ & 1.3 &  IUCF, Wisconsin & \cite{Pr98} \\
250.0 & 41 $A_{xx}$ & 2.5 &  IUCF, Wisconsin & \cite{Pr98} \\
250.0 & 41 $A_{yy}$ & 2.5 &  IUCF, Wisconsin & \cite{Pr98} \\
250.0 & 41 $A_{xz}$ & 2.5 &  IUCF, Wisconsin & \cite{Pr98} \\
280.0 & 41 $P$ & 1.3 &  IUCF, Wisconsin & \cite{Pr98} \\
280.0 & 41 $A_{xx}$ & 2.5 &  IUCF, Wisconsin & \cite{Pr98} \\
280.0 & 41 $A_{yy}$ & 2.5 &  IUCF, Wisconsin & \cite{Pr98} \\
280.0 & 41 $A_{xz}$ & 2.5 &  IUCF, Wisconsin & \cite{Pr98} \\
294.4 & 40 $P$ & 1.3 &  IUCF, Wisconsin & \cite{Pr98} \\
294.4 & 40 $A_{xx}$ & 2.5 &  IUCF, Wisconsin & \cite{Pr98} \\
294.4 & 40 $A_{yy}$ & 2.5 &  IUCF, Wisconsin & \cite{Pr98} \\
294.4 & 40 $A_{xz}$ & 2.5 &  IUCF, Wisconsin & \cite{Pr98} \\
310.0 & 40 $P$ & 1.3 &  IUCF, Wisconsin & \cite{Pr98} \\
310.0 & 40 $A_{xx}$ & 2.5 &  IUCF, Wisconsin & \cite{Pr98} \\
310.0 & 40 $A_{yy}$ & 2.5 &  IUCF, Wisconsin & \cite{Pr98} \\
310.0 & 40 $A_{xz}$ & 2.5 &  IUCF, Wisconsin & \cite{Pr98} \\
350.0 & 40 $P$ & 1.3 &  IUCF, Wisconsin & \cite{Pr98} \\
350.0 & 40 $A_{xx}$ & 2.5 &  IUCF, Wisconsin & \cite{Pr98} \\
350.0 & 40 $A_{yy}$ & 2.5 &  IUCF, Wisconsin & \cite{Pr98} \\
350.0 & 40 $A_{xz}$ & 2.5 &  IUCF, Wisconsin & \cite{Pr98} \\
\end{tabular}
\label{tab_ppdat}
\end{table}

\pagebreak

\begin{table}
\caption{After-1992 $np$ data below 350 MeV included in the 1999 $np$ database.
Error refers to the normalization error. This table contains 524 observables
and 20 normalizations resulting in a total of 544 data.}
\begin{tabular}{ccccr}
                   
   $T_{lab}$ (MeV)
 & \# observable
 & Error (\%)
 & Institution(s)
 & Reference

\\
\hline                                                                          
3.65--11.6 & 9 $\Delta \sigma_T$ & None & TUNL & \cite{Wi95} \\
4.98--19.7 & 6 $\Delta \sigma_L$ & None & TUNL & \cite{Ra99} \\
4.98--17.1 & 5 $\Delta \sigma_T$ & None & TUNL & \cite{Ra99} \\
14.11      & 6 $\sigma$ & 0.7  & T\"ubingen & \cite{BM97} \\
15.8       & 1 $D_t$ & None & Bonn & \cite{Cl98} \\
16.2       & 1 $\Delta \sigma_T$ & None & Prague & \cite{Br96} \\
16.2       & 1 $\Delta \sigma_L$ & None & Prague & \cite{Br97} \\
175.26     & 84 $P$ & Float$^a$ & TRIUMF & \cite{Da96} \\
203.15     & 100 $P$ & 4.7 & TRIUMF & \cite{Da96} \\
217.24     & 100 $P$ & 4.5 & TRIUMF & \cite{Da96} \\
260.0     & 8 $R_t$ & 3.0 & PSI & \cite{Ah98} \\
260.0     & 8 $A_t$ & 3.0 & PSI & \cite{Ah98} \\
260.0     & 3 $A_t$ & 3.0 & PSI & \cite{Ah98} \\
260.0     & 8 $D_t$ & 3.0 & PSI & \cite{Ah98} \\
260.0     & 3 $D_t$ & 3.0 & PSI & \cite{Ah98} \\
260.0     & 8 $P  $ & 2.0 & PSI & \cite{Ah98} \\
260.0     & 3 $P  $ & 2.0 & PSI & \cite{Ah98} \\
261.00     &  88 $P$ & 4.1 & TRIUMF & \cite{Da96} \\
312.0     &  24 $P$ & 4.0 & SATURNE& \cite{Ba93} \\
312.0     &  11 $A_{zz}$ & 4.0 & SATURNE& \cite{Ba94} \\
318.0     & 8 $R_t$ & 3.0 & PSI & \cite{Ah98} \\
318.0     & 8 $A_t$ & 3.0 & PSI & \cite{Ah98} \\
318.0     & 5 $A_t$ & 3.0 & PSI & \cite{Ah98} \\
318.0     & 8 $D_t$ & 3.0 & PSI & \cite{Ah98} \\
318.0     & 5 $D_t$ & 3.0 & PSI & \cite{Ah98} \\
318.0     & 8 $P  $ & 2.0 & PSI & \cite{Ah98} \\
318.0     & 5 $P  $ & 2.0 & PSI & \cite{Ah98} \\
\end{tabular}
$^a$ This data set is floated because all current phase shift analyses and
$np$ potentials predict a norm that is about 4 standard deviations off
the experimental normalization error of 4.9\%, 
\label{tab_npdat}
\end{table}

\pagebreak

\begin{table}
\caption{$\chi^2$/datum for the CD-Bonn potential, the Nijmegen phase
shift analysis~\protect\cite{Sto93}, 
and the Argonne $V_{18}$ potential~\protect\cite{WSS95}
in regard to various databases discussed in the text.}
\begin{tabular}{lccc}
                   
 & CD-Bonn 
 & Nijmegen 
 & Argonne 
\\

 & potential
 & phase shift analysis
 & $V_{18}$ potential
\\
\hline 
\hline 
\multicolumn{4}{c}{\bf proton-proton data} \\
1992 $pp$ database (1787 data) & 1.00 & 1.00 & 1.10 \\
After-1992 $pp$ data (1145 data) & 1.03 & 1.24 & 1.74 \\
1999 $pp$ database (2932 data) & 1.01 & 1.09 & 1.35 \\
\hline 
\multicolumn{4}{c}{\bf neutron-proton data} \\
1992 $np$ database (2514 data) & 1.03 & 0.99 & 1.08 \\
After-1992 $np$ data (544 data) & 0.99 & 0.99 & 1.02 \\
1999 $np$ database (3058 data) & 1.02 & 0.99 & 1.07 \\
\hline 
\multicolumn{4}{c}{\bf {\boldmath $pp$} and {\boldmath $np$} data} \\
1992 $NN$ database (4301 data) & 1.02 & 0.99 & 1.09 \\
1999 $NN$ database (5990 data) & 1.02 & 1.04 & 1.21 \\
\end{tabular}
\label{tab_chi2}
\end{table}

\pagebreak

\begin{table}
\caption{Deuteron properties.}
\begin{tabular}{lllc}
           & CD-Bonn & Empirical & Reference(s) \\
\hline
Binding energy $B_d$ (MeV) 
& 2.224575 & 2.224575(9) & \cite{LA82} \\
Deuteron effective range $\rho_d=\rho(-B_d,-B_d)$ (fm)
& 1.765&  1.765(9) 
& \cite{KMS84,STS95,ER83}\\
Asymptotic $S$ state $A_S$ (fm$^{-1/2}$)
&0.8846& 0.8846(9) 
& \cite{STS95,ER83}\\
Asymptotic $D/S$ state $\eta$        
& 0.0256& 0.0256(4) & \cite{RK90}\\
Matter radius $r_d$ (fm)    
& 1.966 & 1.971(6) & \cite{MSZ95}\\
Quadrupole moment $Q_d$ (fm$^2$)
& 0.270$^a$ & 0.2859(3) & \cite{BC79,ER83}\\
$D$-state probability $P_D$ (\%)    & 4.85  \\
\end{tabular}
$^a$ Without meson current contributions and relativistic corrections.
\label{tab_deu}
\end{table}

\pagebreak

\begin{table}
\caption{Parameters of the scalar isoscalar bosons, $\sigma_1$
and $\sigma_2$, for the $pp$ $T=1$ potential. An asterix denotes
the default which are the $^1S_0$ parameters.
The boson masses $m_{\sigma_1}$ and $m_{\sigma_2}$ are in units of MeV.}
\begin{tabular}{ccc}
              
 & $g^2_{\sigma_1}/4\pi$
  ($m_{\sigma_1}$)
 & $g^2_{\sigma_2}/4\pi$
  ($m_{\sigma_2}$)
\\
\hline 
$\bbox{^1S_0}$ & $\bbox{4.24591}$ $\bbox{(452)}$ & $\bbox{17.61}$ $\bbox{(1225)}$  \\
$^3P_0$ & 7.866 (560) & $\ast$ ($\ast$) \\
$^3P_1$ & 2.303 (424) & $\ast$ ($\ast$) \\
$^3P_2$ & 4.166 (470) & 24.80 ($\ast$) \\
$^1D_2$ & 2.225 (400) & 190.7 ($\ast$) \\
$^3F_2$, $^3F_3$ & 1.5 ($\ast$) & 56.21, 74.44 (793) \\
$^3F_4$, $^3H_4$ & 3.8 ($\ast$) & $\ast$ ($\ast$) \\
$^1G_4$ & $\ast$ ($\ast$) & --- \\
$^3H_5$ & $\ast$ ($\ast$) & --- \\
\end{tabular}
\label{tab_par1}
\end{table}

\vspace*{5cm}

\begin{table}
\caption{Parameters of the scalar isoscalar bosons, $\sigma_1$
and $\sigma_2$, for the $T=0$ $np$ potential. An asterix denotes
the default which are the $^3S_1$ parameters.
The boson masses $m_{\sigma_1}$ and $m_{\sigma_2}$ are in units of MeV.}
\begin{tabular}{ccc}
              
 & $g^2_{\sigma_1}/4\pi$
  ($m_{\sigma_1}$)
 & $g^2_{\sigma_2}/4\pi$
  ($m_{\sigma_2}$)
\\
\hline 
$\bbox{^3S_1}$ & $\bbox{0.51673}$ $\bbox{(350)}$ & $\bbox{14.01164}$ $\bbox{(793)}$  \\
$^1P_1$, $^3D_2$ & 0.81, 0.53 ($\ast$) & 71.5, 154.5 (1225) \\
$^3D_1$ & 0.575 ($\ast$) & --- \\
$^3D_3$ & 3.4 (452) & --- \\
$^1F_3$ & 0.73 ($\ast$) & --- \\
$^3G_3$ & 0.29 ($\ast$) & --- \\
$^3G_4$ & 0.62 ($\ast$) & --- \\
$^3G_5$, $^3I_5$ & 0.96 ($\ast$) & --- \\
$^1H_5$ & $\ast$ ($\ast$) & --- \\
\end{tabular}
\label{tab_par2}
\end{table}

\pagebreak

\begin{table}
\caption{Coupling constants of the scalar isoscalar bosons, $\sigma_1$
and $\sigma_2$, for the $T=1$ $np$ and $nn$ potentials. 
Note that these are not free parameters
(except for $^1S_0$ $np$).
The boson masses are the same as for the $pp$ $T=1$ potential 
(Table~\ref{tab_par1}).}
\begin{tabular}{ccccc}
   
 & \multicolumn{2}{c}{\bf --- neutron-proton ---} 
 & \multicolumn{2}{c}{\bf --- neutron-neutron ---} 
\\
              
 & $g^2_{\sigma_1}/4\pi$
 & $g^2_{\sigma_2}/4\pi$
 & $g^2_{\sigma_1}/4\pi$
 & $g^2_{\sigma_2}/4\pi$
\\
\hline 
$^1S_0$ & 3.96451  & 22.50007
        & 4.26338 & 17.54   \\
$^3P_0$ & 7.866  & 5.8 
        & 7.892 & 16.747  \\
$^3P_1$ & 2.346  & 19.22 
        & 2.326 & 17.61  \\
$^3P_2$ & 4.194  & 24.562 
        & 4.180 & 24.737  \\
$^1D_2$ & 2.236  & 189.7 
        & 2.241 & 190.7  \\
$^3F_2$, $^3F_3$ & 1.573, 1.53  & 56.21, 74.85 
                 & 1.522, 1.53 & 56.28, 74.44  \\
$^3F_4$, $^3H_4$ & 3.8115, 3.85  & 17.61  
                 & 3.81, 3.83  & 17.61   \\
$^1G_4$ & 4.27591 & --- 
        & 4.284  & --- \\
$^3H_5$ & 4.24591 & --- & 4.24591 & --- \\
\end{tabular}
\label{tab_par3}
\end{table}

\pagebreak

\begin{table}
\caption{Deuteron wave functions.}
\begin{tabular}{rrrrrr}
\multicolumn{1}{c}{$r$ (fm)}  &
\multicolumn{1}{c}{$u(r)$ (fm$^{-1/2}$)} &
\multicolumn{1}{c}{$w(r)$ (fm$^{-1/2}$)} &
\multicolumn{1}{c}{$r$ (fm)}  &
\multicolumn{1}{c}{$u(r)$ (fm$^{-1/2}$)} &
\multicolumn{1}{c}{$w(r)$ (fm$^{-1/2}$)} 
\\ \hline 
 $   0.100E-01$ & $    0.304061E-02$ & $   -0.137276E-05$ & $   0.270E+01$ & $    0.457550E+00$ & $    0.107219E+00$  \\
 $   0.200E-01$ & $    0.607313E-02$ & $   -0.895215E-05$ & $   0.280E+01$ & $    0.448837E+00$ & $    0.102572E+00$  \\
 $   0.300E-01$ & $    0.909444E-02$ & $   -0.249495E-04$ & $   0.290E+01$ & $    0.440064E+00$ & $    0.980768E-01$  \\
 $   0.400E-01$ & $    0.121048E-01$ & $   -0.492312E-04$ & $   0.300E+01$ & $    0.431275E+00$ & $    0.937453E-01$  \\
 $   0.500E-01$ & $    0.151065E-01$ & $   -0.804275E-04$ & $   0.320E+01$ & $    0.413778E+00$ & $    0.855923E-01$  \\
 $   0.600E-01$ & $    0.181029E-01$ & $   -0.116610E-03$ & $   0.340E+01$ & $    0.396552E+00$ & $    0.781235E-01$  \\
 $   0.700E-01$ & $    0.210984E-01$ & $   -0.155526E-03$ & $   0.360E+01$ & $    0.379727E+00$ & $    0.713176E-01$  \\
 $   0.800E-01$ & $    0.240975E-01$ & $   -0.194813E-03$ & $   0.380E+01$ & $    0.363387E+00$ & $    0.651366E-01$  \\
 $   0.900E-01$ & $    0.271050E-01$ & $   -0.232058E-03$ & $   0.400E+01$ & $    0.347583E+00$ & $    0.595344E-01$  \\
 $   0.100E+00$ & $    0.301255E-01$ & $   -0.264871E-03$ & $   0.420E+01$ & $    0.332343E+00$ & $    0.544623E-01$  \\
 $   0.200E+00$ & $    0.620193E-01$ & $    0.155643E-03$ & $   0.440E+01$ & $    0.317678E+00$ & $    0.498721E-01$  \\
 $   0.300E+00$ & $    0.993876E-01$ & $    0.335071E-02$ & $   0.460E+01$ & $    0.303592E+00$ & $    0.457178E-01$  \\
 $   0.400E+00$ & $    0.143869E+00$ & $    0.108936E-01$ & $   0.480E+01$ & $    0.290078E+00$ & $    0.419565E-01$  \\
 $   0.500E+00$ & $    0.194545E+00$ & $    0.235574E-01$ & $   0.500E+01$ & $    0.277126E+00$ & $    0.385487E-01$  \\
 $   0.600E+00$ & $    0.248454E+00$ & $    0.409068E-01$ & $   0.520E+01$ & $    0.264721E+00$ & $    0.354587E-01$  \\
 $   0.700E+00$ & $    0.301841E+00$ & $    0.612808E-01$ & $   0.540E+01$ & $    0.252849E+00$ & $    0.326540E-01$  \\
 $   0.800E+00$ & $    0.351374E+00$ & $    0.824033E-01$ & $   0.560E+01$ & $    0.241491E+00$ & $    0.301056E-01$  \\
 $   0.900E+00$ & $    0.394806E+00$ & $    0.102176E+00$ & $   0.580E+01$ & $    0.230629E+00$ & $    0.277874E-01$  \\
 $   0.100E+01$ & $    0.431072E+00$ & $    0.119165E+00$ & $   0.600E+01$ & $    0.220245E+00$ & $    0.256761E-01$  \\
 $   0.110E+01$ & $    0.460046E+00$ & $    0.132683E+00$ & $   0.650E+01$ & $    0.196252E+00$ & $    0.211717E-01$  \\
 $   0.120E+01$ & $    0.482213E+00$ & $    0.142633E+00$ & $   0.700E+01$ & $    0.174846E+00$ & $    0.175676E-01$  \\
 $   0.130E+01$ & $    0.498370E+00$ & $    0.149285E+00$ & $   0.750E+01$ & $    0.155759E+00$ & $    0.146616E-01$  \\
 $   0.140E+01$ & $    0.509415E+00$ & $    0.153089E+00$ & $   0.800E+01$ & $    0.138747E+00$ & $    0.123010E-01$  \\
 $   0.150E+01$ & $    0.516222E+00$ & $    0.154545E+00$ & $   0.850E+01$ & $    0.123589E+00$ & $    0.103699E-01$  \\
 $   0.160E+01$ & $    0.519579E+00$ & $    0.154136E+00$ & $   0.900E+01$ & $    0.110084E+00$ & $    0.877993E-02$  \\
 $   0.170E+01$ & $    0.520158E+00$ & $    0.152287E+00$ & $   0.950E+01$ & $    0.980525E-01$ & $    0.746281E-02$  \\
 $   0.180E+01$ & $    0.518524E+00$ & $    0.149356E+00$ & $   0.100E+02$ & $    0.873354E-01$ & $    0.636565E-02$  \\
 $   0.190E+01$ & $    0.515138E+00$ & $    0.145638E+00$ & $   0.105E+02$ & $    0.777891E-01$ & $    0.544705E-02$  \\
 $   0.200E+01$ & $    0.510374E+00$ & $    0.141367E+00$ & $   0.110E+02$ & $    0.692859E-01$ & $    0.467438E-02$  \\
 $   0.210E+01$ & $    0.504533E+00$ & $    0.136728E+00$ & $   0.115E+02$ & $    0.617120E-01$ & $    0.402170E-02$  \\
 $   0.220E+01$ & $    0.497856E+00$ & $    0.131864E+00$ & $   0.120E+02$ & $    0.549660E-01$ & $    0.346826E-02$  \\
 $   0.230E+01$ & $    0.490539E+00$ & $    0.126886E+00$ & $   0.125E+02$ & $    0.489573E-01$ & $    0.299734E-02$  \\
 $   0.240E+01$ & $    0.482736E+00$ & $    0.121877E+00$ & $   0.130E+02$ & $    0.436055E-01$ & $    0.259535E-02$  \\
 $   0.250E+01$ & $    0.474573E+00$ & $    0.116901E+00$ & $   0.135E+02$ & $    0.388386E-01$ & $    0.225120E-02$  \\
 $   0.260E+01$ & $    0.466150E+00$ & $    0.112004E+00$ & $   0.140E+02$ & $    0.345929E-01$ & $    0.195582E-02$ 
\end{tabular}
\label{tab_dwaves}
\end{table}

\pagebreak

\begin{table}
\caption{Coefficients for the parametrized
 deuteron wave functions ($n=11$).}
\begin{tabular}{rrr}
   $j$     &
\multicolumn{1}{c}{$C_{j}$ (fm$^{-1/2}$)}  &
\multicolumn{1}{c}{$D_{j}$ (fm$^{-1/2}$)}
\\ \hline 
      1  &$   0.88472985D+00 $&$     0.22623762D-01$\\
      2  &$  -0.26408759D+00 $&$    -0.50471056D+00$\\
      3  &$  -0.44114404D-01 $&$     0.56278897D+00$\\
      4  &$  -0.14397512D+02 $&$    -0.16079764D+02$\\
      5  &$   0.85591256D+02 $&$     0.11126803D+03$\\
      6  &$  -0.31876761D+03 $&$    -0.44667490D+03$\\
      7  &$   0.70336701D+03 $&$     0.10985907D+04$\\
      8  &$  -0.90049586D+03 $&$    -0.16114995D+04$\\
      9  &$   0.66145441D+03 $&
           \multicolumn{1}{c}{Eq.~(\ref{eq_dbound2})}\\
     10  &$  -0.25958894D+03 $&
           \multicolumn{1}{c}{Eq.~(\ref{eq_dbound2})}\\
     11  & \multicolumn{1}{c}{Eq.~(\ref{eq_dbound1})}
         & \multicolumn{1}{c}{Eq.~(\ref{eq_dbound2})}
\end{tabular}
\label{tab_dwpar}
\end{table}

\pagebreak

\begin{figure}
\vspace*{-4.0cm}
\hspace*{-1.0cm}
\epsfig{figure=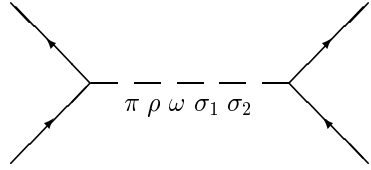,width=15cm}
\vspace*{-11.0cm}
\caption{One-boson exchange Feynman diagrams that define 
the CD-Bonn $NN$ potential.}
\label{fig_OBE}
\end{figure}


\begin{figure}
\hspace*{2.0cm}
\epsfig{figure=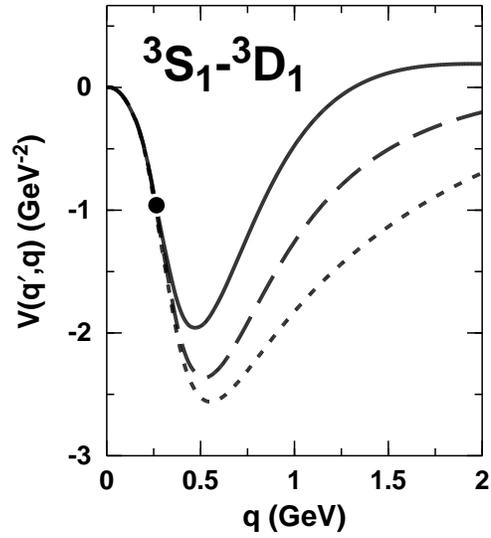,height=14cm}
\vspace*{-3.0cm}
\caption{Half off-shell $^3S_1$$-$$^3D_1$ potential.
The on-shell momentum $q'$ is held fixed at 265 MeV
(equivalent to 150 MeV lab.\ energy),
while the off-shell momentum $q$ runs from zero
to 2000 MeV.
The on-shell point ($q=265$ MeV) is marked by a solid dot.
The solid curve is the relativistic OBE amplitude of $\pi
+ \rho$ exchange.
When the relativistic
OPE amplitude, Eq.~(\ref{eq_OPEPrel}), is replaced by the static/local
approximation, Eq.~(\ref{eq_OPEPnr}),
the dashed curve is obtained, and
when this approximation is also used for the one-$\rho$
exchange, the dotted curve results.}
\label{fig_pot3SD1}
\end{figure}

\pagebreak

\begin{figure}
\vspace*{-3.0cm}
\hspace*{-1.7cm}
\epsfig{figure=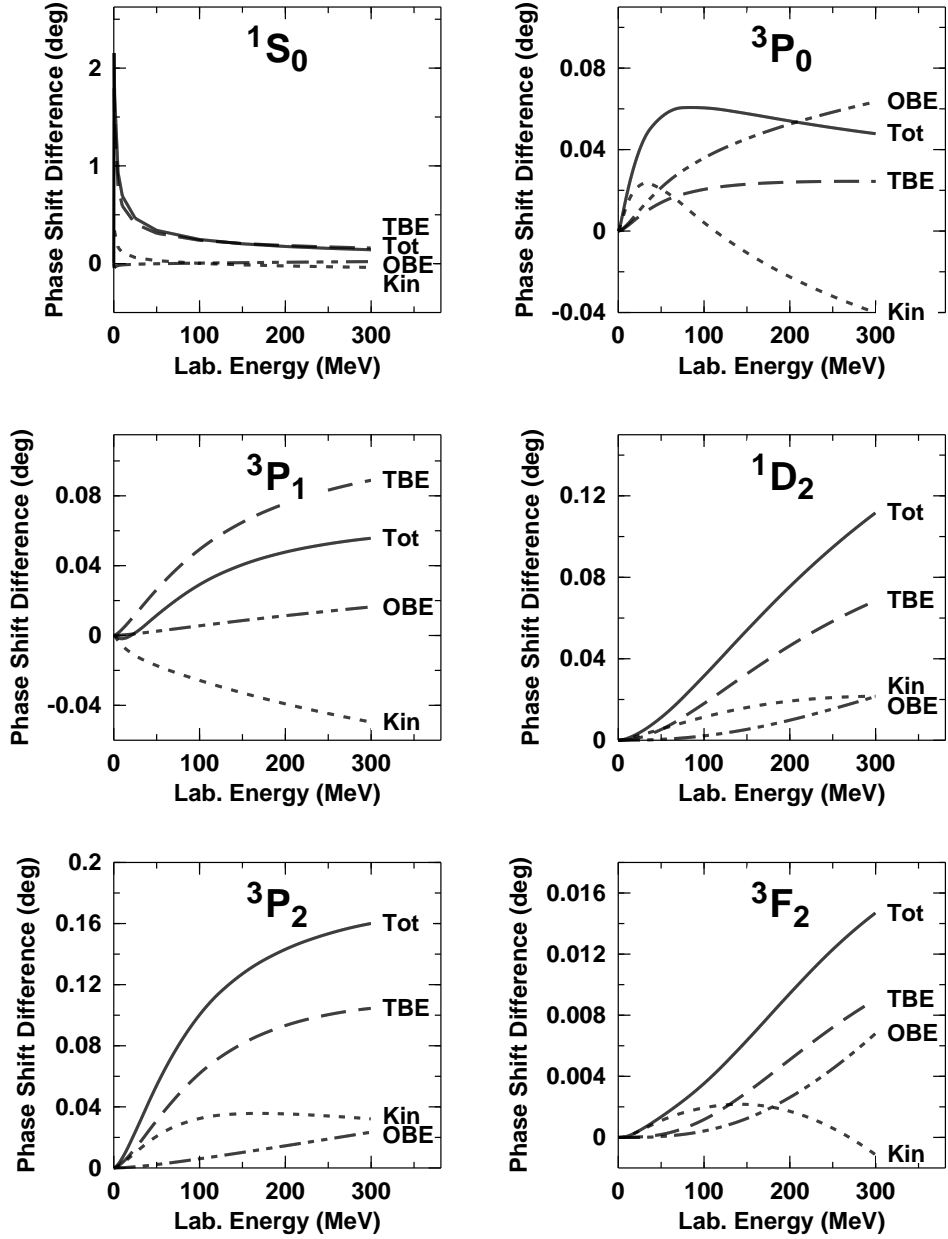,height=24cm}
\vspace*{-3.0cm}
\caption{
Differences $\delta_{nn} - \delta_{pp}$ 
due to the impact of
nucleon mass splitting on kinematics (dotted line labeled `Kin'),
one-boson exchange diagrams (dashed double-dotted, OBE), 
and two-boson exchanges (dashed, TBE). The solid line (`Tot') represents
the total. Notice that each frame has a different scale.}
\label{fig_CSB}
\end{figure}

\pagebreak

\begin{figure}
\vspace*{-3.0cm}
\hspace*{-1.7cm}
\epsfig{figure=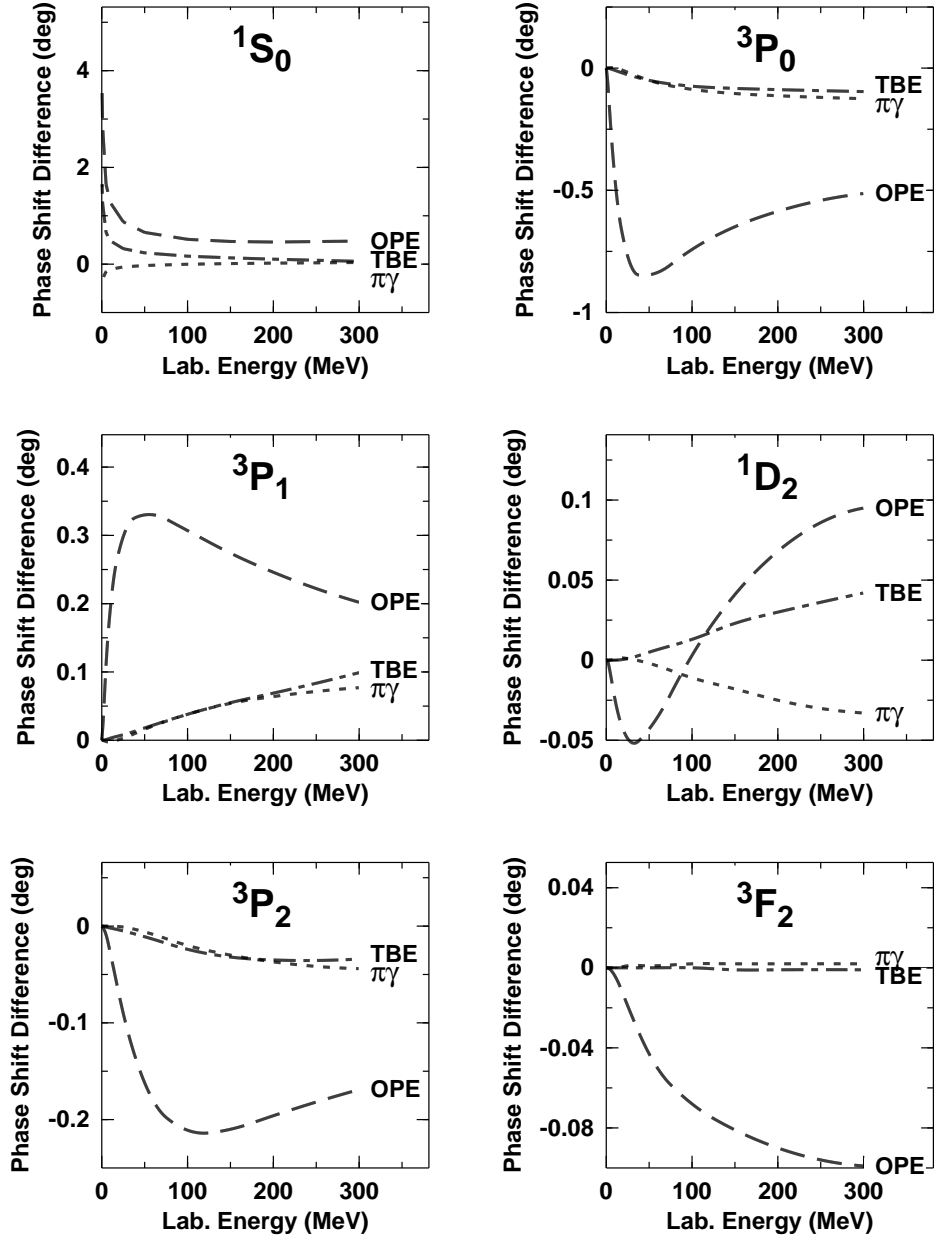,height=24cm}
\vspace*{-3.0cm}
\caption{
Differences 
$\delta_{np} - \delta_{pp}$ 
as produced by various CIB mechanisms.
Shown are the contributions from OPE (dashed curve),
TBE (dashed-dotted), and irreducible $\pi\gamma$ exchange (dotted).}
\label{fig_CIB1}
\end{figure}

\pagebreak

\begin{figure}
\vspace*{-3.0cm}
\hspace*{-1.7cm}
\epsfig{figure=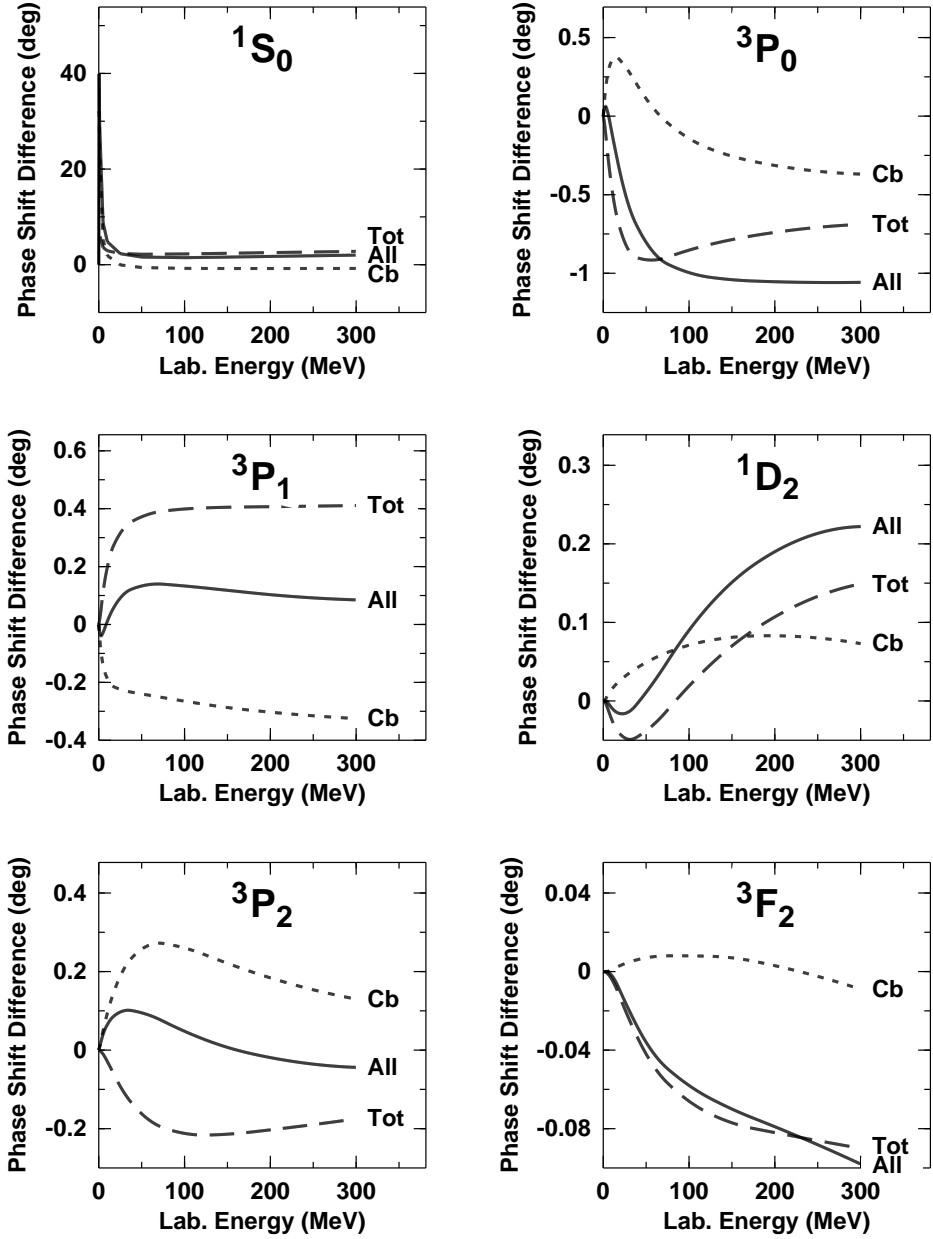,height=24cm}
\vspace*{-3.0cm}
\caption{
The difference 
$\delta_{np} - \delta_{pp}$ 
due to the charge-dependence of the strong force (dashed curve labeled `Tot')
and $(\delta_{pp}-\delta_{pp}^C)$
due to the Coulomb force (dotted, Cb). The sum of both is represented
by the solid line labeled `All'.}
\label{fig_CIB2}
\end{figure}

\pagebreak

\begin{figure}
\vspace*{-3.0cm}
\hspace*{-1.7cm}
\epsfig{figure=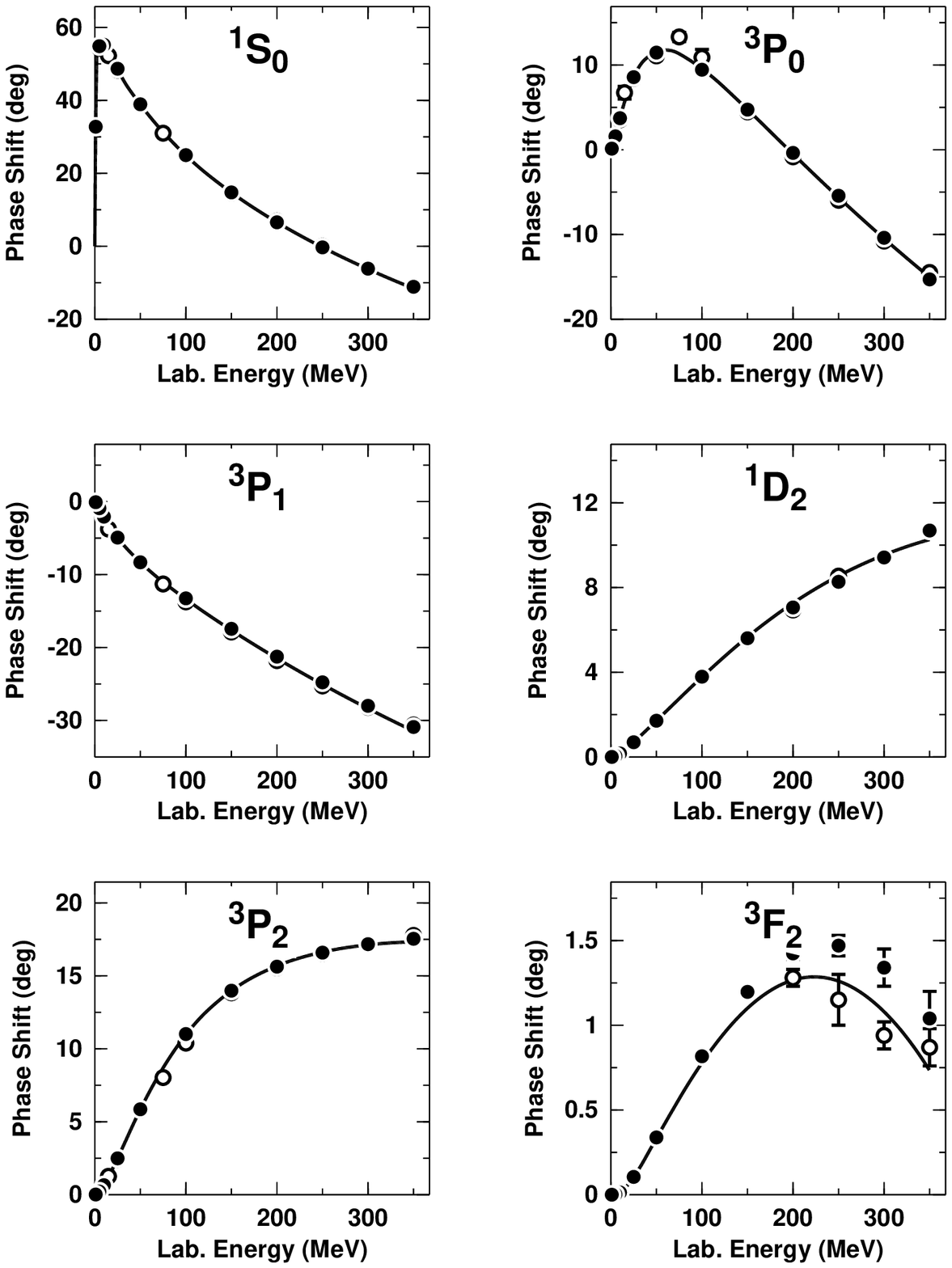,height=24cm}
\vspace*{-3.0cm}
\caption{$pp$ phase parameters in partial waves with $J\leq 4$. 
The solid line represents the predictions by the CD-Bonn potential. 
The solid dots and open circles are the results from the Nijmegen
multi-energy $pp$ phase shift analysis~\protect\cite{Sto93} and the VPI
single-energy $pp$ analysis SM99~\protect\cite{SM99}, respectively.}
\label{fig_phpp1}
\end{figure}

\pagebreak

\vspace*{-3.0cm}
\hspace*{-1.7cm}
\epsfig{figure=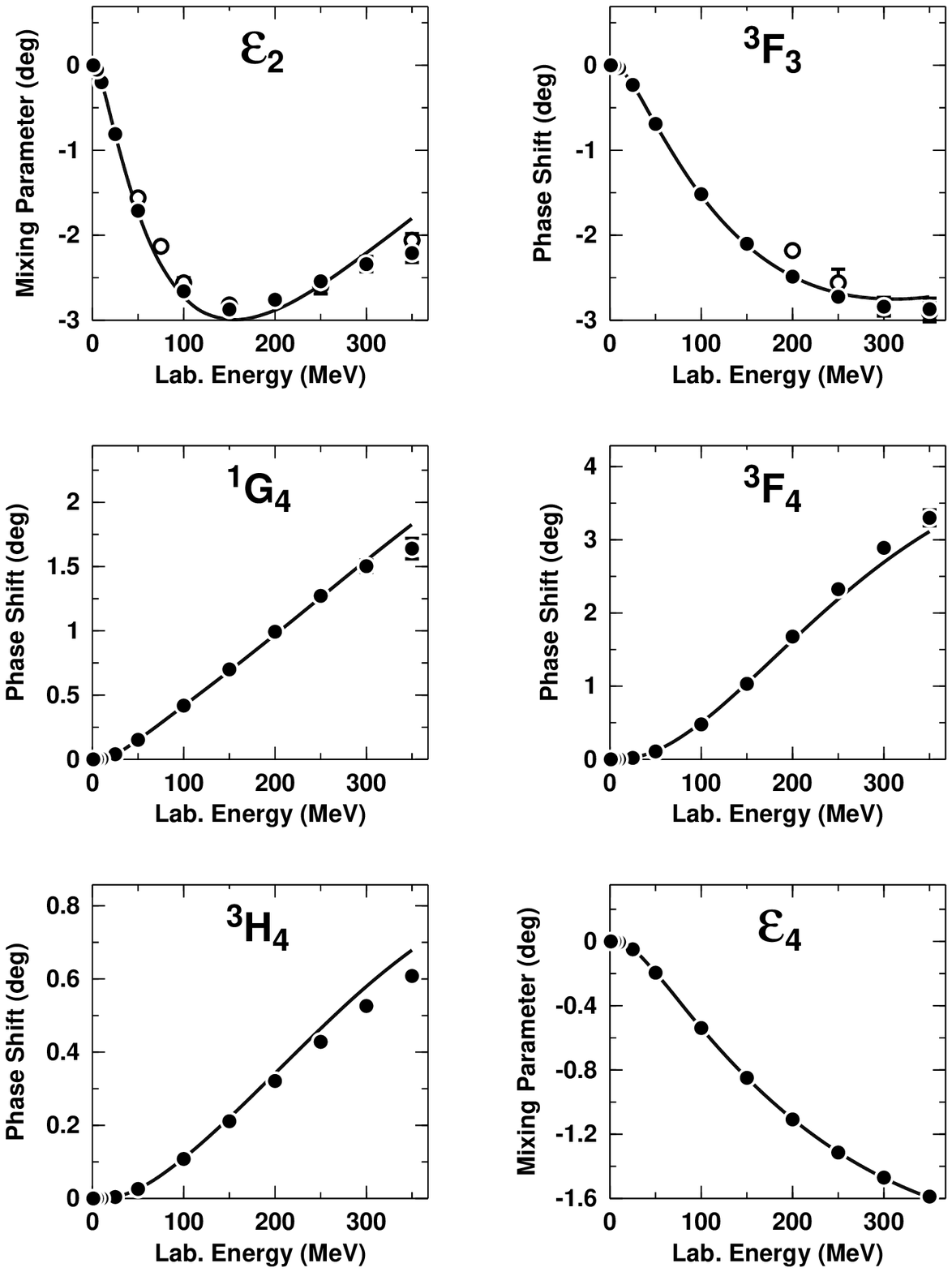,height=24cm}
\vspace*{-3.0cm}
\begin{center}
Fig.~\ref{fig_phpp1} continued.
\end{center}

\pagebreak

\begin{figure}
\vspace*{-3.0cm}
\hspace*{-1.7cm}
\epsfig{figure=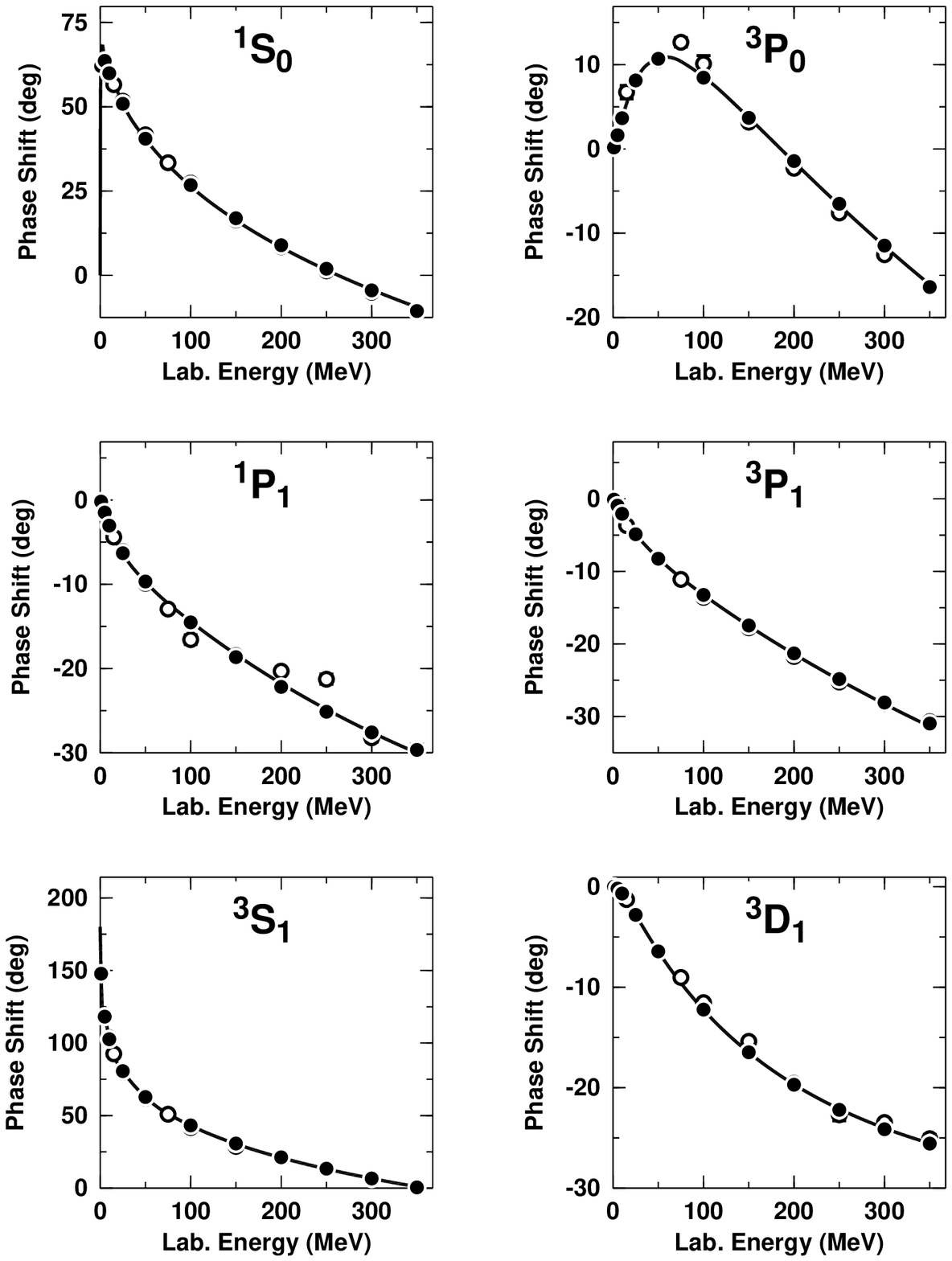,height=24cm}
\vspace*{-3.0cm}
\caption{$np$ phase parameters in partial waves with $J\leq 4$. 
The solid line represents the predictions by the CD-Bonn potential. 
The solid dots and open circles are the results from the Nijmegen
multi-energy $np$ phase shift analysis~\protect\cite{Sto93} and the VPI
single-energy $np$ analysis SM99~\protect\cite{SM99}, respectively.}
\label{fig_ph1}
\end{figure}

\pagebreak

\vspace*{-3.0cm}
\hspace*{-1.7cm}
\epsfig{figure=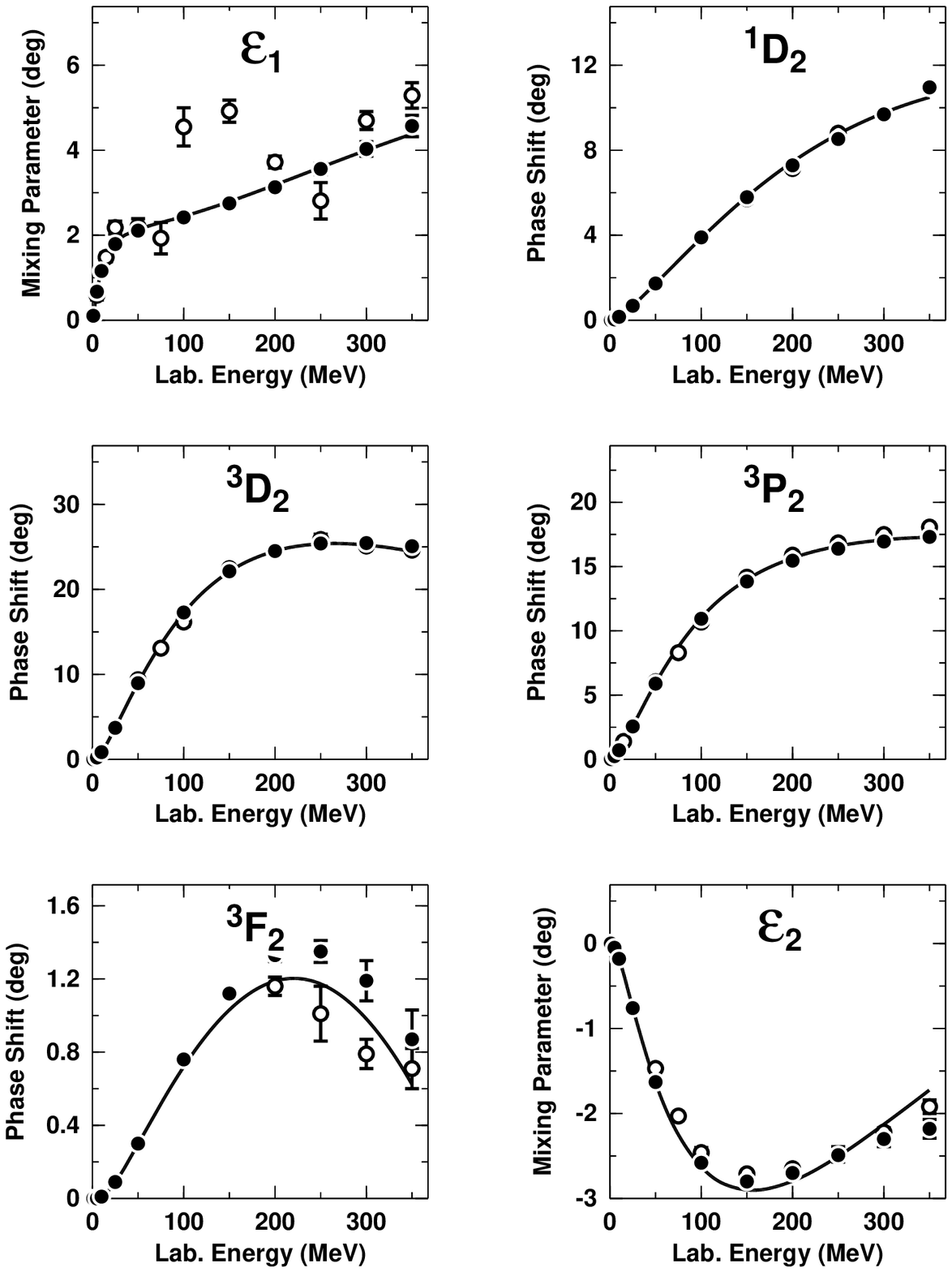,height=24cm}
\vspace*{-3.0cm}
\begin{center}
Fig.~\ref{fig_ph1} continued.
\end{center}

\pagebreak

\vspace*{-3.0cm}
\hspace*{-1.7cm}
\epsfig{figure=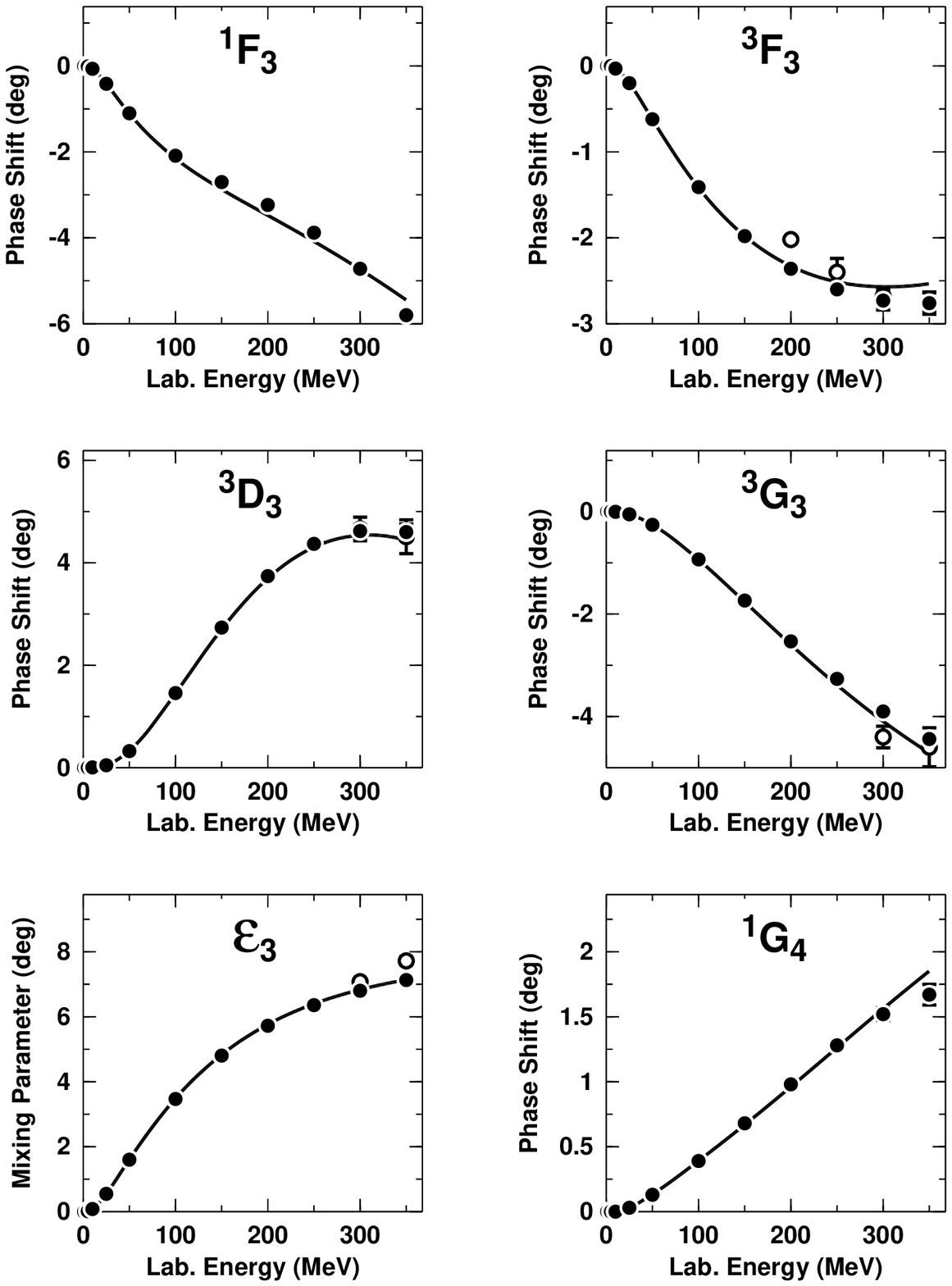,height=24cm}
\vspace*{-3.0cm}
\begin{center}
Fig.~\ref{fig_ph1} continued.
\end{center}

\pagebreak

\vspace*{-3.0cm}
\hspace*{-1.7cm}
\epsfig{figure=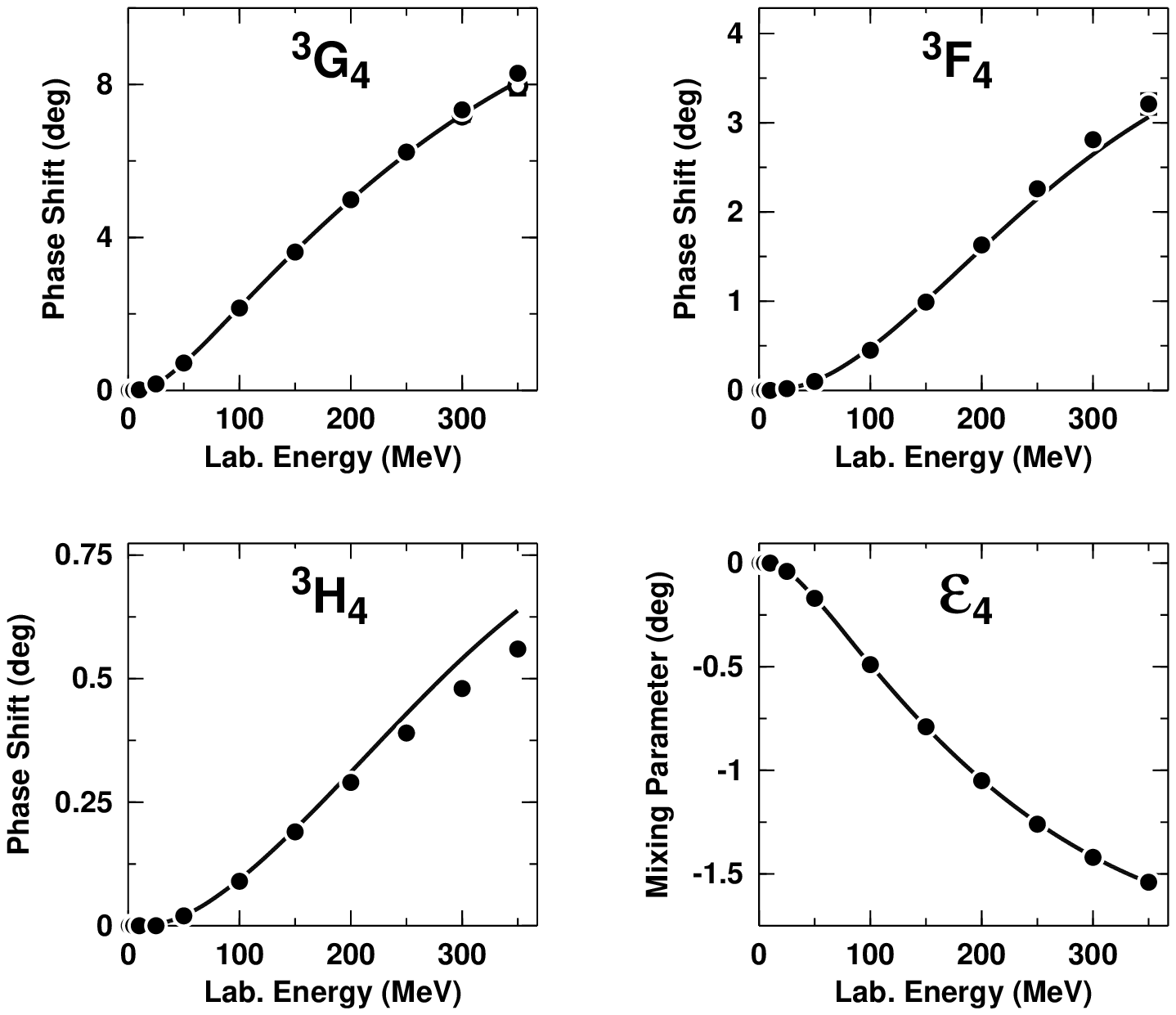,height=24cm}
\vspace*{-9.0cm}
\begin{center}
Fig.~\ref{fig_ph1} continued.
\end{center}

\pagebreak

\begin{figure}
\vspace*{-2cm}
\hspace*{2.0cm}
\epsfig{figure=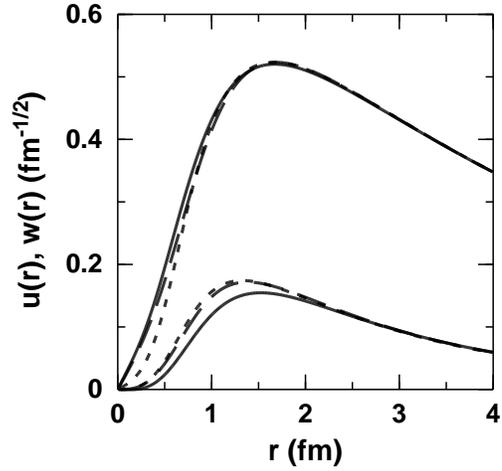,height=14cm}
\vspace*{-3.0cm}
\caption{Deuteron wave functions.
The family of large curves are $u(r)$ and the family of
small curves are $w(r)$. The solid lines represent the
wave functions generated from the CD-Bonn potential, while
the dashed and dotted lines are from the Nijmegen-I~\protect\cite{Sto94} and
Argonne $V_{18}$~\protect\cite{WSS95} potentials, respectively.}
\label{fig_dwaves}
\end{figure}

\begin{figure}
\vspace*{-2cm}
\hspace*{2.0cm}
\epsfig{figure=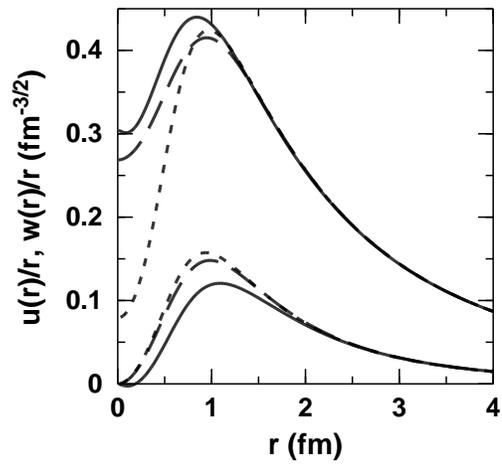,height=14cm}
\vspace*{-3.0cm}
\caption{The deuteron wave functions of Fig.~\protect\ref{fig_dwaves}
in an alternative representation.
The family of large curves are $\bbox{u(r)/r}$ and the family of
small curves are $\bbox{w(r)/r}$.}
\label{fig_dwaves2}
\end{figure}


\begin{references}
\bibitem{foot2}
The entries OPE and TBE in Tables~\ref{tab_CIB1S0} and \ref{tab_CIB}
are the predictions by the revised Bonn Full Model (that uses
$g^2_\pi/4\pi=13.6$ instead of 14.4). These predictions are typically
about 5\% smaller than what was reported in Ref.~\cite{LM98b}.
\bibitem{foot4} A user-friendly computer code for the CD-Bonn potential
can be obtained from the author upon request.
\end{references}
\end{document}